\documentclass[aip,jcp,amsmat,amssymb,amsfonts,twocolumn,superscriptaddress]{revtex4}

\usepackage{graphicx}
\usepackage{dcolumn}
\usepackage{bm}
\usepackage{hyperref}
\usepackage{color}

\newcommand{\bra}[1]{\big\langle#1\big|}
\newcommand{\ket}[1]{\big|#1\big\rangle}

\begin{document}

\title{Critical comparison of electrode models in density functional theory based quantum transport calculations}
\author{D. Jacob}
\affiliation{Max-Planck-Institut f\"ur Mikrostrukturphysik, Weinberg 2, 06120 Halle, Germany}
\author{J. J. Palacios}
\affiliation{Dpto. de F\'isica de la Materia Condensada, Universidad Aut\'onoma de Madrid, Campus de Cantoblanco, 28049 Madrid, Spain}

\date{\today}

\begin{abstract}
We study the performance of two different electrode models in  quantum transport
calculations based on density functional theory: Parametrized Bethe lattices and 
quasi-one dimensional wires or nanowires. A detailed account of implementation details in both 
cases is given. From the systematic study of nanocontacts made of 
representative metallic elements, we can conclude that parametrized electrode models 
represent an excellent compromise between computational cost and electronic structure 
definition as long as the aim is to compare with experiments where the precise atomic structure of the 
electrodes is not relevant or defined with precision. The results obtained using parametrized Bethe lattices 
are essentially similar to the ones obtained with quasi one dimensional electrodes for large enough 
sections of these, adding a natural smearing to the transmission curves that mimics the true nature of polycrystalline
electrodes.
The latter are more demanding from the computational point of view, but present the advantage of expanding 
the range of applicability of transport calculations to situations where the electrodes have a well-defined 
atomic structure, as is case for carbon nanotubes, graphene nanoribbons or semiconducting nanowires. 
All the analysis is done with the help of codes developed by the authors 
which can be found in the quantum transport toolbox {\sc Alacant} and are publicly available.
\end{abstract}


\maketitle
\section{Introduction}
\label{intro}

One of the most active research fields in Nanoscience is the one focusing on understanding and controlling the 
charge transport between bulk electrodes when these are connected by an atomic- or a molecular-size region and a bias
voltage is applied between them\cite{Agrait:pr:03},
in other words, understanding and controlling the formation and concomittant resistance of nanoscopic bridges.
More than 10 years of experimental along with theoretical work is finally taking us to an unprecedented
level of control and comprehension of these systems.  On the theoretical side, we have witnessed 
the marriage of theoretical quantum transport basics and density functional theory, giving birth
to one of the most active and fructiferous fields in theoretical nanoscience
\cite{Lang:prb:95,PhysRevLett.84.979,PhysRevB.63.121104,Palacios:prb:01,Xue:jcp:01,Palacios:prb:02,
Brandbyge:prb:02,Heurich:prl:02,Di-Ventra:prb:02,PhysRevLett.90.106801,Fujimoto:prb:03,Louis:prb:03,
Xue:prb:03:I,Xue:prb:03:II,Basch2003a,Jelinek:prb:03,Ke:prb:04,Hirose:prb:04,Xue:prb:04:70:8,Liang:prb:04,
Rocha:prb:04,Frederiksen:prl:04,Thygesen:prb:05,Evers:prb:04,Tada04,Nara2004,Ferretti:prl:05,Toher:prl:05,
Fujimoto:prb:05,Asari:prb:05,Wu05,Choi2005,Ke2005b,Ke2005a,Bagrets:prb:06,Smogunov:prb:06,Jiang06,Havu2006,
Hod2006,Rocha:prb:06,Nakamura2006a,Prociuk2006,Mera2007,Mizuseki2007,Nakamura2007,Qian2007b,Thygesen2007,
Pauly08,ISI:000257822200012,ISI:000257325900010,ISI:000256777300014,ISI:000256242300016,ISI:000254542800199,
Bredow2008,Hyldgaard2008,Kondo2008a,Li2008,Mowbray2008,Smeu2008,Thygesen2008a,Yoshizawa2008,Zhao2008,
ISI:000267183200024,Gutierrez2009,Lopez-Bezanilla2009,Wang2009,Jacob:prl:09}.

For nanoscopic conductors every atom counts and the transport properties are strongly dependent on the 
detailed atomic arrangement.
Thus, in order to make theoretical predictions that can be compared with experimental
results, it is 
important, to have a reliable description of, first, the atomic structure of the conductor 
and, second, the accompanying electronic structure. This can be achieved most conveniently 
with the aid of {\em ab initio} electronic structure methods based on atomic orbitals such 
as, e.g., {\sc Gaussian}\cite{Gaussian:03}, 
{\sc Crystal}\cite{Crystal:06} or {\sc Siesta}\cite{Ordejon:prb:96}. These codes implement
density functional theory (DFT) to obtain an effective mean-field
description of the electronic structure  
of, in our case, the nanoscopic bridge. This is typically done through
the effective  Kohn-Sham one-body Hamiltonian that
takes into account the electron-electron interactions at a static mean-field
level. 

A central challenge in the theoretical description of these systems is that 
the electronic structure of the atomic- or molecular-size conductor
is altered by the coupling to the bulk electrodes. Thus, in calculating the 
electronic structure the coupling to the (semi-infinite) electrodes has to be 
taken into account. This poses the difficult problem of dealing with an infinite 
system without translation invariance. In addition, one should strictly carry out the electronic and atomic
structure calculation out of equilibirum,
as imposed by the applied voltage. This is usually done by making 
use of the one-body Green's functions (GFs) and the so-called partitioning technique, 
as will be explained in the following sections. 

Clearly, while the detailed atomic and electronic structure of the nanoscopic bridge plays a crucial role, 
farther away from the bridge these become less important.
Besides, the exact atomic structure of the bulk electrodes, as encountered in real experiments, is 
not known and cannot be controlled with precision beyond a few contact atoms\cite{PhysRevLett.100.175502}. This lack of 
control lies behind the statistical deviations observed in measurable quantitites such as the conductance. 
This brings us to the important question of how to model the bulk electrodes, mantaining a compromise between 
realism and computational effort. 
Several possibilities of how to model the electrodes have been presented in the literature, with 
every model having advantages and disadvantages\cite{Lang:prb:95,Palacios:prb:01,Brandbyge:prb:02,Heurich:prl:02}. 
Here we explore the use of two types 
of electrode models: (i) parametrized tight-binding (TB) Bethe lattices\cite{Palacios:prb:01,Palacios:prb:02} 
and (ii) perfect nanowires of finite
section, stressing their weaknesses and strengths as models to represent the reality.
Another related question which we partially address in this paper is to what extent the particular shape or atomic
arrangement of the electrodes near the bridge
introduces variations in the conductance and how these depend on the chemical nature of the atoms.
The results presented below 
are all obtained with codes developed by the authors over the years 
which can be found in the publicly available quantum transport toolbox ALACANT\cite{ALACANT}.

\section{Non-equilibrium Green's functions and Landauer formalisms}
\label{sec:NEGF}

In the following, and for completeness' sake, we give a summary of the central aspects to the
non-equilibrium Green's function (NEGF) and Landauer formalisms formulated for a non-orthogonal 
localized atomic basis set. Although most of the details can be found in the early literature
\cite{PhysRevB.63.121104,Palacios:prb:01,Xue:jcp:01,Palacios:prb:02,Brandbyge:prb:02,Heurich:prl:02,
Louis:prb:03,Xue:prb:03:I,Xue:prb:03:II,Ke:prb:04,Rocha:prb:04,Thygesen:prb:05,Ferretti:prl:05}, here 
we discuss in depth some of those that are not usually addressed and become essential for a correct 
implementation of the above mentioned formalisms.

\begin{figure}
  \begin{center}
    \includegraphics[width=0.6\linewidth]{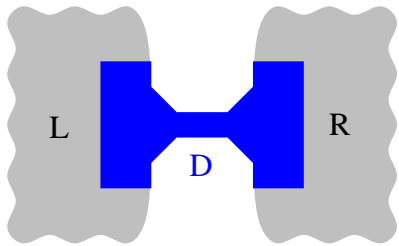}
    \caption{
      Sketch of the transport problem. The system is divided
      into three parts: Left electrode (L), device (D), and right
      electrode (R).
      \label{fig:LDR}
  }
  \end{center}
\end{figure}

We divide the system into three parts, as shown in Fig. \ref{fig:LDR}: 
The semi-infinite left (L) and right (R) electrodes or, hereon, leads, and the intermediate 
region between the two leads hereon called device (D) which contains the 
a central, narrow region where most of the scattering takes place (e.g., a nanoscopic constriction of the same
material as the leads or a trapped molecule). We assume that the leads are only coupled to the 
scattering region but not to each other. In a localized atomic basis 
set the Hamiltonian $\bf H$ of the system is given by:
\begin{equation}
  \label{eq:HLDR}
  {\bf H} = \left( 
    \begin{array}{ccc}
      {\bf H}_{\rm L}  & {\bf H}_{\rm LD} & {\bf 0} \\
      {\bf H}_{\rm DL} & {\bf H}_{\rm D}  & {\bf H}_{\rm DR} \\
      {\bf 0}         & {\bf H}_{\rm RD} & {\bf H}_{\rm R} 
    \end{array}
  \right)
\end{equation}
Since atomic basis sets are usually non-orthogonal we also have
to take into account the overlap between the atomic orbitals given
by the matrix $\bf S$:
\begin{equation}
  \label{eq:SLDR}
  {\bf S} = \left( 
    \begin{array}{ccc}
      {\bf S}_{\rm L}  & {\bf S}_{\rm LD} & {\bf 0} \\
      {\bf S}_{\rm DL} & {\bf S}_{\rm D}  & {\bf S}_{\rm DR} \\
      {\bf 0}         & {\bf S}_{\rm RD} & {\bf S}_{\rm R} 
    \end{array}
  \right)
\end{equation}

In order to deal with the problem of an infinite system without 
translation invariance, it is convenient to make use of the one-body 
Green's functions as explained, e.g., in the book by Economou\cite{Economou:book:83}. 
The one-body Green's function (GF) is defined 
as the resolvent operator of the one-body Schr\"odinger equation:
\begin{equation}
  (z - \hat{H}) \hat{G}(z) = \hat{I}.
\end{equation}
where $z$ is, in general, a complex number and $\hat{I}$ is the identity. 

If $z$ does not coincide with an eigenvalue $\epsilon_k$ of 
the Hamiltonian  $\hat{H}$ the GF operator has the following 
simple solution:
$\hat G(z) =  (z-\hat H)^{-1} \mbox{ for } z \ne \epsilon_k$.

Obviously, for $z=\epsilon_k$ the GF operator has a pole and 
is thus not well defined. In this case one can define two GFs 
by a limiting procedure: The {\it retarded} GF is defined as 
$\hat G^{(+)}(E) := \lim_{\eta \to 0} \hat G(E+i\eta)$,
and the {\it advanced} GF is defined as 
$\hat G^{(-)}(E) := \lim_{\eta \to 0} \hat G(E-i\eta)$
where $E$ is a real number (the energy). The retarded (advanced) GF can be 
analytically continued into the upper (lower) complex plane. 
Moreover, away from the poles of $\hat{G}(z)$, i.e., for 
$z\ne\epsilon_k$ both definitions coincide with $\hat{G}(z)$:
$\hat G^{(+)}(z)=\hat G^{(-)}(z)=\hat G(z)$. 

Because of the non-orthogonality of the basis set it is
convenient to define the following Green's function matrix
\begin{equation}
  (z\mathbf{S} - \mathbf{H}) \mathbf{G}(z) = \mathbf{1}.
\end{equation}
Note that this Green's function matrix is not the standard one 
which is simply defined by the matrix elements of the GF operator 
$\bra{i}\hat{G}(z)\ket{j}$. However, the latter GF matrix
is inconvenient to handle in the case of non-orthogonal basis sets
(see Appendix \ref{app:NOBS} for a complete discussion).

Using the technique explained in App. \ref{app:partitioning} it can 
be shown that the GF of the device region is given by the following 
matrix:
\begin{equation}
  \label{eq:GD}
  \mathbf{G}_{\rm D}(z)=
  \left(z \mathbf{S}_{\rm D}-\mathbf{H}_{\rm D}-\mathbf{\Sigma}_{\rm L}(z)-\mathbf{\Sigma}_{\rm R}(z) \right)^{-1}
\end{equation}
where $\mathbf{\Sigma}_L$ and $\mathbf{\Sigma}_R$ are the so-called lead self-energies
which describe the coupling of the device to the semi-infinite L and R leads. These
self-energies can be calculated from the Green's functions of the (isolated) 
leads, $\mathbf{g}_\alpha(z)=(z\mathbf{S}_\alpha-\mathbf{H}_\alpha)^{-1}$:
\begin{equation}
  \label{eq:SigmaLR}
  \mathbf{\Sigma}_\alpha(z) = (z\mathbf{S}_{{\rm D}\alpha}-\mathbf{H}_{{\rm D}\alpha}) 
  \, \mathbf{g}_\alpha(z) \,
  (z\mathbf{S}_{{\rm D}\alpha}^\dagger-\mathbf{H}_{{\rm D}\alpha}^\dagger)
\end{equation}
where the index $\alpha$ denotes the electrode L or R, and we have exploited
the hermiticity of the Hamiltonian and the overlap matrix, i.e.,
$\mathbf{H}_{{\rm D}\alpha}^\dagger=\mathbf{H}_{\alpha{\rm D}}$ and 
$\mathbf{S}_{{\rm D}\alpha}^\dagger=\mathbf{S}_{\alpha{\rm D}}$.

All quantities of interest such as the density of states (DOS), 
charge density, current $I$, and zero-bias as well as differential conductance $dI/dV$
can be calculated from the GF matrix of the device region $\mathbf{G}_{\rm D}$ and 
the lead self-energies $\mathbf{\Sigma}_{\rm L}$ and $\mathbf{\Sigma}_{\rm R}$.
For instance, in the case of an effective Kohn-Sham one-body Hamiltonian, as considered here,
the current through the nanoscopic conductor is given by the famous 
Landauer formula\cite{Landauer:philmag:70}:
\begin{eqnarray}
  \label{eq:Landauer}
  I &=& \frac{2e}{h} \int dE \, \left[ f(E-\mu_L)-f(E-\mu_R) \right] \, T(E) ,
\end{eqnarray}
where $f$ represents the Fermi distribution function, $\mu_\alpha$ the left and right chemical
potentials, and the transmission function, $T(E)$, can be 
calculated from the retarded and advanced GFs by the Caroli expression\cite{Caroli:jphysc:71}:
\begin{equation}
  \label{eq:Transm}
  T(E) = {\rm Tr}\left[ 
    \mathbf{\Gamma}_{\rm L}(E) \mathbf{G}_{\rm D}^{(-)}(E)
    \mathbf{\Gamma}_{\rm R}(E) \mathbf{G}_{\rm D}^{(+)}(E)
  \right].
\end{equation}
Here $\mathbf{\Gamma}_{\rm L}$ and
$\mathbf{\Gamma}_{\rm R}$ are the so-called coupling matrices which are
defined as
\begin{eqnarray}
  \mathbf\Gamma_{\rm L}(E) 
  &:=& i \, \left(\mathbf\Sigma_{\rm L}^{(+)}(E) - \mathbf\Sigma_{\rm L}^{(-)}(E)\right)
  \\
  \mathbf\Gamma_{\rm R}(E) 
  &:=& i \, \left( \mathbf\Sigma_{\rm R}^{(+)}(E) - \mathbf\Sigma_{\rm R}^{(-)}(E) \right).
\end{eqnarray}
Note that, since the self-energy matrices are usually symmetric,
the coupling matrices are just (twice) the imaginary parts of the self-energies.

At zero temperature the zero-bias conductance is now just given by the 
transmission function at the Fermi level $\mu$ (i.e. the electrochemical 
potential at zero temperature):
\begin{equation}
  G = \frac{2e^2}{h} \times T(\mu)
\end{equation}
Hence, the transmission function is the central quantity for calculating
the electronic transport properties of nanoscopic conductors. 
It is worth noting at this point that
there is a controversy on the use of (Kohn-Sham) DFT to calculate the transmission function. In addition to the 
obvious fact that there is no mathematical support to the use of DFT out of equilibrium,
it has been recently argued that
  a DFT description of the device region can never yield the right value of the zero-bias conductance, not even 
using the exact exchange-correlation potential in case this was known. In fact, the corrections to the DFT
zero-bias transmission calculated using standard functionals can be important in high resistance cases. 
We refer the interested reader to Ref. \cite{Burke:prb:06} for a full discussion of these
issues which are beyond the scope of this work.

\section{Electrode models}

In the ALACANT toolbox two different codes can be found, differing in the way the bulk electrodes are implemented. 
In the first the electrodes can be described by a 
parametrized TB Bethe lattice (BL) model with the coordination 
and parameters appropriate for the chosen electrode material.  
In the second  the electrodes are approximated by finite section wires described with a Kohn-Sham Hamiltonian, 
usually computed at the same level as that of the scattering region or device.

\subsection{Bethe Lattice electrodes}
\label{sec:BL}
\begin{figure}
  \begin{center}
    \includegraphics[width=0.95\linewidth]{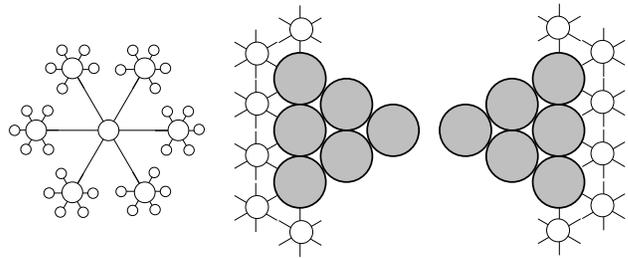}
  \end{center}
  \caption{
    Left: Finite section of Bethe lattice (BL) with 
    coordination 6. All atoms of the BL have the same coordination as in the 
    corresponding crystalline structure giving rise to short range order. But 
    there is no long range order in the BL due to the absence of closed loops.
    Right: 2D cartoon of a nanocontact (big grey circles) with the 
    first atoms of the BL (small white circles) attached to the 
    outer planes of the nanocontact. 
  }
  \label{fig:bl-scheme}
\end{figure}

A Bethe lattice, sometimes also called Caley tree, is generated by 
connecting a site with $N$ nearest-neighbors in directions that
can be those of a particular crystalline lattice. The new $N$ sites are 
each one of them connected to $N-1$ different sites and so on and so forth.
The generated lattice has the local topology of an actual lattice (number of neighbors
and crystal directions) but has no rings, and thus does not describe
the  long range order characteristic of real crystals. The left hand 
side of Fig. \ref{fig:bl-scheme} shows the first three layers of 
a BL with coordination 6. 

The advantage of choosing a BL over other models resides on the one hand 
in the lack of long-range order which mimics the poly-crystallinity of real 
electrodes. On the other hand the BL captures the short-range order since 
the local coordination of an atom is that of an atom in the bulk crystal
of the corresponding material. 

\begin{figure*}
  \begin{tabular}{ccc}
    \includegraphics[width=0.33\linewidth]{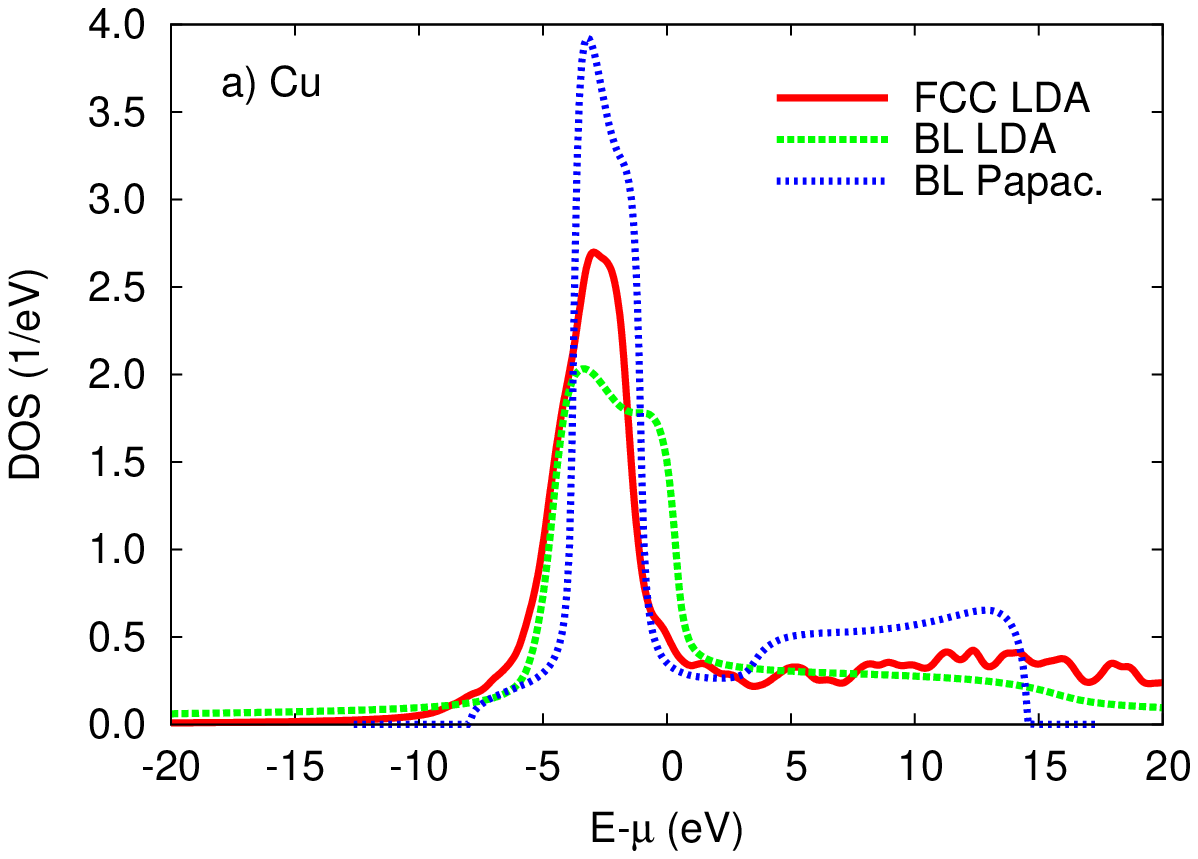} &
    \includegraphics[width=0.33\linewidth]{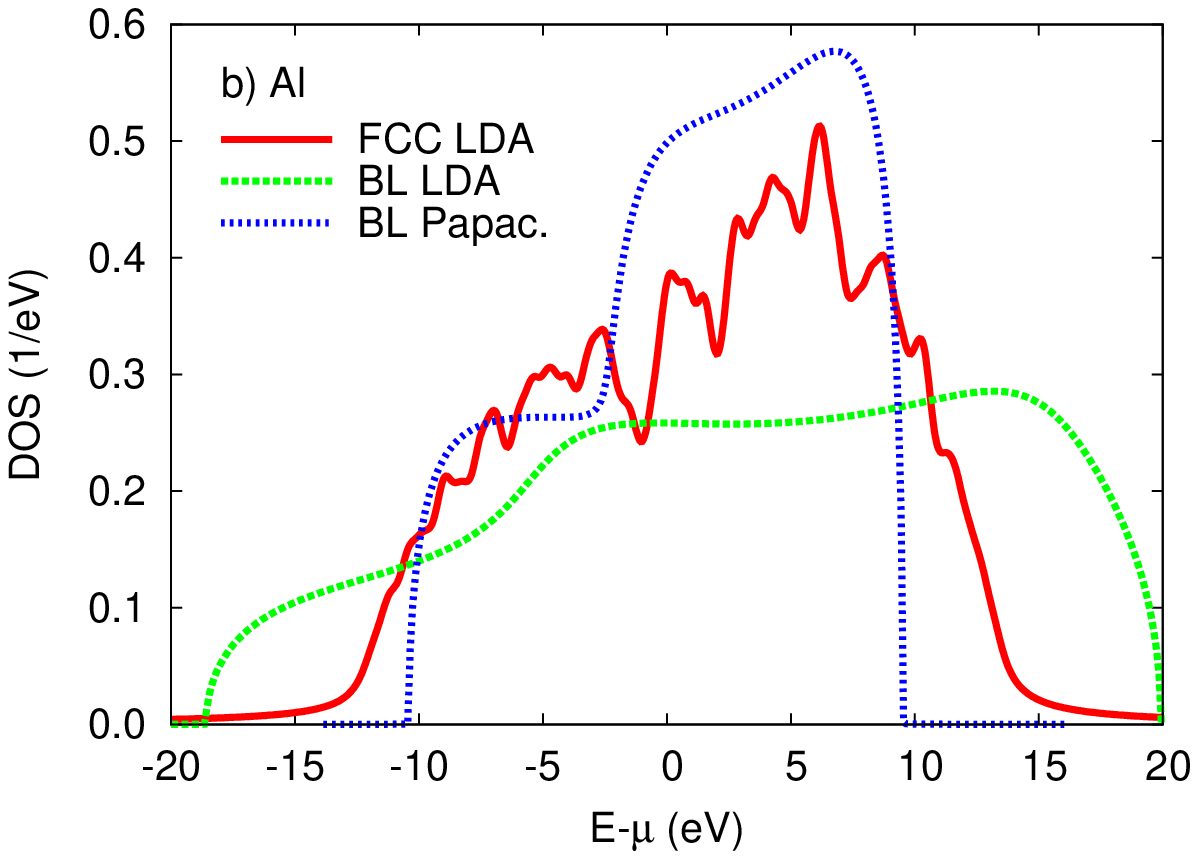} &
    \includegraphics[width=0.33\linewidth]{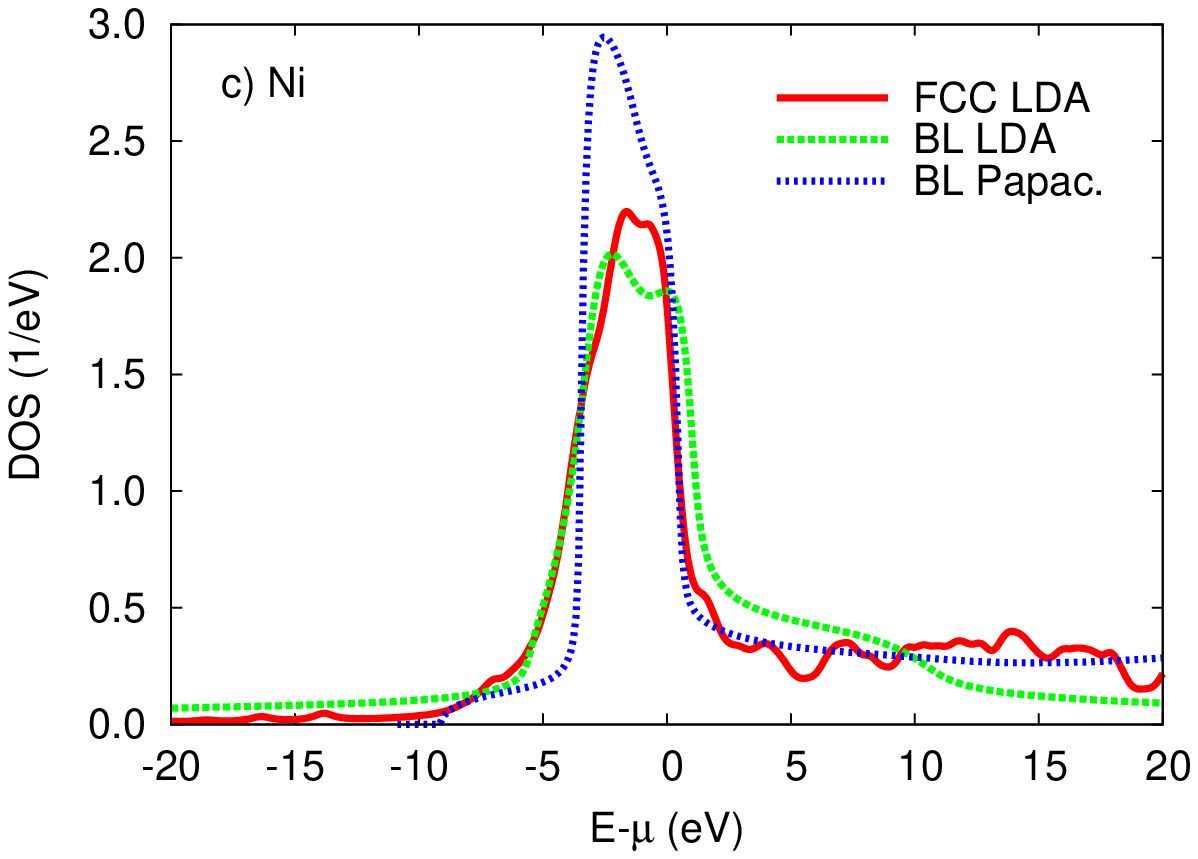} 
  \end{tabular}
  \caption{
    Comparison of Bethe Lattice DOS resulting from a parametrization
    obtained from an LDA Hamiltonian (green dashed lines) and from a Papaconstantopoulus 
    parametrization (blue dotted lines) with the DOS for a real FCC lattice calculated 
    in LDA (red continuous lines) for the three electrode materials 
    Cu (a), Al (b) and Ni (c).
  }
  \label{fig:bl-dos}
\end{figure*}

The right hand side of Figure \ref{fig:bl-scheme} illustrates schematically 
how the device (represented here by a single-element nanocontact) is connected to the BL electrodes: 
For a given chosen atom, typically in the outer planes of the device, a branch of the BL is added in the 
direction $\mathbf\tau_i$ of any missing bulk atom (including those missing in the 
same plane). The directions in which tree branches are added are indicated by white 
small circles which represent the first atoms of the branch in that direction. 
This corresponds to adding a BL self-energy $\mathbf\Sigma_{\mathbf \tau_i}$
to the atom in that direction (see App. \ref{app:bethe-lattices} for a derivation
of this formula):
\begin{equation}
  \mathbf\Sigma_{\tau_i}(E) ={\mathbf H}_{\mathbf \tau_i}
  \left [ E{\mathbf I}-{\mathbf H}_0-
    ({\mathbf \Sigma}_{T}(E)-
    {\mathbf \Sigma}_{\bar{\mathbf \tau_i}}(E))\right ]^{-1}
  {\mathbf H}_{\mathbf\tau_i}^{\dagger}
\end{equation}
where $\mathbf\Sigma_{T}$ is the sum over all self-energies
in all directions and the bar in $\bar\tau_i$ indicates the 
opposite direction of $\tau_i$, i.e. $\bar\tau_i\equiv-\tau_i$.
The electrode self-energies $\mathbf\Sigma_{\rm L}$ and $\mathbf\Sigma_{\rm R}$ are
thus obtained by summing up the directional BL self-energies 
$\mathbf\Sigma_{\tau_i}$ in all directions $\tau_i$ missing on 
that particular atom $\mathbf\Sigma_{a_{\rm L/R},\mathbf \tau_i}$ for all the atoms 
connected to that electrode:
\begin{equation}
  \mathbf\Sigma_{\rm L/R}(E) = 
  \sum_{\mbox{\small all atoms {\it a}}_{\rm L/R}\atop\mbox{\small connected to L/R}} 
  \sum_{\mbox{\small all missing}\atop\mbox{\small directions}\,\mathbf\tau_i} 
  \mathbf\Sigma_{a_{\rm L/R},\mathbf\tau_i}(E)
\end{equation}
Assuming that the most important structural details of the electrode are already included 
in the central cluster, the BLs should have no other relevance than that of introducing 
a generic bulk electrode for a given metal. 

In order to compare the results of the BL model with the actual electronic
structure of the corresponding real crystal lattice, we calculate the bulk 
DOS of the BL from the imaginary part of the local GF:
\begin{equation}
  \rho_0(E) = -\frac{1}{\pi}{\rm Im} {\rm Tr}[\mathbf{G}_{0}(E)],
\end{equation}
where the local Green's function $\mathbf{G}_0$ of the Bethe lattice is
given by
\begin{equation}
  \mathbf{G}_{0}(E) = ( E - \mathbf{H}_0 - \mathbf{\Sigma}_{T}(E) )^{-1}.
\end{equation}
In Fig. \ref{fig:bl-dos} we compare the bulk DOS of BL models using different parametrizations 
with electronic structure calculations in the local density approximation (LDA) for the three 
different electrode materials considered here (Cu, Al and Ni). The BLs have coordination 12, 
corresponding to the FCC crystalline structure of the bulk materials. 
On the one hand we have taken the TB parameters directly from the nearest-neighbor hoppings 
and on-site energies of the LDA Kohn-Sham Hamiltonian of the FCC crystal (ignoring the overlap).
where the calculations have been carried out with the help of the CRYSTAL code. On the other hand 
we have taken the TB parameters established by Papaconstantopoulus and coworkers\cite{Papacon-web}.
As can also be seen from Fig. \ref{fig:bl-dos}, the BL construction results in all cases in a smooth DOS 
which reproduces the basic features of the one corresponding to a mono-crystalline solid. As can be 
seen in Fig. \ref{fig:bl-dos}, depending on the type of material, the use of one set of parameters 
or another can result in a more accurate description.  However, the choice of parameters should
not be determinant in the final conductance results as long as the device is sufficiently large.

As usual we have assumed an orthonormal basis set for the Bethe lattice. On the other hand, 
the basis set of the device region is typically non-orthogonal. Hence, the question 
arises of how to match the two different levels of modeling. A straightforward approach 
is to {\it orthogonalize} the device basis set, for example with the L\"owdin orthogonalization 
scheme through the transformation 
$\mathbf{H}_{\rm D}^\prime = {\mathbf{S}_{\rm D}}^{-1/2} \mathbf{H}_{\rm D} {\mathbf{S}_{\rm D}}^{-1/2}$.
Equivalently, one can {\it de-orthogonalize} the self-energies $\mathbf\Sigma_{\rm L}$ 
and $\mathbf\Sigma_{\rm R}$: 
$\mathbf{\Sigma}_{\rm L/R}^\prime = {\mathbf{S}_{\rm D}}^{1/2} \mathbf{\Sigma}_{\rm L/R} {\mathbf{S}_{\rm D}}^{1/2}$.

Alternatively, one can obtain the TB parameters in a non-orthogonal basis
set, and take into account the overlap between orbitals on neighbouring atoms in the
calculation of the BL self-energies. In this case the Dyson equation for the calculation
of the BL self-energy is trivially modified as follows:
\begin{eqnarray}
  \lefteqn{\mathbf\Sigma_{\tau_i}(E) = ({\mathbf H}_{\mathbf \tau_i}-E{\mathbf S}_{\mathbf \tau_i})}
  \\
  && \times \left [ E{\mathbf S}_0-{\mathbf H}_0-
    ({\mathbf \Sigma}_{T}(E)-
    {\mathbf \Sigma}_{\bar{\mathbf \tau_i}}(E))\right ]^{-1}
  ({\mathbf H}_{\mathbf\tau_i}^{\dagger}-E{\mathbf S}_{\mathbf \tau_i}^\dagger)
  \nonumber
\end{eqnarray}
where ${\mathbf S}_{\mathbf \tau_i}$ is the overlap matrix between orbitals
on neighboring atoms in the direction $\mathbf\tau_i$.

In case of a non-orthogonal basis set one has to take into account
the non-diagonal part of the GF between different sites of the BL
when computing the BL DOS:
\begin{equation}
  \rho_0(E) =\
  -\frac{1}{\pi}{\rm Im} {\rm Tr}\left[\mathbf{G}_{0}(E)\mathbf{S}_0 + \sum_{\tau_i} \mathbf{G}_{0,\tau_i}(E)\mathbf{S}_{\tau_i} \right].
  \nonumber
\end{equation}
This is most easily done by extending the unit cell Hamiltonian of the BL with all 
nearest neighbour sites, computing the GF $\mathbf{G}_{\rm X0}(E)$ 
of the extended unit cell (X0), and then taking the partial trace for the central 
site $0$ of the matrix product $\mathbf{G}_{\rm X0}(E)\mathbf{S}_{\rm X0}$.
\begin{equation}
  \rho_0(E) = -\frac{1}{\pi}{\rm Im} {\rm Tr}_0[\mathbf{G}_{\rm X0}(E)\mathbf{S}_{\rm X0}].
\end{equation}

\begin{figure}
  \begin{center}
    \includegraphics[width=0.7\linewidth]{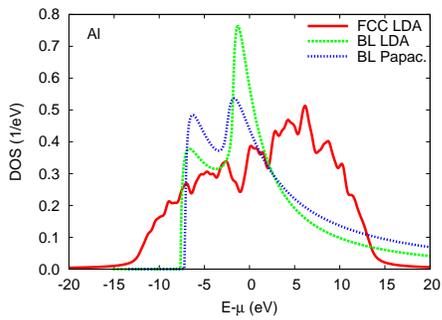}
  \end{center}
  \caption{
    Bethe Lattice DOS with different parametriztions taking into account overlap 
    between atomic orbitals compared to DOS of real FCC lattice calculated with LDA 
    (red continuous lines) for Al. The BL parameters have been taken either directly 
    from the LDA Kohn-Sham Hamiltonian (green dashed lines) or from the Papaconstantopoulus 
    TB parametrization with overlap (blue dotted lines). 
  }
  \label{fig:bl-dos-ovl}
\end{figure}

In Fig. \ref{fig:bl-dos-ovl} we compare the bulk DOS of BL models with different parametrizations
(taking into account the overlap between orbitals on neighbouring atoms) with LDA electronic structure 
calculations for the case of Al. As before we have taken
the TB parameters either directly from the nearest-neighbor hoppings and on-site energies of the LDA 
Kohn-Sham Hamiltonian of the FCC crystal (this time taking into account the overlap) or we have taken 
the TB parameters established by Papaconstantopoulus and coworkers (this time with overlap).
Now the BL DOS does not resemble the DOS of a real crystal lattice anymore: The band
width now becomes semi-infinite extending infinitely to positive energies. This artifact
can only be healed by scaling down the overlap considerably. We therefore
conclude that the introduction  of non-orthogonal basis sets in the description of the BL does not seem to be
very useful for mimicking the DOS of real materials, being preferable the self-energy deorthogonalization 
procedure described above.

\subsection{Nanowire electrodes}
\label{sec:nanowire}
\begin{figure}
  \begin{center}
    \includegraphics[width=0.6\linewidth]{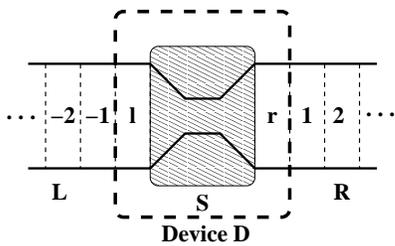}
    \caption{
      Sketch of transport problem for the case of one-dimensional
      nanowires as electrodes. The system is divided
      into 3 parts: Left electrode (L), device (D) and right
      electrode (R).
      \label{fig:LDR1D}
  }
  \end{center}
\end{figure}

The second type of model for the leads consists of finite-section wires where
the electronic structure is described at the same computational level as that of the device.
As indicated in Fig. \ref{fig:LDR1D}, we subdivide the one-dimensional leads into unit cells 
which must be chosen sufficiently large so that the coupling between 
non-neighboring unit cells can be neglected.
In general a unit cell consists of several primitive unit cells of the crystal.
The Hamiltonian matrix of the left lead ${\bf H}_{\rm L}$ can be subdivided into
sub-matrices in the following manner:

\begin{eqnarray}
  \label{eq:H_L}
  \mathbf{H}_{\rm L} &=& 
  \left(
    \begin{array}{ccccc}
      \ddots & \ddots         & \ddots         &                & \mathbf 0   \\
      \,     & \mathbf H_1^\dagger & \mathbf H_0         & \mathbf H_1         &        \\
      \,     &                & \mathbf H_1^\dagger & \mathbf H_0         & \mathbf H_1 \\
      \mathbf 0   &                &                & \mathbf H_1^\dagger & \mathbf H_0
    \end{array}
  \right)
\end{eqnarray}
Analogously the Hamiltonian of the right lead is given by the following
matrix:
\begin{equation}
  \label{eq:H_R}
  \mathbf{H}_{\rm R} 
  = \left(
    \begin{array}{ccccc}
      \mathbf H_0         & \mathbf H_1         &        &        & \mathbf 0   \\
      \mathbf H_1^\dagger & \mathbf H_0         & \mathbf H_1 &        &        \\
      \,             & \mathbf H_1^\dagger & \mathbf H_0 & \mathbf H_1 &        \\
      \mathbf 0           &                & \ddots & \ddots & \ddots
    \end{array}
  \right)
\end{equation}
In a similar way, the overlap inside the leads is given by the matrices
\begin{equation}
  \label{eq:S_L}
  \mathbf{S}_{\rm L} = 
  \left(
    \begin{array}{ccccc}
      \ddots & \ddots         & \ddots         &                & \mathbf 0   \\
      \,     & \mathbf S_1^\dagger & \mathbf S_0         & \mathbf S_1         &        \\
      \,     &                & \mathbf S_1^\dagger & \mathbf S_0         & \mathbf S_1 \\
      \mathbf 0   &                &                & \mathbf S_1^\dagger & \mathbf S_0  
    \end{array}
  \right)
\end{equation}
and
\begin{equation}
  \label{eq:S_R}
  \mathbf{S}_{\rm R} 
  = \left(
    \begin{array}{ccccc}
      \mathbf S_0         & \mathbf S_1         &        &        & \mathbf 0   \\
      \mathbf S_1^\dagger & \mathbf S_0         & \mathbf S_1 &        &        \\
      \,             & \mathbf S_1^\dagger & \mathbf S_0 & \mathbf S_1 &        \\
      \mathbf 0           &                & \ddots & \ddots & \ddots
    \end{array}
  \right)
\end{equation}

Furthermore, the unit cell of each lead that is immediately connected to the scattering region 
(unit cell ``$l$'' for the left and unit cell ``$r$'' for the right lead)  is included into 
the device part of the system. So the Hamiltonian of the device region reads
\begin{equation}
  \label{eq:Device}
  \mathbf{H}_{\rm D} 
  = \left( 
    \begin{array}{ccc}
      \mathbf H_l     & \mathbf H_{l,S} & \mathbf 0_{l,r} \\
      \mathbf H_{S,l} & \mathbf H_S     & \mathbf H_{S,r} \\
      \mathbf 0_{r,l} & \mathbf H_{r,S} & \mathbf H_r   
    \end{array} \right)
\end{equation}
and the overlap matrix is given by:
\begin{equation} 
  \mathbf{S}_{\rm D} 
  = \left( 
    \begin{array}{ccc}
      \mathbf S_l     & \mathbf S_{l,S} & \mathbf 0_{l,r} \\
      \mathbf S_{S,l} & \mathbf S_S     & \mathbf S_{S,r} \\
      \mathbf 0_{r,l} & \mathbf S_{r,S} & \mathbf S_r  
    \end{array} 
  \right)
\end{equation}

Since only the $l$- and $r$-parts of the device region are immediately 
connected to the two semi-infinite nanowire electrodes the self-energy 
matrices  $\mathbf\Sigma_{\rm L}$ and $\mathbf\Sigma_{\rm R}$ that describe
the coupling of the device region to the electrodes L and R are given by 
matrices that are different from zero only in the $l$- and $r$-parts, respectively:
\begin{eqnarray} 
  \mathbf\Sigma_{\rm L}(z) 
   = \left( 
     \begin{array}{ccc}
      \mathbf\Sigma_l(z) & \mathbf 0_{l,S} & \mathbf 0_{l,r} \\
      \mathbf 0_{S,l}  & \mathbf 0_S     & \mathbf 0_{S,r} \\
      \mathbf 0_{r,l}  & \mathbf 0_{r,S} & \mathbf 0_r
    \end{array} \right)
\end{eqnarray}
and
\begin{eqnarray} 
  \mathbf\Sigma_{\rm R}(z)
   = \left( 
     \begin{array}{ccc}
      \mathbf 0_l     & \mathbf 0_{l,S} & \mathbf 0_{l,r} \\
      \mathbf 0_{S,l} & \mathbf 0_S     & \mathbf 0_{S,r} \\
      \mathbf 0_{r,l} & \mathbf 0_{r,S} & \mathbf\Sigma_r(z)
    \end{array} \right)
\end{eqnarray}
As shown in App. \ref{app:self-energy-1D}, the non-zero submatrices $\mathbf\Sigma_l$ 
and $\mathbf\Sigma_r$ can be calculated from the Hamiltonian and 
overlap sub-matrices of the two leads by the following Dyson equations:
\begin{eqnarray}
  \label{eq:DysonL}
  \mathbf{{\Sigma}}_{l}(z) &=& 
  (\mathbf{H}_1^\dagger - z \, \mathbf{S}_1^\dagger) \, 
  (z\,\mathbf{S}_0-\mathbf{H}_0-\mathbf{{\Sigma}}_{l}(z))^{-1}
  \nonumber\\
  && \hspace{2em} \times (\mathbf{H}_1 - z\,\mathbf{S}_1)
  \\ \nonumber \\
  \label{eq:DysonR}
  \mathbf{{\Sigma}}_{r}(z) &=&
  (\mathbf{H}_1 - z\,\mathbf{S}_1) \,
  (z\,\mathbf{S}_0 - \mathbf{H}_0 - \mathbf{{\Sigma}}_{r}(z))^{-1}
  \nonumber \\
  && \hspace{2em} \times (\mathbf{H}_1^\dagger - z \, \mathbf{S}_1^\dagger).
\end{eqnarray}

\section{Self-consistent electronic structure of the device}

In the context of standard DFT electronic structure calculations of finite or periodic systems
the self-consistent mean-field
electronic potential and the density matrix (or Kohn-Sham wave functions) are determined by the sole input of the atomic
structure of the cluster or cell, through the chosen exchange-correlation functional. 
In the context of quantum transport, where the
systems are infinite, but present no translational invariance, 
the ``boundary conditions'' imposed by the electrodes play an additional and important role.
The electronic structure of the device
region also depends on the model chosen to represent the electrodes and the details of how to carry out
the self-consistency may vary, particularly when out of equilibrium.
We discuss now two different alternatives: {\it The embedded cluster approach}, associated to the use of
BLs (see Sec. \ref{sec:BL}), and the {\it supercell approach}, 
where the electrodes are described by nanowires (see Sec. \ref{sec:nanowire}).

\subsection{Embedded cluster approach}

In the embedded cluster approach 
the electronic structure of the infinite system is calculated self-consistently 
only within a finite-size region --the scattering or device region containing the 
nanoconstriction or molecule-- while the electronic structure of the 
rest of the system (i.e., the two bulk electrodes) is fixed from the very beginning to 
that of a simplified parametrized BL model (see Sec. \ref{sec:BL}).
The basic premise here is to set up a device region sufficiently large. In other words, a sufficiently wide 
section of the bulk electrodes must be included in the device region so that the 
interface resistance between the BL and the device, the former being described at a lowest level of approximation than
the latter, does not contribute significantly to the overall resistance which is essentially determined by
the intrinsic resistance of the device.

In equilibrium the two leads or electrodes must have the same electrochemical potential. If 
the leads are made of different materials with different work functions,
an overall charge transfer must occur somewhere. This gives rise to an electric field, 
shifting the band structures of the two leads relative to each other, and 
subsequently aligning the electrochemical potentials of the two leads. 
The net effect of the charge transfer
on the electronic structure of the bulk electrode material outside the device can be 
taken into account by simply shifting the electrochemical potentials (and of course the band 
structure) of the two materials to a common electrochemical potential. Leads of the same material but presenting different 
crystallographic orientations might also present different work functions, but the BL model cannot
account for this difference. Notice that we
 have refrained from being too specific about where the charge transfer takes place.
A localized charge transfer in the device region, typically a one- or quasi-one-dimensional system,
cannot be entirely responsible for the
electrochemical alignment far away from the device for 
obvious electrostatic reasons.  One must be cautious with the extent of the region necessary for this transfer
to take place. Only for infinite two-dimensional interfaces between different materials the charge accumulates
strictly at the interface. In the opposite limit of one-dimensional systems in point contact, the charge transfer
must extend logarithmically into the bulk\cite{Leonard:prl:99}. Regardless of where the charge transfer takes place, anyway,
thermodynamical equilibrium must be reached.
 
Taking a common electrochemical potential $\mu$, the rigid electrostatic shifts are 
$\Delta_{\rm L}=\mu-\mu_{\rm L}^0$ and $\Delta_{\rm R}=\mu-\mu_{\rm R}^0$ where $\mu_{\rm L}^0$
and $\mu_{\rm R}^0$ are the electrochemical potentials (or work functions) of the materials
of the left and right electrode, respectively. The lead Hamiltonians corrected by the 
electrostatic shift are thus given by $\mathbf{H}_{\rm L}+\Delta_{\rm L}\mathbf{S}_{\rm L}$ 
and $\mathbf{H}_{\rm R}+\Delta_{\rm R}\mathbf{S}_{\rm R}$.
Now the Kohn-Sham Hamiltonian of the entire system (leads+device) is given by 
\begin{eqnarray}
  \label{eq:KS-HLDR}
  \lefteqn{\mathbf{H}_{\rm KS}-\mu\mathbf{S}  = }
  \\
  && \left(
  \begin{array}{ccc}
    \mathbf{H}_{\rm L}-\mu_{\rm L}^0\mathbf{S}_{\rm L}   & \mathbf{H}_{\rm LD}-\mu_{\rm L}^0\mathbf{S}_{\rm LD}  & \mathbf{0}         \\
    \mathbf{H}_{\rm DL}-\mu_{\rm L}^0\mathbf{S}_{\rm DL} & \mathbf{H}_{\rm D}-\mu\mathbf{S}_{\rm D}             & \mathbf{H}_{\rm DR}-\mu_{\rm R}^0\mathbf{S}_{\rm DR} \\
    \mathbf{0}                                        & \mathbf{H}_{\rm RD}-\mu_{\rm R}^0\mathbf{S}_{\rm RD}  & \mathbf{H}_{\rm R}-\mu_{\rm R}^0\mathbf{S}_{\rm R}
  \end{array}
  \right)
  \nonumber
\end{eqnarray}
where we have included the electrochemical potential $\mu$ of the entire system. 
The corresponding overlap matrix that takes into account the 
non-orthogonality of the atomic orbitals has of course the same form as in eq.
\ref{eq:SLDR}. 

\begin{figure}
  \begin{center}
    \includegraphics[width=0.98\linewidth]{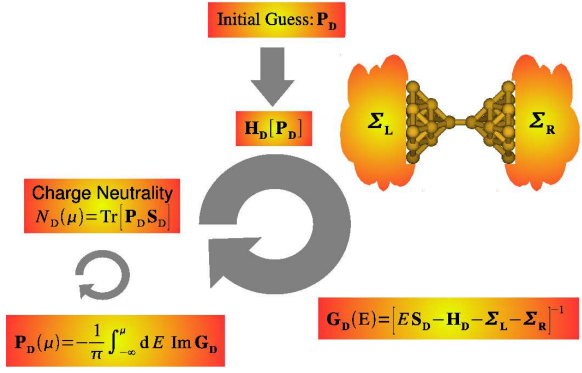}
  \end{center}
  \caption{
    Diagram illustrating the self-consistent procedure for KS based
    NEGF formalism as explained in the text.
  }
  \label{fig:ALACANT-SCF}
\end{figure}

The Kohn-Sham Hamiltonian of the entire system depends only on the electron density 
$n_{\rm D}(\vec{r})$ of the device region since the electronic structure of the rest 
of the system is kept fixed (apart from the electrostatic shifts $\Delta_{\rm L}$
and $\Delta_{\rm R}$ which depend on the electrochemical potential $\mu$): 
$\mathbf{H}_{\rm KS}=\mathbf{H}_{\rm KS}[n_{\rm D}]$.
The electron density $n_{\rm D}(\vec{r})$ can be obtained from the density matrix of 
the device region:\cite{footnote1}
\begin{equation}
  n_{\rm D}(\vec r)  = \sum_{\alpha,\beta \in \rm D} \phi_\alpha(\vec r) \;\mathbf{P}_{\alpha\beta}\; \phi_\beta^\ast(\vec r)
\end{equation}
which in turn is found by integrating (most conveniently done by analytic 
continuation to the complex plane) the device part of the Green's function:
\begin{equation}
  \mathbf{P}_{\rm D}(\mu)  = -\frac{1}{\pi} {\rm Im} \int_{-\infty}^{\mu} {\rm d}E \, \mathbf{G}_{\rm D}^{(+)}(E) 
\end{equation}
where the device Green's function is given by
\begin{equation}
  \label{eq:KS-G_D}
  \mathbf{G}_{\rm D}^{(+)}(E) = \left( E\mathbf{S}_{\rm D} - \mathbf{H}_{\rm D} -\mathbf{\Sigma}_{\rm L}(E)-\mathbf{\Sigma}_{\rm R}(E) \right)^{-1}
\end{equation}

Now in order to obtain the electrochemical potential of the entire system, one has to impose
charge neutrality in the entire system. Since the metallic leads outside the device 
region are charge neutral themselves,
it suffices to impose charge neutrality within the device region:
\begin{eqnarray}
  \label{eq:charge-neutrality}
  N_{\rm D}(\mu) &=& {\rm Tr}[\mathbf{P}_{\rm D}(\mu)\mathbf{S}_{\rm D}]
  \nonumber\\
  &=& -\frac{1}{\pi} {\rm Im} \int_{-\infty}^\mu {\rm d}E\, {\rm Tr}[\mathbf{G}_{\rm D}^{(+)}(E)\;\mathbf{S}_{\rm D}]
\end{eqnarray}
Since $\mathbf{G}_{\rm D}$ via $\mathbf{H}_{\rm D}$ is a functional of the electron density $n_{\rm D}(\vec r)$, 
the electronic structure of the device region can now be determined self-consistently, already having taken 
into consideration
the effect of the leads through the selfenergies in Eq.  \ref{eq:KS-G_D}. 
For a practical implementation of this procedure we have created an interface to 
the quantum chemistry code {\sc Gaussian03}, taking thus advantage of the various DFT implementations that can be found 
in such a code.
Figure \ref{fig:ALACANT-SCF} shows a schematic picture of the self-consistent 
calculation of the electronic structure of the device region as described above.

Out of equilibrium, i.e., for a finite bias $eV=\mu_L-\mu_R$ one has to use
the NEGF technique. In this case there
is an additional contribution to the density matrix which can be calculated 
by integrating the so-called lesser GF $\mathbf{G}^{<}$ within the bias window:
\begin{equation}
  \mathbf{P}_{\rm D}^{\rm neq}(\mu_{\rm L},\mu_{\rm R}) = \mathbf{P}_{\rm D}^{\rm eq}(\mu_{\rm R})
  -\frac{i}{2\pi} \int_{\mu_{\rm R}}^{\mu_{\rm L}} dE \, \mathbf{G}^{<}_{\rm D}(E),
\end{equation}
where we have assumed a positive difference between L and R electrochemical 
potentials, $\mu_L-\mu_R>0$. As for the equilibrium case, charge neutrality must be 
imposed, e.g., by setting $\mu_{\rm R}$ to the appropriate value.
The lesser GF can be easily calculated from the retarded and advanced GFs:
\begin{eqnarray}
  \label{eq:Gless}
  \mathbf{G}^<_{\rm D}(E) &=& i \mathbf{G}_{\rm D}^{(+)}(E)[ f(E-\mu_{\rm L}) \mathbf\Gamma_{\rm L}(E) 
  \nonumber\\
  && + f(E-\mu_{\rm R}) \mathbf\Gamma_{\rm R}(E)] \, \mathbf{G}_{\rm D}^{(-)}(E)
\end{eqnarray}
For a full discussion of the actual implementation of these expressions see Ref. \cite{Louis:prb:03}.
We stress here we are taking into account the electron-electron interactions in the device region 
by an effective mean-field description at the level of DFT. Thus the device Hamiltonian
is an effective Kohn-Sham one-body Hamiltonian. In this approximation to the true many-body problem
Eqs. \ref{eq:Landauer} and \ref{eq:Transm} remain valid even out of equilibrium and at finite 
temperature.

\subsection{Supercell approach}

\begin{figure}
  \includegraphics[width=0.98\linewidth]{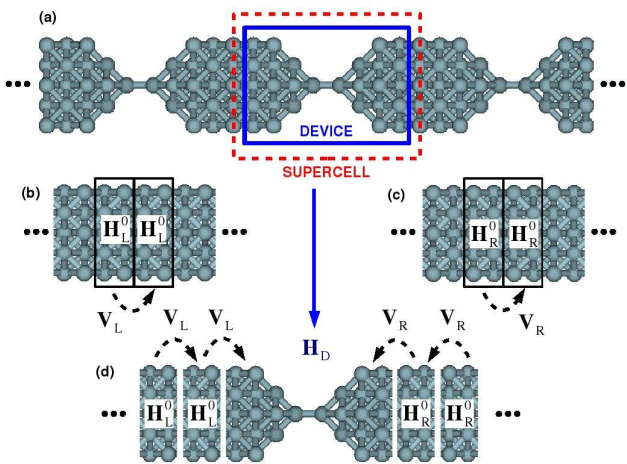}
  \caption{
    Illustration of the supercell approach to calculate the electronic structure 
    of the device and of the leads: (a) One-dimensional periodic system to calculate
    the electronic structure of the device region. (b,c) Infinite nanowires to
    calculate the electronic structure of the left (L) and right (R) semi-infinite 
    leads. (d) Sketch of the setup of the physical system: The device region (D) is 
    suspended between two semi-infinite leads L and R.
  }
  \label{fig:supercell}
\end{figure}

In the supercell approach we calculate the electronic 
structure of the device region and the electrodes with {\it ab initio}
electronic structure programs for periodic systems that use localized basis sets such as {\sc Crsytal} or {\sc Siesta}.
We first define a one-dimensional periodic 
system consisting of the device region as the unit cell, as shown in Fig. \ref{fig:supercell}(a). 
It is crucial that the device part D contains a sufficiently large portion of the nanowire 
electrodes so as to guarantee that the electronic structure of the device region and thus the 
Hamiltonian $\mathbf{H}_{\rm D}$ is the same as the electronic structure of the device 
between two semi-infinite nanowires.  In other words, we seek to
connect the two leads L and R far enough away from the scattering 
region where the electronic structure has relaxed to that of a bulk (i.e., infinite) nanowire. 

In a similar way, the unit cell Hamiltonian matrix $\mathbf{H}^{\rm L/R}_0$ and
hopping  matrices between unit cells for left and right
leads $\mathbf{H}^{\rm L/R}_1$ are extracted from periodic calculations of infinite nanowires of
finite width [see Fig. \ref{fig:supercell}(b,c)].  The lead self-energies
$\mathbf\Sigma_{\rm L}$, $\mathbf\Sigma_{\rm R}$ which describe the coupling 
of the device region D to the semi-infinite nanowires L and R in the situation 
depicted in Fig. \ref{fig:supercell}(d) can now be calculated by the Dyson equations 
(\ref{eq:DysonL}) and (\ref{eq:DysonR}).

Since the electronic structure of the electrodes has been calculated for 
perfect nanowires, the effect of an eventual charge transfer within the
device region has not been taken into account yet. As pointed out
before, the net effect of the charge transfer, mostly within the device region,
on the electronic structure of the bulk electrode material outside the 
device can be taken into account by simply shifting the electrochemical potentials 
(and of course the band structure) of the two materials to a common 
electrochemical potential. Hence, the Kohn-Sham Hamiltonian of the entire system
is also given by Eq. (\ref{eq:KS-HLDR}) where now the common electrochemical potential
$\mu$ is the one obtained from the supercell calculation of the device region,
and $\mu_{\rm L}^0$ and $\mu_{\rm R}^0$ are the electrochemical potentials obtained from
the electronic structure calculations of the infinite nanowires.
The Green's function of the device region is given by eq. (\ref{eq:KS-G_D}) where the
self-energies $\mathbf{\Sigma}_{\rm l}$ and $\mathbf{\Sigma}_{\rm r}$ are
now obtained from the Dyson equations (\ref{eq:DysonL}) and (\ref{eq:DysonR})
but with the energies shifted by the electrostatic shifts $\Delta_{\rm L}=\mu-\mu_{\rm L}^0$ and 
$\Delta_{\rm R}=\mu-\mu_{\rm R}^0$:
\begin{eqnarray}
  \lefteqn{\mathbf{{\Sigma}}_{\rm l}(z) =
  \left(\mathbf{H}_1^\dagger - (z + \Delta_{\rm L}) \, \mathbf{S}_1^\dagger\right) }
  \\
  && \times \left( (z+\Delta_{\rm L})\,\mathbf{S}_0-\mathbf{H}_0-\mathbf{{\Sigma}}_{\rm l}(z) \right)^{-1}
  \left( (\mathbf{H}_1-(z+\Delta_{\rm L})\,\mathbf{S}_1) \right)
  \nonumber\\
  \nonumber\\
  \lefteqn{\mathbf{{\Sigma}}_{\rm r}(z) =
  \left(\mathbf{H}_1 - (z + \Delta_{\rm R}) \, \mathbf{S}_1 \right) }
  \\
  && \times \left( (z+\Delta_{\rm R})\,\mathbf{S}_0-\mathbf{H}_0-\mathbf{{\Sigma}}_{\rm r}(z) \right)^{-1}
  \left(\mathbf{H}_1^\dagger-(z+\Delta_{\rm R})\,\mathbf{S}_1^\dagger \right)
  \nonumber
\end{eqnarray}
By this procedure we have connected the device region D with two {\it semi-infinite}
nanowires that have the electronic structure of bulk, i.e. {\it infinite}, nanowires far from the device.

\section{Comparison of electrode models}

\begin{figure}
  \begin{tabular}{ccc}
    Sequence I & Sequence II & Sequence III \\
    \includegraphics[width=0.85\linewidth,angle=270]{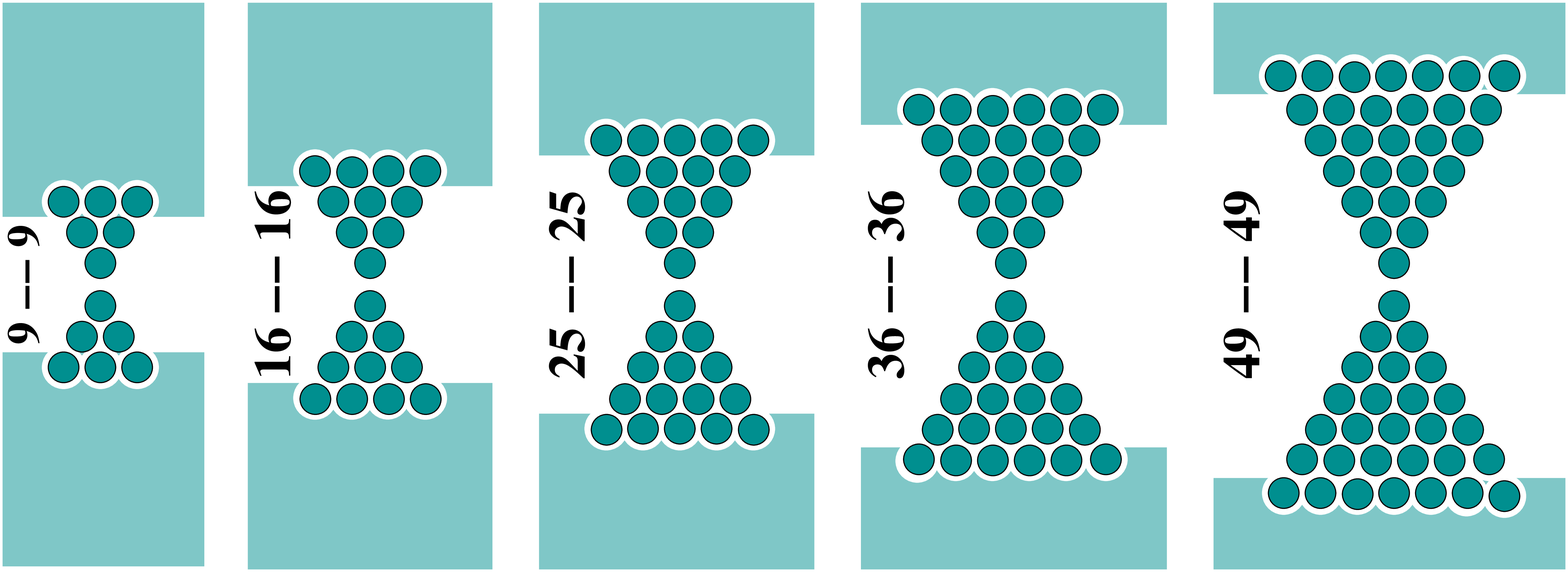} &
    \includegraphics[width=0.85\linewidth,angle=270]{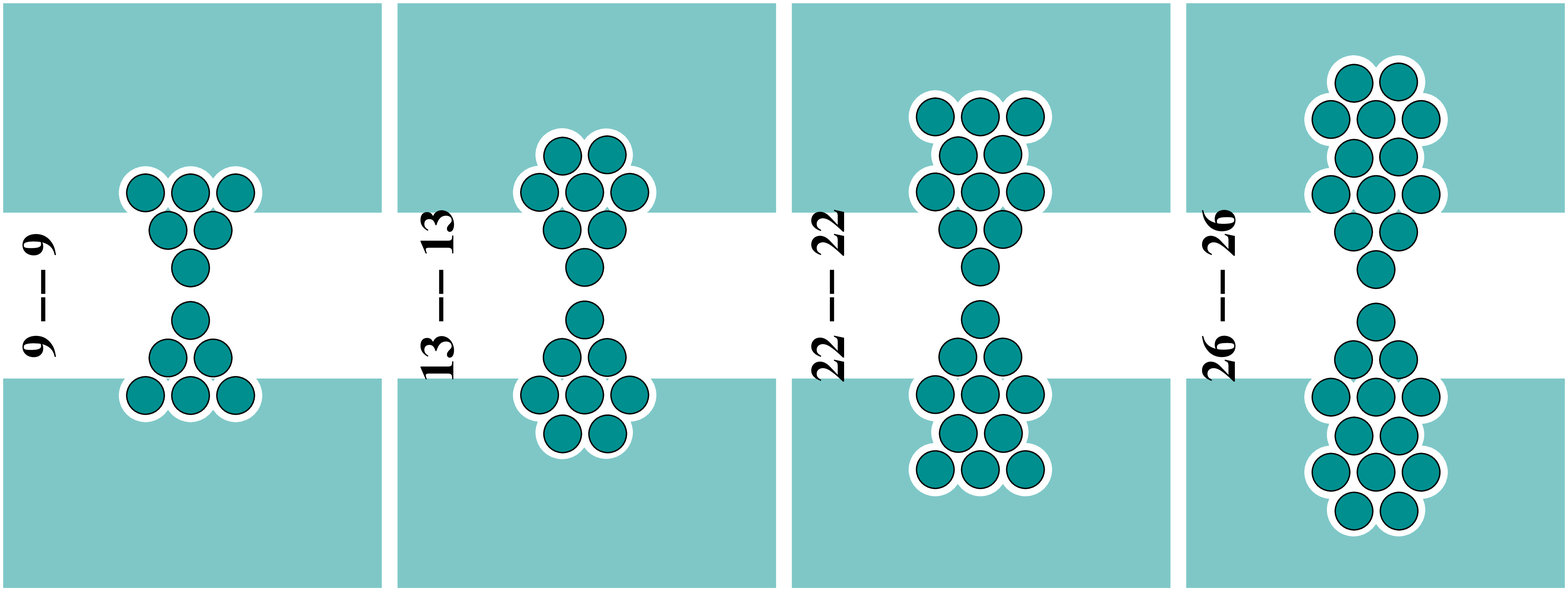} &
    \includegraphics[width=0.85\linewidth,angle=270]{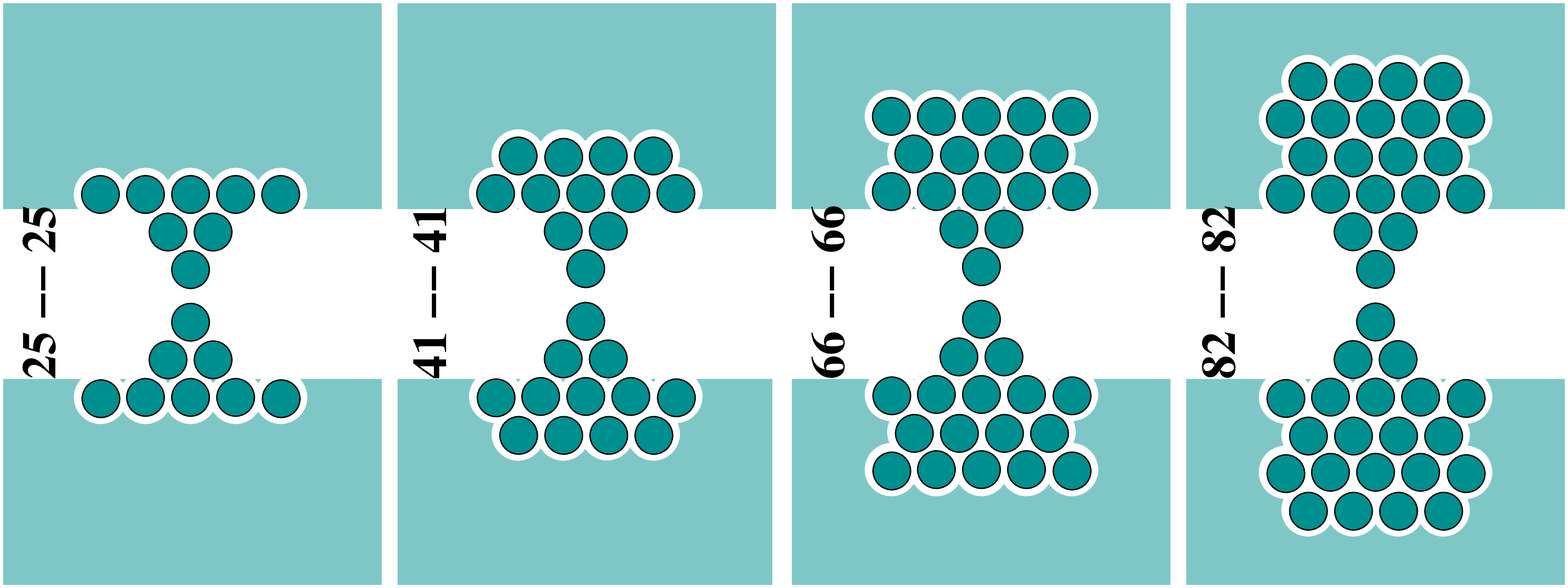}
  \end{tabular}
  \caption{
    Sketch of the pyramidal nanocontact geometries used in the {\it ab initio} 
    quantum transport calculations with Bethe lattice electrodes. Each column
    represents a sequence where the amount of bulk electrode material in the
    device region is increased in a disctinct way. 
  }
  \label{fig:bl-geometries}
\end{figure}

We now compare results obtained with the two electrode models, i.e., the
Bethe lattice and the nanowire electrodes and accompanying different implementations of the self-consistent
procedure. We have chosen for this study an archetypal metallic 
nanocontact model formed by two pyramidal tips joined by the apex atoms. For the {\em ab initio} calculation
of the device region and the nanowire electrodes we use LDA
and the minimal valence basis set by Christian and coworkers \cite{Hurley:jcp:86}. Should one be interested 
in a more quantitative study of the conductance of these systems it would be convenient to increase the 
size of the basis set, but this is not the main aim of this work.  For the Bethe lattice
we take the tight-binding parametrization by Papaconstantopoulos\cite{Papacon-web}, 
obtained by fitting tight-binding parametrizations to DFT calculations. Differences between the different parametrizations
discussed above are essentially irrelevant.

In Fig. \ref{fig:bl-geometries} we show three different sequences of increasing size 
for the embedded cluster calculations with Bethe lattice electrodes. The narrowest 
section or contact atomic structure is maintained throughout the sequence.
In sequence I, the pyramidal form 
is maintained for the entire device while it is increased in size.
In this case the Bethe lattices are only connected to the base layers of the pyramids.
In sequence II and III, only the tips maintain the pyramidal
form. The rest of the device region are finite sections of 001 surfaces of 
the fcc crystal lattice.  In each step of a sequence an atomic layer is added. 
The difference between sequences II and III is the width
of the finite sections of bulk electrode included in the device region.

\begin{figure}
  \begin{center}
    \includegraphics[width=\linewidth]{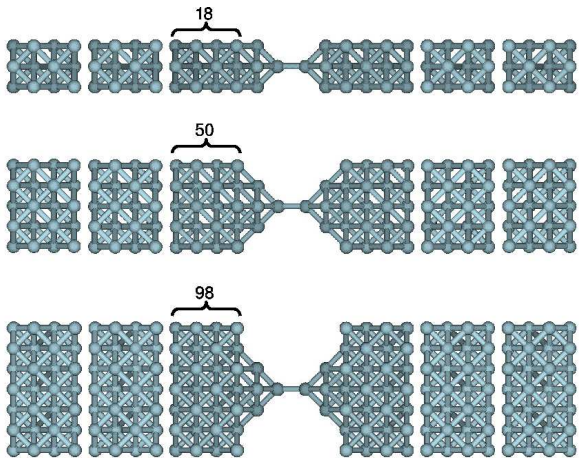}
  \end{center}
  \caption{
    Sketch of the nanocontact geometries and the corresponding
    nanowire electrodes used in the {\it ab initio} quantum 
    transport calculations with nanowire electrodes. 
  }
  \label{fig:nw-geometries}
\end{figure}

In Fig. \ref{fig:nw-geometries} the sequence of model geometries for the
case of nanowire electrodes is shown. The device region is also composed of two pyramidal
tips and also contains the unit cells used for the computation of the
semi infinite nanowire self-energies. In each step the section of the nanowire
is increased.

\subsection{An s-type conductor: Cu}

\begin{figure*}
  \begin{tabular}{cc}
    \includegraphics[width=0.5\linewidth]{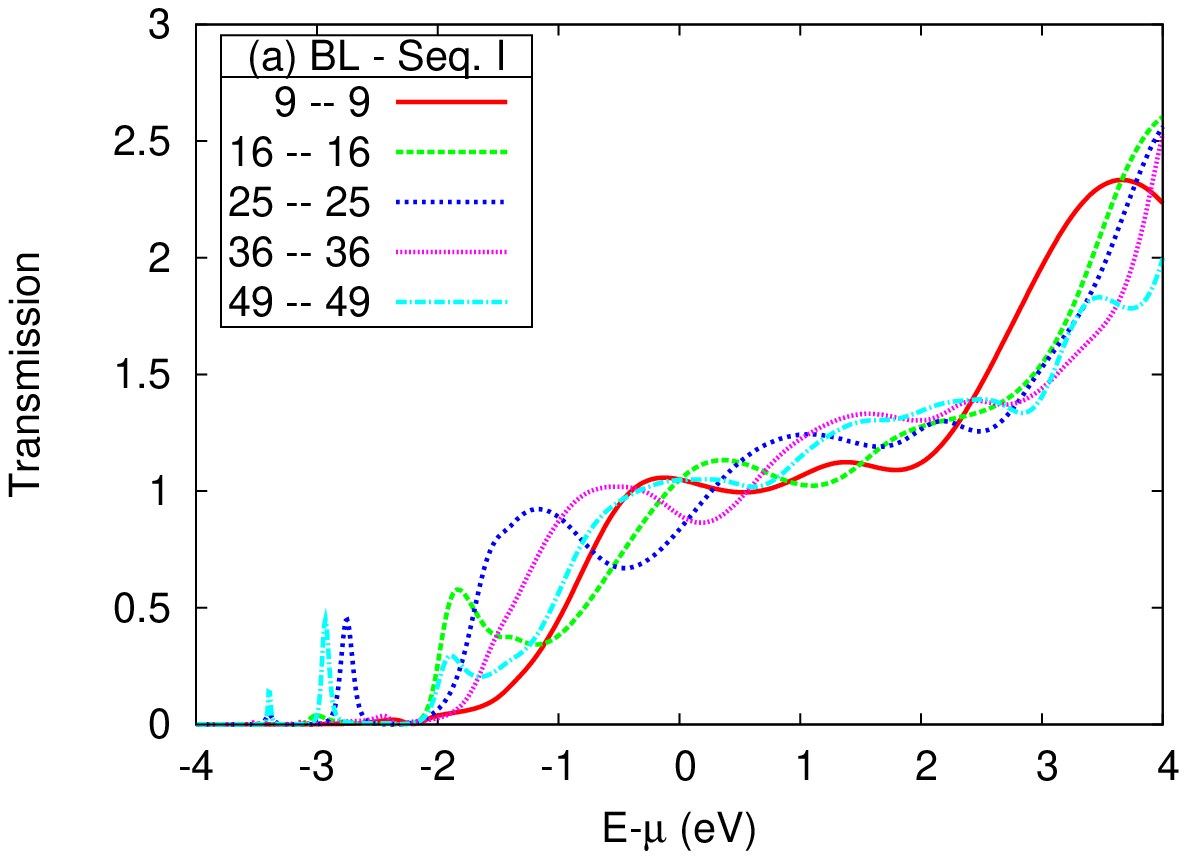} &
    \includegraphics[width=0.5\linewidth]{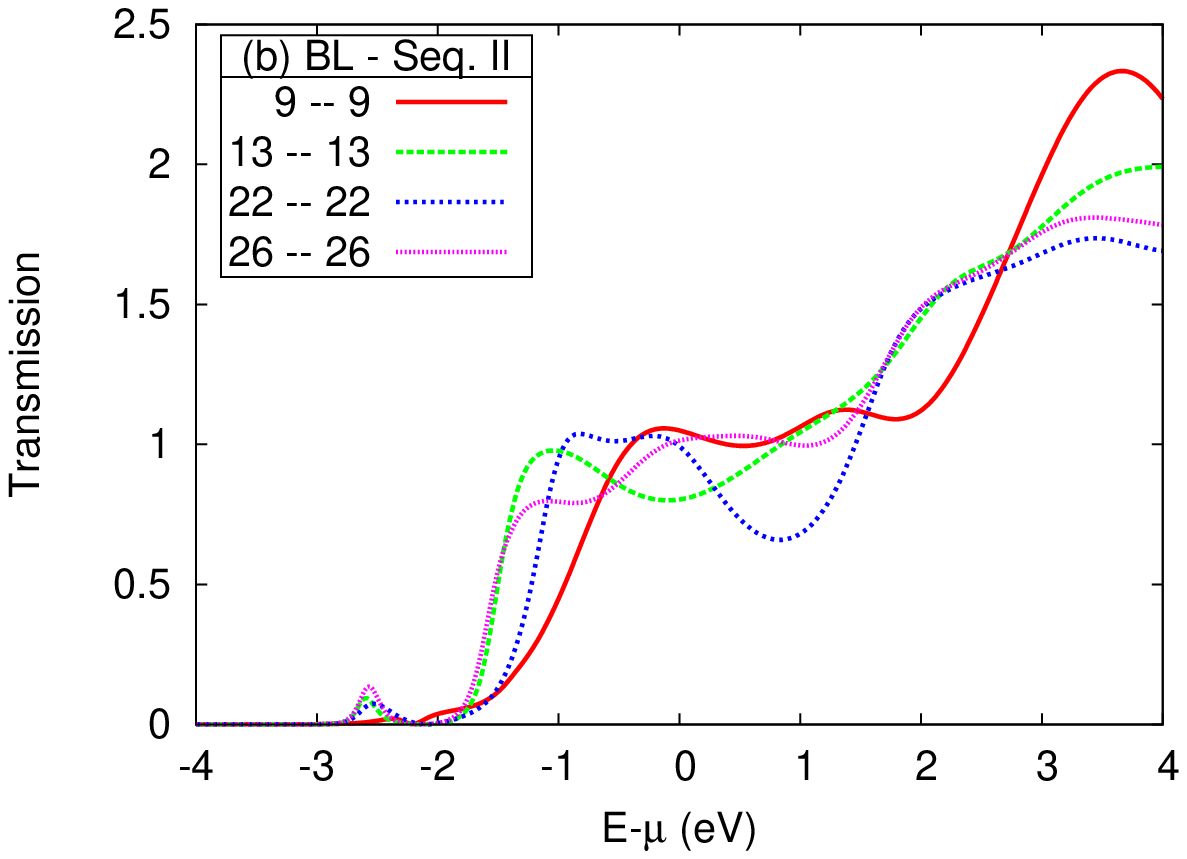} \\
    \includegraphics[width=0.5\linewidth]{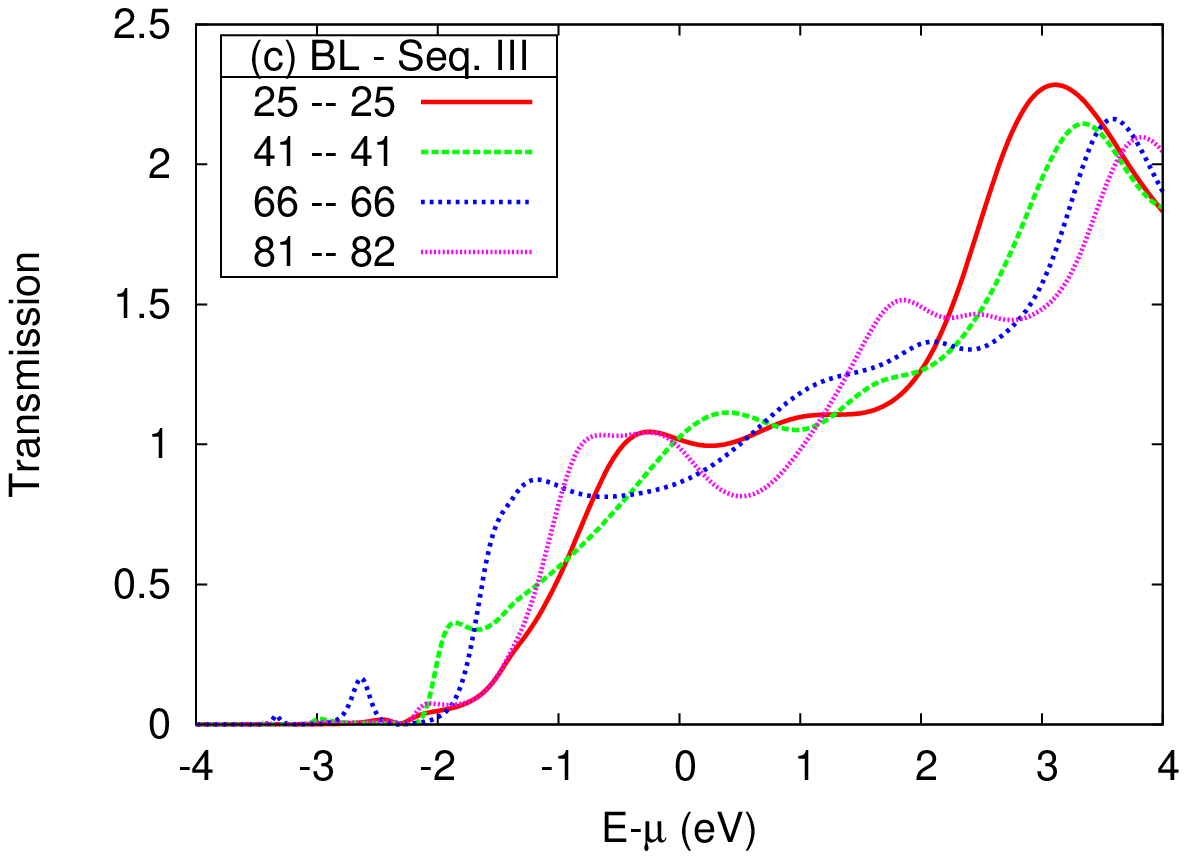} & 
    \includegraphics[width=0.5\linewidth]{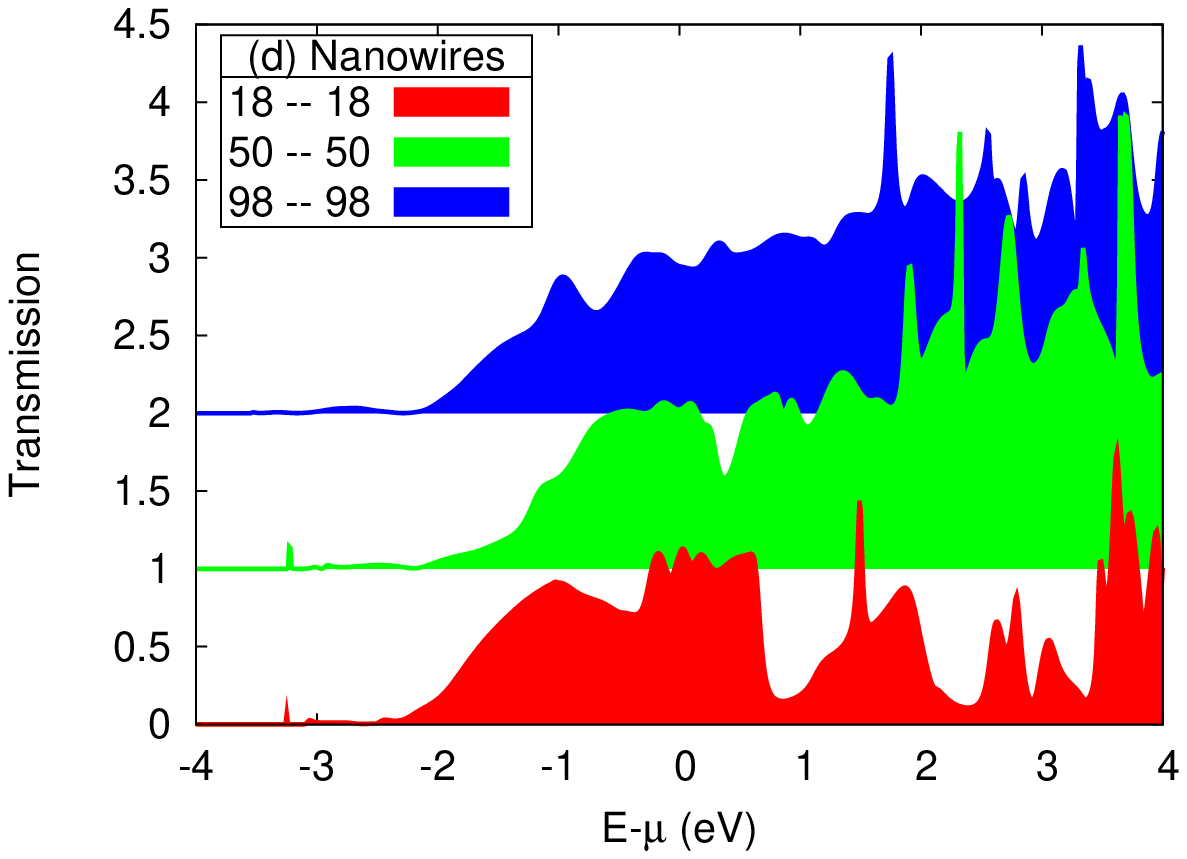}
  \end{tabular}
  \caption{
    Transmission functions for Cu nanocontacts with the same tip geometry
    but for different electrode models. 
    (a)-(c) Transmission functions calculated with Bethe lattice electrodes
    for different amounts of bulk electrode material included into the device 
    region according according to the three sequences of geometries shown in
    Fig. \ref{fig:bl-geometries}.
    (d) Transmission functions calculated with nanowire electrodes of different
    diameters according to Fig. \ref{fig:nw-geometries} (the individual transmission 
    curves have been offset by 1 in order to distinguish them from each other).
  }
  \label{fig:cu-results}
\end{figure*}
First we have studied the relatively simple situation of an $s$-type material, i.e., 
only $s$-type electrons are contributing to the DOS near the Fermi level 
and hence to the zero-bias conductance. Here we consider Cu which is a 
low-resistivity metal frequently used in nano-electronics and STM experiments.
Figs. \ref{fig:cu-results}(a)-(c) show the transmission functions 
(near the Fermi level) calculated with Bethe lattice 
electrodes. The nanocontacts share the same basic contact geometry but have different 
amounts of bulk electrode material included in the device region according 
to the different geometry sequences explained above and illustrated in 
Fig. \ref{fig:bl-geometries}. As can be seen the individual transmission 
functions vary for different geometries within each sequence and also 
between sequences. However, the overall shapes of the individual
transmission functions are very similar. Near the Fermi level all transmission
functions feature a plateau around one implying a zero-bias conductance of
approximately 1~$G_0$. The origin of this ``quantized'' conductance lies in the 
single open channel composed of Cu $4s$-orbitals which is almost perfectly conducting. The number and orbital
composition of the conducting channels can be done as explained in Ref. \onlinecite{Jacob:prb:06}. 

Fig. \ref{fig:cu-results}(d) shows the transmission functions
for Cu nanocontacts with the same basic contact geometry as before but now
with Cu nanowires of finite section serving as bulk electrodes instead 
of the Bethe lattices. Now the transmission functions are much
spikier than before, especially at higher energies. This is
somewhat reduced by increasing the width of the nanowire electrodes since the density of peaks
increases and begin to merge. The origin of this fine structure in the transmission function lies in the
well-defined conducting channels in the electrodes\cite{Ke2005b}.
Interestingly, the plateau of transmission one near the Fermi level
is clearly visible. Also the overall transmission curves are roughly similar 
to the ones obtained before with the Bethe lattice electrodes, at least
for the bigger nanowires. The computational effort is, however, greatly increased in the latter case.

The relative stability of the $T(E)\approx 1$ plateau near the Fermi level with 
respect to changes in the size and specific form of the device 
region or with respect to the electrode model 
is of course due to the low sensitivity to elastic scattering 
of the very delocalized $s$-type conduction electrons. The variations
of the transmission curves for the different electrode models and device regions
can still be attributed to the different interference patterns of the conduction electrons. 
These results are consistent with experimental evidence which shows a sharp but 
nevertheless finite-width peak in the conductance histogram of Cu 
nanocontacts at 1~$G_0$.

\subsection{An sp-type conductor: Al}

\begin{figure*}
  \begin{tabular}{cc}
    \includegraphics[width=0.5\linewidth]{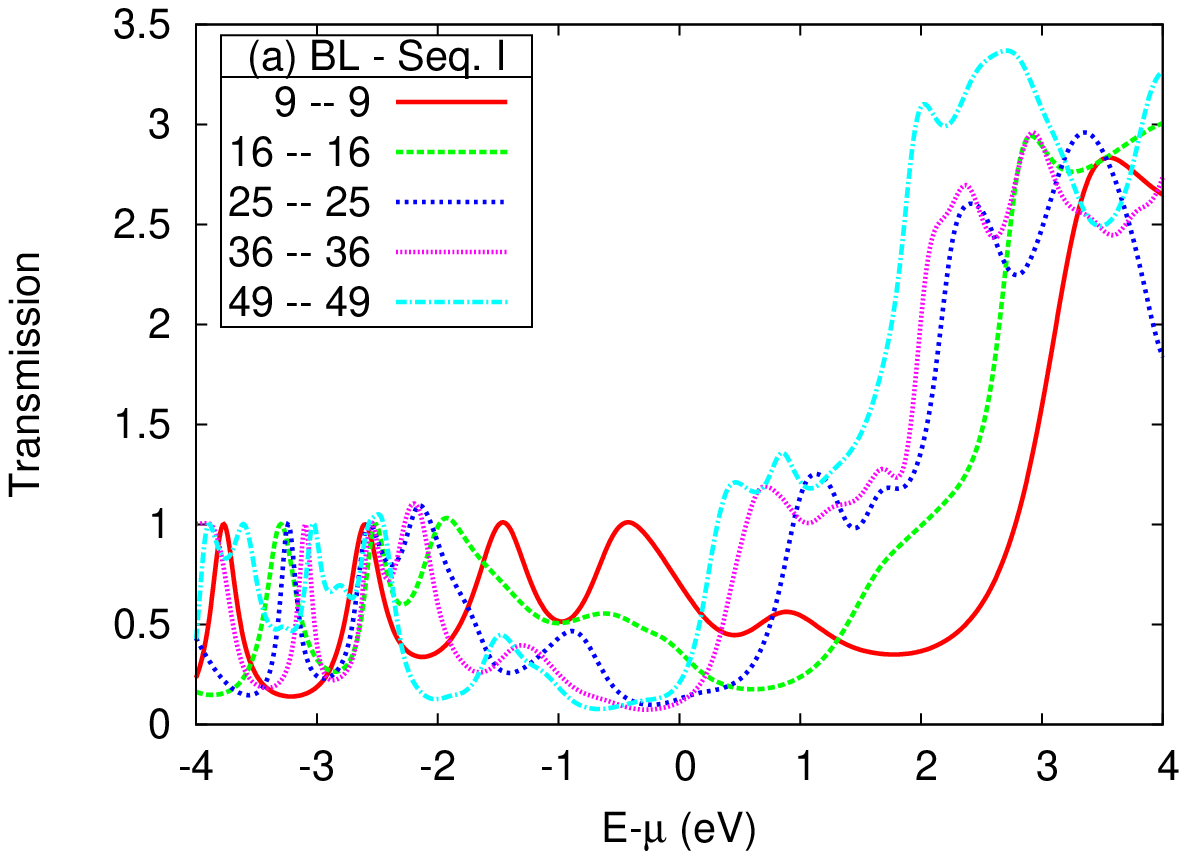} &
    \includegraphics[width=0.5\linewidth]{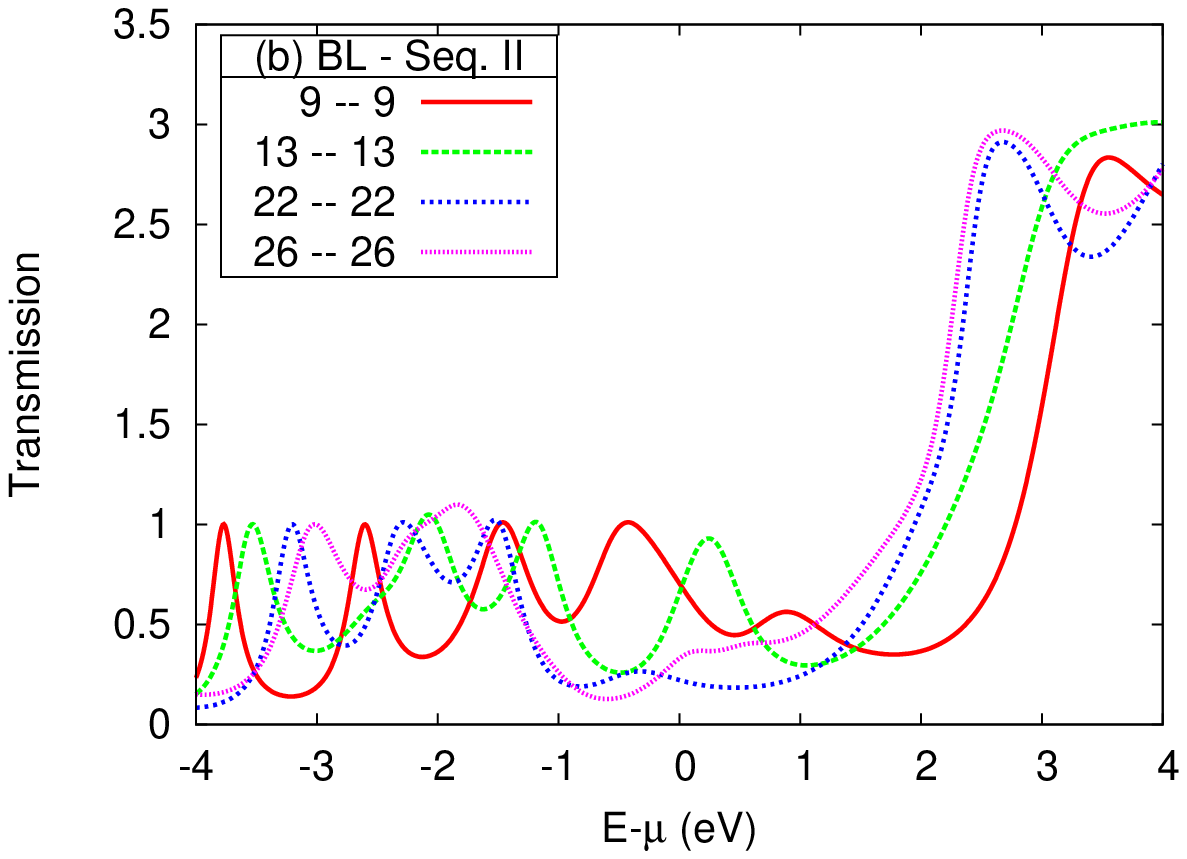} \\
    \includegraphics[width=0.5\linewidth]{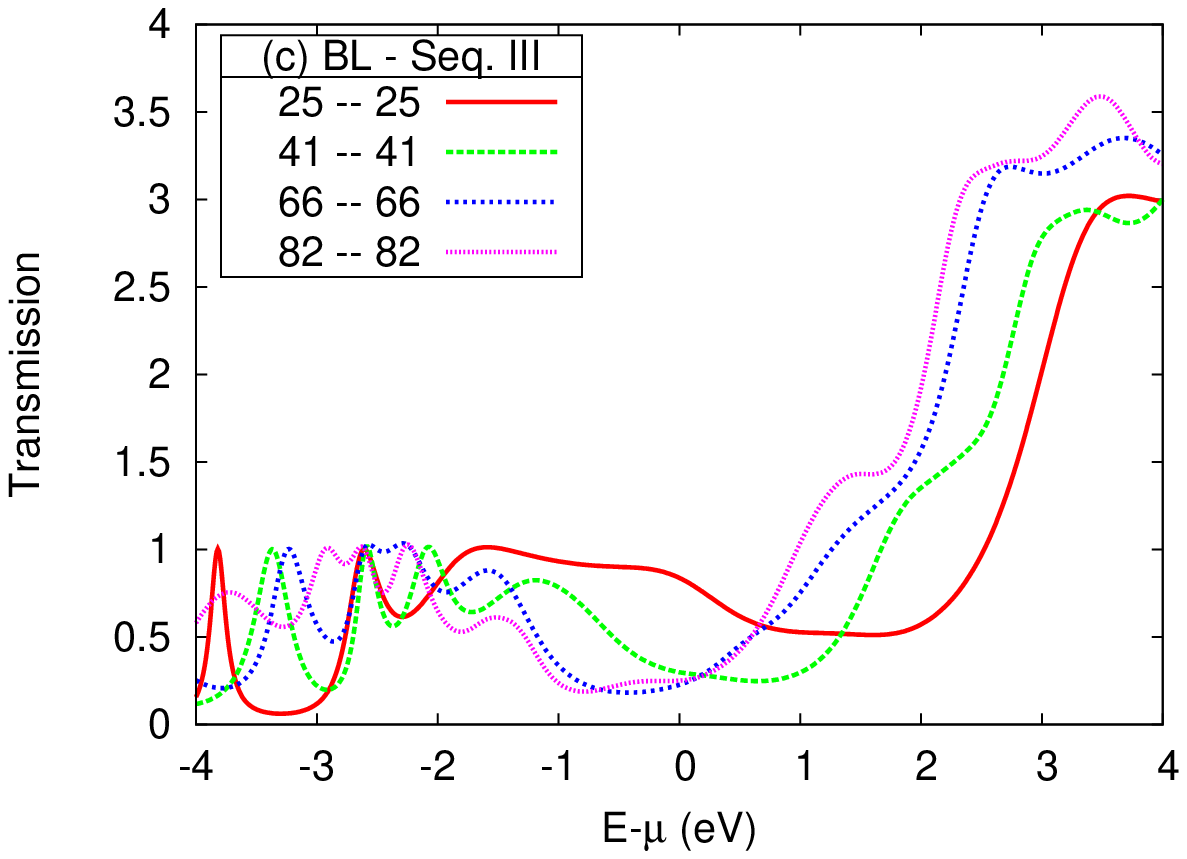} &
    \includegraphics[width=0.5\linewidth]{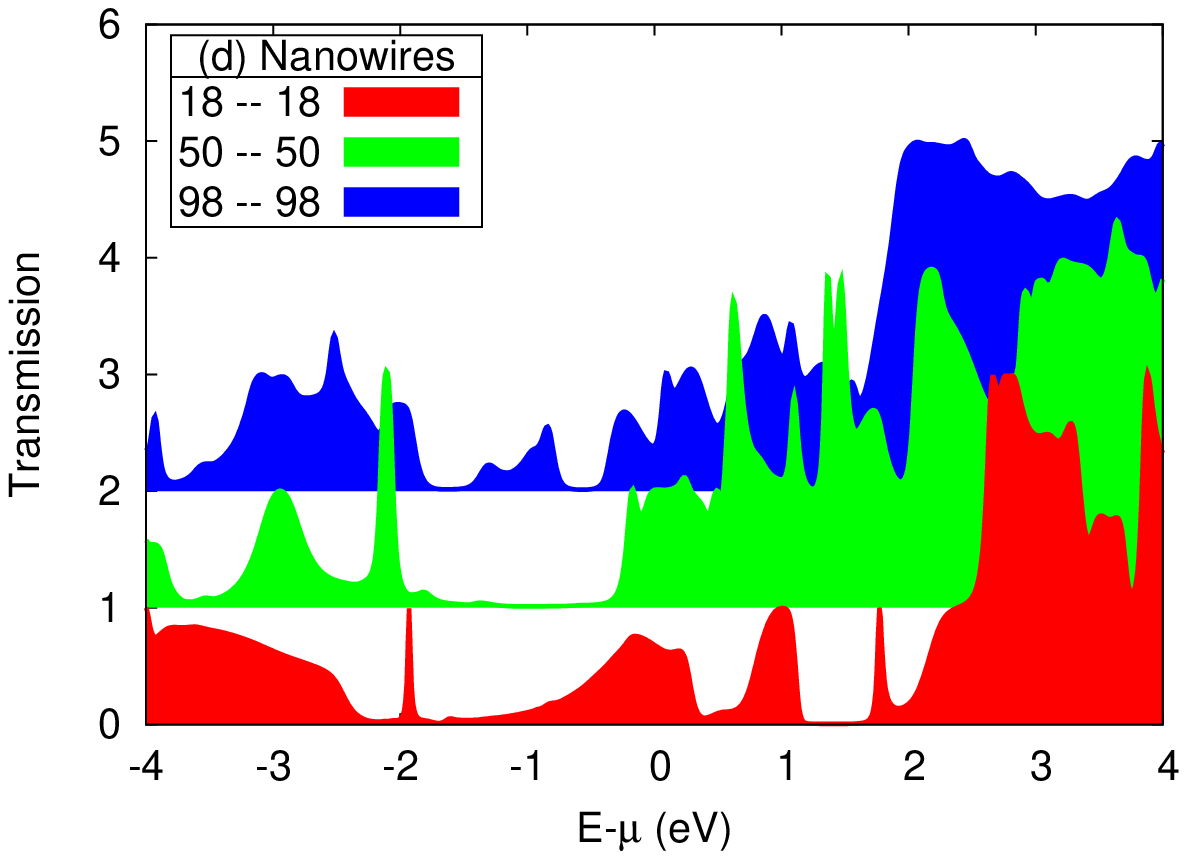}
  \end{tabular}
  \caption{
    Transmission functions for Al nanocontacts with the same tip geometry
    but for different electrode models. 
    (a)-(c) Transmission functions calculated with Bethe lattice electrodes
    for different amounts of bulk electrode material included into the device 
    region according to the three sequences of geometries shown in Fig. 
    \ref{fig:bl-geometries}.
    (d) Transmission functions calculated with nanowire electrodes of different
    diameters according to Fig. \ref{fig:nw-geometries} (the individual transmission 
    curves have been offset by 1 in order to distinguish them from each other).
  }
  \label{fig:al-results}
\end{figure*}

We now turn to a slightly more complicated situation of 
nanocontacts made from Al which is an $sp$-type conductor.
In this case we expect that changes in the geometry of
the nanoconctact should have a bigger effect on the 
transmission than in the case of a simple $s$-type conductor 
since $p$-orbitals contributing now to the conductance 
are more susceptible to elastic scattering than $s$-orbitals
due to their directionality. In fact, this can be appreciated in the different
sequences (see Fig. \ref{fig:al-results}). Now, as the device region increases, the transmission curves present
 larger variability. Furthermore, the conductance at the Fermi level
changes noticeably, approaching a definite value only for large systems where we consistently obtain 
zero-bias conductances below 0.5~$G_0$. This result does not seem to be in agreement
with experimental evidence showing, typically, zero-bias conductances
between 0.5 and 1~$G_0$ for the last conductance plateau before breaking\cite{Yanson:prl:97}. 
We remind the reader, however, that a much more complete statistical analysis is needed to draw 
conclusions in this regard. This analysis was carried out in the past\cite{PhysRevB.72.245405} 
and a good agreement was found between theory and the experimental results. A common feature to 
most transmission curves is a pronounced increase above the Fermi level. This shoulder or peak
originates from the doubly-degenerate $p_x$,$p_y$-channel while the transmission 
through the $s$- and $p_z$-channels is suppressed\cite{Cuevas:prl:98:81}.
Again the transmission functions obtained with the finite-section 
nanowire electrodes present more structure [see Fig. \ref{fig:al-results}(d)], eventually converging to
the overall shape obtained with the Bethe lattice electrodes for large enough section electrodes.

\subsection{An sd-type conductor: Ni}

As a last example we consider the case of a nanocontact made out of Ni, an $sd$ material. This 
case is the more complex from the point of view of the electronic structure since, in principle, 6 orbitals are 
expected to contribute to the conductance at the Fermi level. For simplicity's sake, we consider paramagnetic Ni, where
the two spin channels contribute equally to the current. A realistic theoretical treatment aiming
at understanding the available experimental results of truely magnetic Ni 
has been presented in the past\cite{Jacob:prb:05,Calvo:ieee:08} 
and will not be repeated here. Contrary to what one might expect, the transmission
at the Fermi level does not depend too much on the chosen size of the central region or device when
using Bethe lattice electrodes, keeping a fairly constant value around 3 even for the smallest systems. 
It is, again, in the case
of using nanowire electrodes that becomes apparent that one needs to increase the section of the wires before converging
to a similar number, at the concommitant computational cost.
The orbital analysis indicates that the $s$-channel and two $d$-channels are mainly
the ones responsible for this number. Away from the Fermi level the variations among curves are similar
to the ones in the previous cases, hardly exceeding in magnitude the transmission unity.

\begin{figure*}
  \begin{tabular}{cc}
    \includegraphics[width=0.5\linewidth]{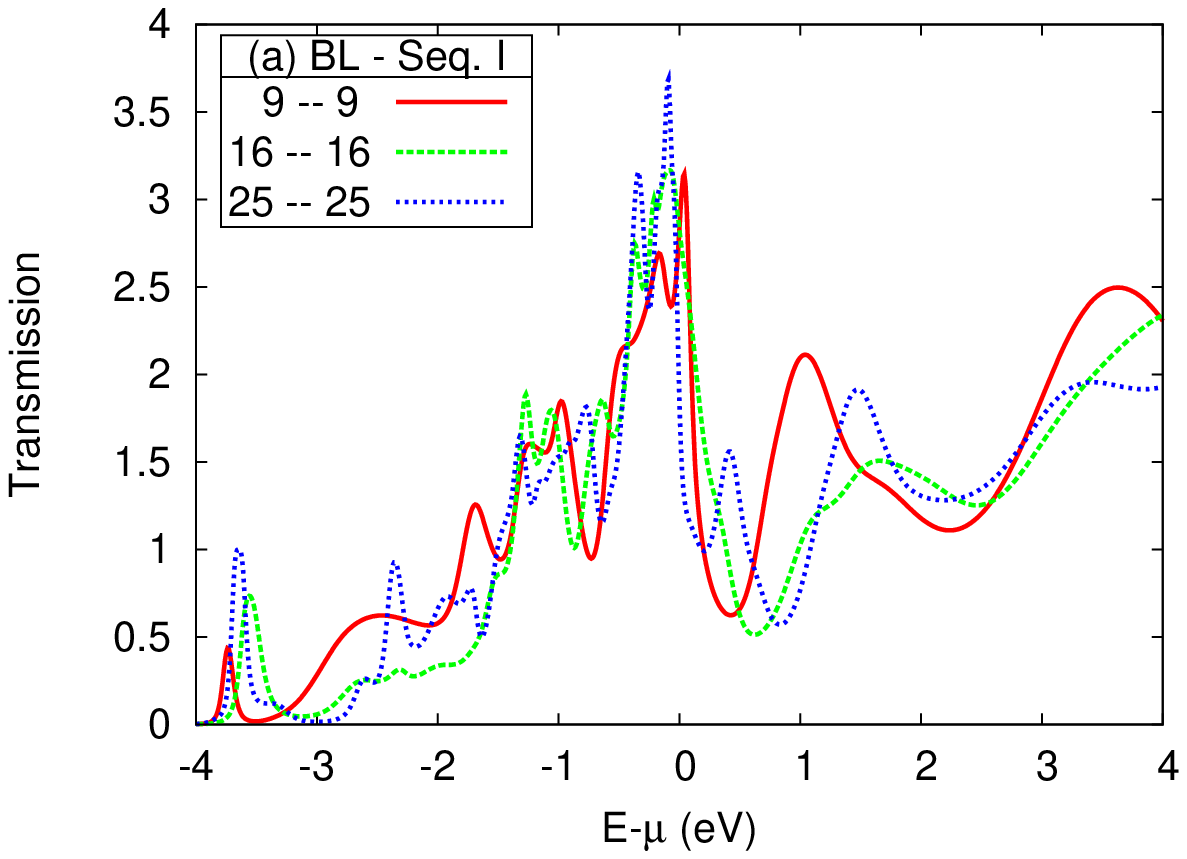} &
    \includegraphics[width=0.5\linewidth]{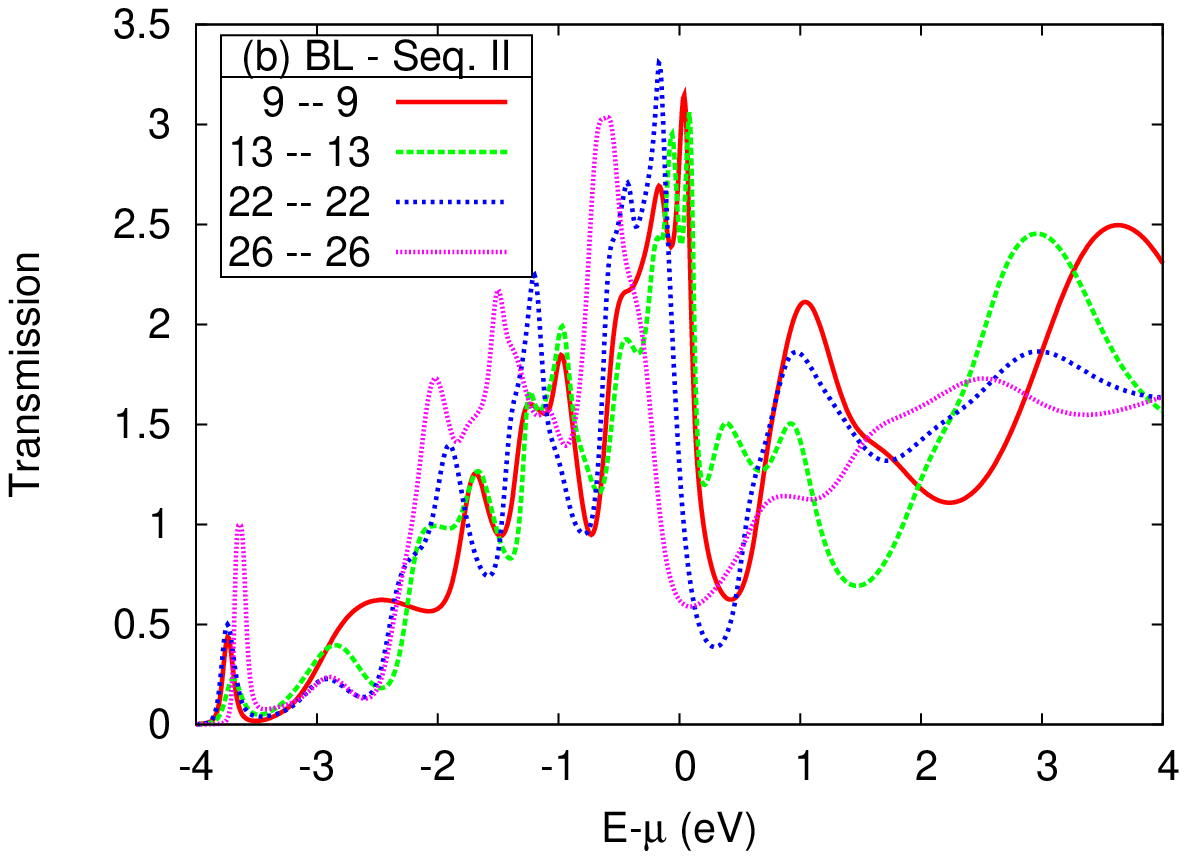} \\
    \includegraphics[width=0.5\linewidth]{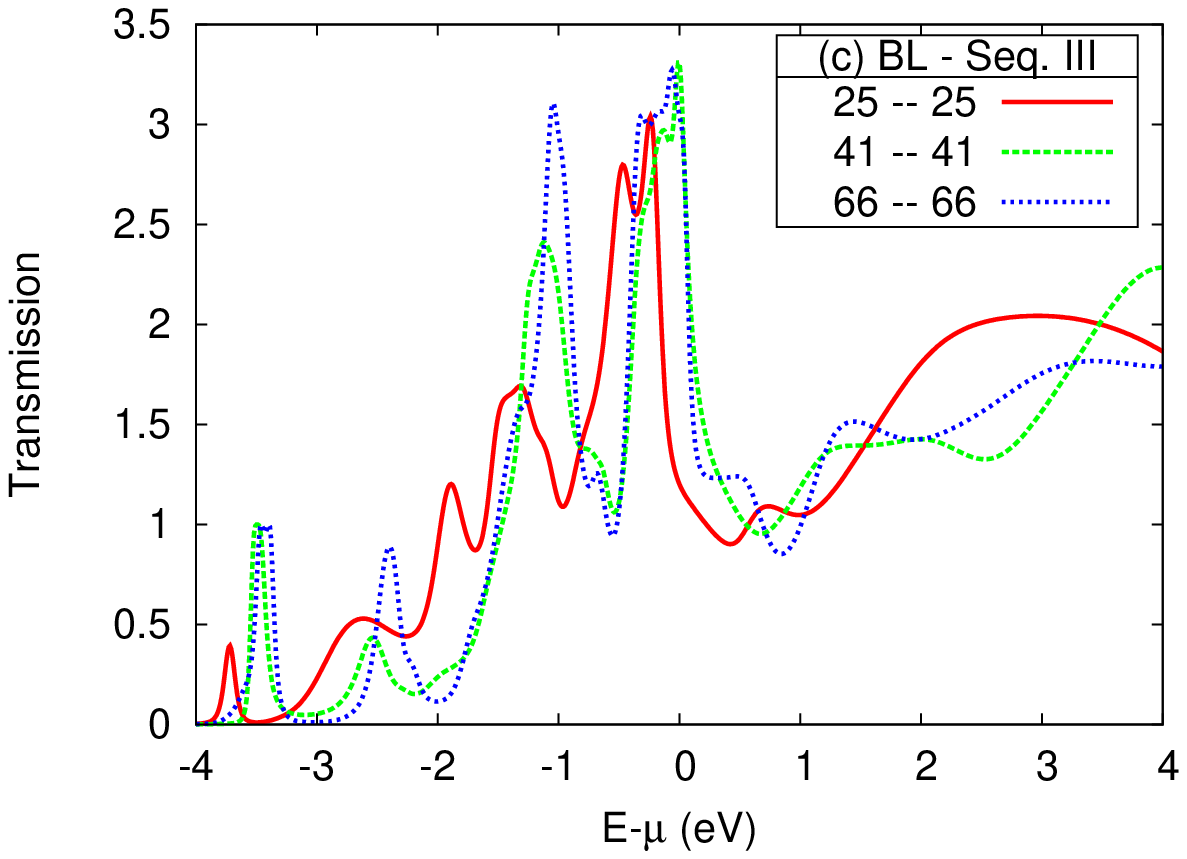} &
    \includegraphics[width=0.5\linewidth]{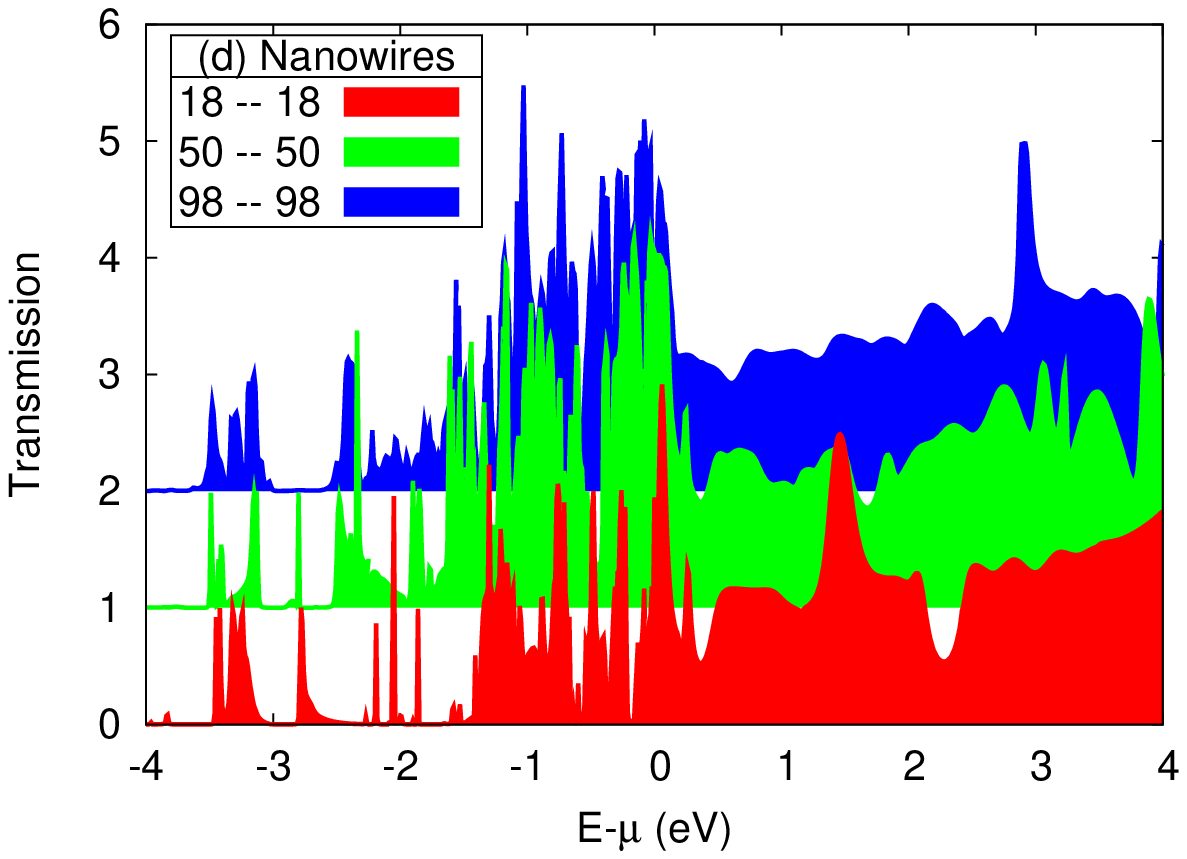}
  
\end{tabular}
  \caption{
    Transmission functions for Ni nanocontacts with the same tip geometry
    but for different electrode models. 
    (a)-(c) Transmission functions calculated with Bethe lattice electrodes
    for different amounts of bulk electrode material included into the device 
    region according to the three sequences of geometries shown in Fig. 
    \ref{fig:bl-geometries}.
    (d) Transmission functions calculated with nanowire electrodes of different
    diameters according to Fig. \ref{fig:nw-geometries} (the individual transmission 
    curves have been offset by 1 in order to distinguish them from each other).
  }
  \label{fig:ni-results}
\end{figure*}

\section{Discussion and Conclusions}
We have presented a detailed account of the theoretical and computational treatment of quantum transport in 
nanostructures. While similar analysis have been reported in the past, ours mainly focuses on implementation 
details usually skipped in the literature, but crucial for those interested in developing codes as the two 
presented here. In addition to the well-known pitfalls in the use of DFT in transport problems, the way this 
implementation is carried out determines, to a good extent, the results or the difficulty in obtaining reliable 
results that can be compared with experiments. We have thus made a critical comparison between to archetypal 
types of electrodes, which is a source of discrepancy and controversy: parametrized vs. {\em ab initio}. Without 
pretending our two codes to be representative of all the other developed by many groups, we can conclude that 
the use of parametrized electrodes presents two advantages with respect to a more faithful description of the 
electronic structure of the electrodes. First, the variability between transmission curves is greatly reduced 
even for small devices or central regions when compared to the use of nanowire electrodes. Second, the computational 
cost in the calculation of the self-energy in the former case can be orders of magnitude smaller than in the latter, 
particularly for large section wires. The use of semiinfinite wires as electrodes is, nevertheless, essential to 
properly understand scattering in a variety of systems such as true semiconducting nanowires, atomic chains\cite{Jacob08}, 
carbon nanotubes, or graphene nanoribbons\cite{Munoz-Rojas06-1}.

\section*{Acknowledgements}
This work has been partially funded by Spanish MICINN under Grants Nos. FIS2010-21883-C02-02 and CONSOLIDER CSD2007-00010.

\begin{appendix}

\section{Representation of operators in non-orthogonal basis sets}
\label{app:NOBS}

The natural definition for the matrix $\mathbf A$ of a one-body operator $\hat A$
in a non-orthogonal basis set (NOBS) $\left\{\ket{\alpha}\right\}$ is simply by its matrix elements:
\begin{equation}
  \label{eq:NOB-Mat}
  \mathbf A = ( A_{\alpha\beta} ) = \left( \bra{\alpha} \hat A \ket{\beta} \right).
\end{equation}
However, the representation of an operator in a NOBS is not that simple:
\begin{equation}
  \label{eq:NOB-Op}
  \hat A = \sum_{\alpha,\beta} \ket{\alpha} (\mathbf{S}^{-1} \mathbf{A} \mathbf{S}^{-1} )_{\alpha\beta} \bra{\beta}
\end{equation}
where $\mathbf{S}=(S_{\alpha\beta})=\left(\langle\alpha\mid\beta\rangle\right)$ is the overlap matrix. 
It is easy to see that this definition leads results in the matrix elements $A_{\alpha\beta}$ 
defined above. Then the identity operator in the NOBS representation is given by
\begin{equation}
  \label{eq:NOB-id}
  \hat{\rm I} = \sum_{\alpha,\beta} \ket{\alpha} (\mathbf{S}^{-1})_{\alpha\beta} \bra{\beta},
\end{equation}
which is also easy to proof.

Now we define a second matrix 
\begin{equation}
  \label{eq:MatTilde}
  \tilde{\mathbf{A}} := \mathbf{S}^{-1} \mathbf{A} \mathbf{S}^{-1},
\end{equation}
which is the matrix that appears above in the representation of the operator in a NOBS. 

One should take care when using the representation of an operator in a NOBS. For example, the matrix element
of an operator between two non-orthogonal orbitals $\ket{\alpha}, \ket{\beta}$ can be zero, 
$\bra{\alpha}\hat{A}\ket{\beta}=0$, but the corresponding matrix element of the matrix $\tilde{\mathbf{A}}$ does 
not necessarily vanish due to the multiplication with the inverse of the overlap matrix on both sides. 
Thus there is actually a non-zero contribution of the two orbitals to the operator although the corresponding 
matrix element of the operator is zero 

Orthogonalizing the basis set by the L\"owdin orthogonalization scheme \cite{Szabo:book:89}, the 
matrices $\mathbf{A}$ and $\tilde{\mathbf{A}}$ are transformed to the matrix $\mathbf{A}^\perp=\left(\bra{i}\hat{A}\ket{j}\right)$ 
defined in the new orthogonal basis set $\left\{\ket{i}\right\}$ according to:
\begin{equation}
  \mathbf{A}^\perp = \mathbf{S}^{-1/2} \mathbf{A} \mathbf{S}^{-1/2} = \mathbf{S}^{+1/2} \tilde{\mathbf{A}} \mathbf{S}^{+1/2}.
\end{equation}
Though there are also other orthogonalization schemes, the L\"owdin scheme is particularly useful in the
context of quantum chemistry methods based on atomic orbitals as the center of the orthogonalized orbital 
remains centered on the same atom as the original non-orthogonal orbital.

\section{Partitioning method}
\label{app:partitioning}

As explained in Sec. \ref{sec:NEGF} we model the transport 
problem by dividing the system in three
parts. Two semi-infinite leads (L) and (R) with bulk electronic structure
are connected to a finite region called device (D).
In a local basis set the Hamiltonian and the overlap matrix of the system are given
by (\ref{eq:HLDR}) and (\ref{eq:SLDR}). Dividing the F matrix into sub-matrices 
in a similar manner we obtain the following matrix equation:
\begin{widetext}
\begin{eqnarray*}
  \label{eq:GTilde-Matrix-eq}
  \lefteqn{
    \left(
      \begin{array}{ccc}
        z\,\mathbf{ S}_{\rm L} -\mathbf{ H}_{\rm L}  & z\,\mathbf{ S}_{\rm LD}-\mathbf{ H}_{\rm LD} & \mathbf{ 0}_{\rm RL}                 \\
        z\,\mathbf{ S}_{\rm DL}-\mathbf{ H}_{\rm DL} & z\,\mathbf{ S}_{\rm D} -\mathbf{ H}_{\rm D}  & z\,\mathbf{ S}_{\rm DR}-\mathbf{ H}_{\rm DR} \\
        \mathbf{ 0}_{\rm RL}                        & z\,\mathbf{ S}_{\rm RD}-\mathbf{ H}_{\rm RD} & z\,\mathbf{ H}_{\rm R} -\mathbf{ H}_{\rm R}     
      \end{array}
    \right) \times }
  \nonumber \\
  \nonumber \\
  & & \hspace{2.2cm} \times \left(
    \begin{array}{ccc}
      \mathbf{ {G}}_{\rm L}(z)  & \mathbf{ {G}}_{\rm LD}(z) & \mathbf{ {G}}_{\rm LR}(z) \\
      \mathbf{ {G}}_{\rm DL}(z) & \mathbf{ {G}}_{\rm D}(z)  & \mathbf{ {G}}_{\rm DR}(z) \\
      \mathbf{ {G}}_{\rm RL}(z) & \mathbf{ {G}}_{\rm RD}(z) & \mathbf{ {G}}_{\rm R}(z)       
    \end{array}
  \right) 
  = \left(
  \begin{array}{ccc}
    \mathbf 1_{\rm L} & \mathbf 0_{\rm LD}& \mathbf 0_{\rm LR} \\
    \mathbf 0_{\rm DL}& \mathbf 1_{\rm D} & \mathbf 0_{\rm DR} \\
    \mathbf 0_{\rm RL}& \mathbf 0_{\rm RD} & \mathbf 1_{\rm R}
  \end{array}
\right).   
\nonumber\\
\end{eqnarray*}
\end{widetext}
This yields 9 equations for the 9 sub-matrices of the GF ${\mathbf G}$. We can resolve this matrix
equation columnwise. Multiplying all rows of $E\mathbf S - \mathbf H$ with the first column of ${\mathbf G}$ 
yields three equations for ${\mathbf G}_{\rm L}$, ${\mathbf G}_{\rm DL}$ and ${\mathbf G}_{\rm RL}$ 
which yield:
\begin{eqnarray*}
  \label{eq:gf:GTilde_L}
  {\mathbf G}_{\rm L}(z) &=& ( z \mathbf S_{\rm L} - \mathbf H_{\rm L} - {\mathbf\Sigma}_{\rm D+R}(z) )^{-1} \\
  \label{eq:gf:GTilde_DL}
  {\mathbf G}_{\rm DL}(z) &=& {\mathbf g}_{\rm D+R}(z) \, (\mathbf H_{\rm DL}-z \mathbf S_{\rm DL}) \, {\mathbf G}_{\rm L}(z) \\
  \label{eq:gf:GTilde_RL}
  {\mathbf G}_{\rm RL}(z) &=& {\mathbf g}_{\rm R}(z) \, ({\mathbf H}_{\rm RD}-z {\mathbf S}_{\rm RD}) \, {\mathbf G}_{\rm DL}(z)
\end{eqnarray*}
Similarly we obtain from multiplication with the second column:
\begin{eqnarray*}
  \label{eq:gf:GTilde_D}
  {\mathbf G}_{\rm D}(z) &=& ( z {\mathbf S}_{\rm D} - {\mathbf H}_{\rm D} - {\mathbf\Sigma}_{\rm L}(z) - {\mathbf\Sigma}_{\rm R}(z) )^{-1} \\
  \label{eq:gf:GTilde_LD}
  {\mathbf G}_{\rm LD}(z) &=& {\mathbf g}_{\rm L}(z) \, (\mathbf H_{\rm LD}- z \mathbf S_{\rm LD}) \, {\mathbf G}_{\rm D}(z) \\
  \label{eq:gf:GTilde_RD}
  {\mathbf G}_{\rm RD}(z) &=& {\mathbf g}_{\rm R}(z) \, (\mathbf H_{\rm RD}- z \mathbf S_{\rm RD}) \, {\mathbf G}_{\rm D}(z)
\end{eqnarray*}
And finally from multiplication with the third column, we obtain:
\begin{eqnarray*}
  \label{eq:gf:GTilde_R}
  {\mathbf G}_{\rm R}(z) &=& ( z \mathbf S_{\rm R} - \mathbf H_{\rm R} - {\mathbf\Sigma}_{\rm D+L}(z) )^{-1} \\
  \label{eq:gf:GTilde_DR}
  {\mathbf G}_{\rm DR}(z) &=& {\mathbf g}_{\rm D+L}(z) \, (\mathbf H_{\rm DR}-z \mathbf S_{\rm DR}) \, {\mathbf G}_{\rm R}(z) \\
  \label{eq:gf:GTilde_LR}
  {\mathbf G}_{\rm LR}(z) &=& {\mathbf g}_{\rm L}(z) \, (\mathbf H_{\rm LD}-z \mathbf S_{\rm LD}) \, {\mathbf G}_{\rm DR}(z)
\end{eqnarray*}

We have introduced the Green's functions of the isolated left and right lead ${\mathbf g}_{\rm L}$ and 
${\mathbf g}_{\rm R}$ and the corresponding self-energies $\Sigma_{\rm L}$ and $\Sigma_{\rm R}$:
\begin{eqnarray*}
  \label{eq:gf:gTilde_L}
  {\mathbf g}_{\rm L}(z) &\equiv& (z \mathbf S_{\rm L} - \mathbf H_{\rm L})^{-1} \\ 
  \label{eq:gf:SigmaTilde_L}
  {\mathbf\Sigma}_{\rm L}(z) &\equiv& (\mathbf H_{\rm DL}-z \mathbf S_{\rm DL}) \, {\mathbf g}_{\rm L}(z) \, (\mathbf H_{\rm LD}-z \mathbf S_{\rm LD}) \\
  \label{eq:gf:gTilde_R}
  {\mathbf g}_{\rm R}(z) &\equiv& (z \mathbf S_{\rm R} - \mathbf H_{\rm R})^{-1} \\
  \label{eq:gf:SigmaTilde_R}
  {\mathbf\Sigma}_{\rm R}(z) &\equiv& (\mathbf H_{\rm DR}-z \mathbf S_{\rm DR}) \, {\mathbf g}_{\rm R}(z) \, (\mathbf H_{\rm RD}-z \mathbf S_{\rm RD})
\end{eqnarray*}

Furthermore, we have defined the Green's function of the device plus the left lead only, ${\mathbf g}_{\rm D+L}$,
of the device plus the right lead only, ${\mathbf g}_{\rm D+R}$, and the corresponding self-energies 
${\mathbf\Sigma}_{\rm D+L}$ and ${\mathbf\Sigma}_{\rm D+R}$ each one representing the coupling of one of the leads to 
the device and the other lead:
\begin{eqnarray*}
  \label{g_D+L}
  {\mathbf g}_{\rm D+L}(z) &\equiv& (z \mathbf S_{\rm D} - \mathbf H_{\rm D} - \Sigma_{\rm L}(z) )^{-1} \\
  \label{g_D+R}
  {\mathbf g}_{\rm D+R}(z) &\equiv& (z \mathbf S_{\rm D} - \mathbf H_{\rm D} - \Sigma_{\rm R}(z) )^{-1} \\
  \label{eq:Sigma_D+L}
  {\mathbf\Sigma}_{\rm D+R}(z) &\equiv& (\mathbf H_{\rm RD}-z \mathbf S_{\rm RD}) {\mathbf g}_{\rm D+L}(z) (\mathbf H_{\rm DR}-z \mathbf S_{\rm DR}) \\
  \label{eq:Sigma_D+R}
  {\mathbf\Sigma}_{\rm D+L}(z) &\equiv& (\mathbf H_{\rm LD}-z \mathbf S_{\rm LD}) {\mathbf g}_{\rm D+R}(z) (\mathbf H_{\rm DL}-z \mathbf S_{\rm DL})
\end{eqnarray*}

\section{Bethe lattices}
\label{app:bethe-lattices}

In this appendix we discuss how self-energies for  Bethe lattices (BL) 
used to describe the leads are calculated. A BL is generated by 
connecting a site with $N$ nearest-neighbors in directions that
could be those of a particular crystalline lattice. The new $N$ sites are 
each one connected to $N-1$ different sites and so on and so forth.
The generated lattice has the actual local topology (number of neighbors
and crystal directions) but has no rings, and thus does not describe
the  long range order characteristic of real crystals. Let 
$n$ be a generic site connected to one preceding neighbor $n-1$
and $N-1$ neighbors of the following shell ($n+i$ with $i=1,..,N-1$). 
For simplicity's sake, we carry out the derivation for an orthogonal basis. 
Following App. \ref{app:self-energy-1D}.
the generalization to the case of a non-orthogonal basis is straigthforward.

Dyson's equation
for an arbitrary non-diagonal Green's function is
\begin{equation}
\label{eq:Dyson-BL}
  (E\mathbf{ I}-\mathbf{ H}_0)\mathbf{ G}_{n,k}=\mathbf{ H}_{n,n-1}\mathbf{ G}_{n-1,k}+
  \sum_{i=1,...,N-1} \mathbf{ H}_{n,i}\mathbf{ G}_{i,k}
\end{equation}
where $k$ is an arbitrary site, $E$ the energy, 
and $\mathbf{ H}_{i,j}$ is a matrix
that incorporates interactions between orbitals at sites $i$ and $j$
(bold capital characters are used to denote matrices). 
$\mathbf{ H}_0$ is a diagonal matrix containing the orbital levels
and $\mathbf{ I}$ is the identity matrix. 
Then, we define a transfer matrix as
\begin{equation}
\label{eq:T-Mat-BL}
\mathbf{ T}_{i-1,i}\mathbf{ G}_{i-1,j}=\mathbf{ G}_{i,j}
\end{equation}
Multiplying Eq. (\ref{eq:Dyson-BL}) by the inverse of $\mathbf{ G}_{n-1,n}$ we obtain,
\begin{equation}
\label{eq:Dyson-TMat-BL}
  (E\mathbf{ I}-\mathbf{ H}_0)\mathbf{ T}_{n-1,n}=\mathbf{ H}_{n,n-1}+
  \sum_{i=1,...,N-1}\mathbf{ H}_{n,i}\mathbf{ T}_{n,i}\mathbf{ T}_{n-1,n}
\end{equation}
Due to the absence of rings the above equation is
valid for any set of lattice sites, and, thus, solving
the BL is reduced to a calculation of a few transfer matrices.
Note that a transfer matrix such as that of Eq. (\ref{eq:T-Mat-BL}) could
also be defined in a crystalline lattice but, in that case it would 
be useless. 

Eq. (\ref{eq:Dyson-TMat-BL}) can be solved iteratively,
\begin{equation}
  \label{eq:Dyson-TMat-BL-2}
  \mathbf{ T}_{n-1,n}=\left[E\mathbf{ I}-\mathbf{ H}_0-\sum_{i=1,...,N-1}
    \mathbf{ H}_{n,i}\mathbf{ T}_{n,i}\right ]^{-1}\mathbf{ H}_{n,n-1}
\end{equation}
If the orbital basis set and the lattice have full
symmetry (including inversion symmetry) the different transfer
matrices can  be obtained from just a single one through
appropriate rotations. However this is not always the case (see below).

Before proceeding any further we define self-energies that can be (and
commonly are) used in place of transfer matrices,
\begin{equation}
  \mathbf{ \Sigma}_{i,j}=\mathbf{ H}_{i,j}\mathbf{ T}_{i,j}
\end{equation}
Eq. (\ref{eq:Dyson-TMat-BL-2}) is then rewritten as,
\begin{equation}
  \mathbf{ \Sigma}_{n-1,n}=\mathbf{ H}_{n-1,n}\left[E\mathbf{ I}-\mathbf{ H}_0
    \sum_{i=1,...,N-1}
    \mathbf{ \Sigma}_{n,i}\right ]^{-1}\mathbf{ H}_{n-1,n}^{\dagger}
\end{equation}
\noindent where we have made use of the general property 
$\mathbf{ H}_{n,n-1}= \mathbf{ H}_{n-1,n}^{\dagger}$.

As discussed hereafter, in a general case of no  symmetry this would be 
a set of $N$ coupled equations ($2N$ if there is no inversion symmetry).
Symmetry can be broken due to either
the spatial atomic arrangement,  the orbitals on
the atoms that occupy each lattice site, or both. When no symmetry exists,
the following procedure has to be followed to obtain 
the self-energy in an arbitrary direction. The method is valid
for any basis set or lattice. Let $\mathbf{ \tau_i}$ be the $N$
nearest-neighbor directions of the lattice we are interested in
and ${\hat H}_\mathbf{ \tau_i}$ the interatomic interaction matrix in 
these directions. To make connection with the notation used above note that
the vector that joins site $n-1$ to site $n$, namely,
$\mathbf{ r_n-r_{n-1}}$ would necessarily be one of the lattice directions
of the set $\mathbf{ \tau_i}$.
The self-energies associated to each direction have to be obtained
from the following set of $2N$ coupled self-consistent equations,
\begin{equation}
  {\mathbf \Sigma}_{\mathbf \tau_i}={\mathbf H}_{\mathbf \tau_i}
  \left [ E{\mathbf I}-{\mathbf H}_0-
    ({\mathbf \Sigma}_{\bar T}-
    {\mathbf \Sigma}_{\bar {\mathbf \tau_i}})\right ]^{-1}
  {\mathbf H}_{\mathbf \tau_i}^{\dagger}
\end{equation}
\begin{equation}
  {\mathbf \Sigma}_{\bar {\mathbf \tau_i}}={\mathbf H}_{\bar {\mathbf \tau_i}}
  \left [ E{\mathbf I}-{\mathbf H}_0-({\mathbf \Sigma}_{T}-
    {\mathbf \Sigma}_{\mathbf \tau_i})\right ]^{-1}
  {\mathbf H}_{\bar {\mathbf \tau_i}}^{\dagger}
\end{equation}
where $i=1,...,N$ and ${\bar {\mathbf \tau_i}}=-{\mathbf \tau_i}$. 
${\mathbf H}_{\mathbf \tau_i}$ is the interatomic
interaction in the ${\mathbf \tau_i}$ direction, and 
${\mathbf \Sigma}_{T}$
and ${\mathbf \Sigma}_{\bar T}$ are the sums of  the self-energy 
matrices entering through all the Cayley tree branches attached to an 
atom and their inverses, respectively, {\it i.e.}  
\begin{equation}
  {\mathbf \Sigma}_{T}=\sum_{i=1}^{N}{\mathbf \Sigma}_{\mathbf \tau_i}
  \hspace{1em}\mbox{ and }\hspace{1em}
  {\mathbf \Sigma}_{\bar T}=\sum_{i=1}^{N}{\mathbf \Sigma}_
  {\bar{\mathbf \tau_i}}
\end{equation}
This set of $2N$ matrix equations has to be solved iteratively.
It is straightforward to check that, in cases of full symmetry,  
it reduces to the single equation.
The local density of states can be obtained from the diagonal
Green's function matrix,
\begin{equation}
  {\mathbf G}_{n,n}=\left [ E{\mathbf I}-{\mathbf H}_0
    -\sum_{i=1,..,N}{\mathbf \Sigma}_{\tau_i}\right ]^{-1}
\end{equation}
\section{Self-energy of a one-dimensional lead}
\label{app:self-energy-1D}

Here we will derive the Dyson equation (\ref{eq:DysonR}) for the calculation of
the self-energy of the semi-infinite right lead. The derivation of the Dyson equation
for the left lead (\ref{eq:DysonL}) goes in a completely analogous way.

The Hamiltonian matrix $\mathbf H_{\rm R}$ of the (isolated) semi-infinite right electrode is defined in eq.
(\ref{eq:H_R}) as:
\begin{equation}
  \mathbf{H}_{\rm R} 
  = \left(\begin{array}{ccccc}
    \mathbf H_0         & \mathbf H_1         &        &        & \mathbf 0   \\
    \mathbf H_1^\dagger & \mathbf H_0         & \mathbf H_1 &        &        \\
    \,             & \mathbf H_1^\dagger & \mathbf H_0 & \mathbf H_1 &        \\
    \mathbf 0           &                & \ddots & \ddots & \ddots
    \end{array}
  \right)
\end{equation}
and the overlap matrix is given in eq. (\ref{eq:S_R}) as:
\begin{equation}
  \label{S_R}
  \mathbf{S}_{\rm R} 
  = \left(\begin{array}{ccccc}
    \mathbf S_0         & \mathbf S_1         &        &        & \mathbf 0   \\
    \mathbf S_1^\dagger & \mathbf S_0         & \mathbf S_1 &        &        \\
    \,             & \mathbf S_1^\dagger & \mathbf S_0 & \mathbf S_1 &        \\
    \mathbf 0           &                & \ddots & \ddots & \ddots
    \end{array}
  \right)
\end{equation}
To obtain the self-energy of the lead we have to calculate the GF
of the lead from its defining equation:
\begin{equation}
  ( z \mathbf S_{\rm R} - \mathbf H_{\rm R} ) {\mathbf g}_{\rm R}(z) = \mathbf 1
\end{equation}

\begin{widetext}
In the same way as the Hamiltonian and the overlap matrix we subdivide the GF matrix ${\mathbf g}_{\rm R}$ 
into sub-matrices corresponding to the unit cells of the lead. Now the above equation for the right lead's
GF reads:
\begin{eqnarray}
  \left(\begin{array}{cccc}
    z \mathbf S_0 - \mathbf H_0  & z \mathbf S_1 - \mathbf H_1    &      &                \\
    z \mathbf S_1^\dagger - \mathbf H_1^\dagger & z \mathbf S_0 - \mathbf H_0                 & z \mathbf S_1 - \mathbf H_1    &                      \\
    \hspace{0.5cm}\ddots              & \hspace{0.5cm}\ddots              & \hspace{0.5cm}\ddots &                      
  \end{array}  
  \right)
  \left( 
    \begin{array}{ccc}
      {\mathbf g}_{1,1} & {\mathbf g}_{1,2} & \ldots \\
      {\mathbf g}_{2,1} & {\mathbf g}_{2,2} & \ldots \\
      \vdots                 & \vdots                 &
    \end{array}
  \right) 
  &=&  
  \left( 
    \begin{array}{ccc}
      \mathbf 1   & \mathbf 0   & \cdots \\
      \mathbf 0   & \mathbf 1   & \ddots \\
      \vdots & \ddots & \ddots
    \end{array}
  \right) 
  \nonumber \\
\end{eqnarray}
As explained in Sec.\ref{sec:nanowire} it suffices to calculate the ``surface'' GF, i.e. ${\mathbf g}_{1,1}$.
From multiplication of the 1st, the 2nd and so on until the $n$-th line of $(z \mathbf S_{\rm R} - \mathbf H_{\rm R})$ with the 1st column of 
${\mathbf g}_{\rm R}(z)$ we get the following chain of equations:
\begin{eqnarray}
  \label{eq:g11}
  (z \mathbf S_0 - \mathbf H_0) \, {\mathbf g}_{1,1}(z) + (z \mathbf S_1 - \mathbf H_1) \, {\mathbf g}_{2,1}(z) &=&  \mathbf 1 
  \\
  (z \mathbf S_1^\dagger - \mathbf H_1^\dagger) \, {\mathbf g}_{1,1}(z) + (z \mathbf S_0 - \mathbf H_0) \, {\mathbf g}_{2,1}(z) 
  + (z \mathbf S_1 - \mathbf H_1) \, {\mathbf g}_{3,1}(z) &=& \mathbf 0
  \\
  &\vdots& 
  \nonumber \\
  (z \mathbf S_1^\dagger - \mathbf H_1^\dagger) \, {\mathbf g}_{n-1,1}(z) + (z \mathbf S_0 - \mathbf H_0) \, {\mathbf g}_{n,1}(z)
  + (z \mathbf S_1 - \mathbf H_1) \, {\mathbf g}_{n+1,1}(z) &=& \mathbf 0
\end{eqnarray}
For $n>1$ the equations for determining ${\mathbf g}_{n,1}(z)$ all have the same structure:
\begin{eqnarray}
   \label{eq:gn1}
  (z \mathbf S_0 - \mathbf H_0) \, {\mathbf g}_{n,1}(z) &=& 
  (\mathbf H_1^\dagger - z \mathbf S_1^\dagger) \, {\mathbf g}_{n-1,1}(z) + (\mathbf H_1 - z \mathbf S_1) \, {\mathbf g}_{n+1,1}(z).
\end{eqnarray}
\end{widetext}

We define a transfer matrix for $n>1$ by:
\begin{eqnarray}
  \label{eq:T-matrix}
  \mathbf{T}_{n-1,n}(z) {\mathbf g}_{n-1,1}(z) = {\mathbf g}_{n,1}(z).
\end{eqnarray}
The transfer matrix thus transfers information from site $n-1$ to site $n$ of 
the lead, i.e. from the left to the right. Multiplying Eq. (\ref{eq:gn1}) by 
$({\mathbf{g}}_{n-1,1})^{-1}$ we obtain:
\begin{eqnarray}
  \lefteqn{(z \mathbf S_0 - \mathbf H_0) \, \mathbf{T}_{n-1,n}(z) = } \\
  && (\mathbf H_1^\dagger - z \mathbf S_1^\dagger) + (\mathbf H_1 - z \mathbf S_1) \, 
  \mathbf{T}_{n,n+1}(z) \, \mathbf{T}_{n-1,n}(z)
  \nonumber
\end{eqnarray}
Reordering we obtain the following iterative equation for the transfer matrices:
\begin{eqnarray}
  \label{eq:Dyson-TMat}
  \lefteqn{ \mathbf{T}_{n-1,n}(z) = }\\
  && (z \mathbf S_0 - \mathbf H_0 - (\mathbf H_1 - z \mathbf S_1) \, \mathbf{T}_{n,n+1}(z) )^{-1}
  \, (\mathbf H_1^\dagger - z \mathbf S_1^\dagger) 
  \nonumber
\end{eqnarray}
Since the electrode is semi-infinite it looks the same from each unit cell 
when looking to the right. Thus a given ${\mathbf{g}}_{n-1,1}$, 
results always in the same ${\mathbf{g}}_{n,1}$ {\it independent} of $n$.
Thus the transfer matrix must be independent of $n$: $\mathbf{T}_{n-1,n}(z)\equiv\mathbf{T}(z)$,
and Eq. (\ref{eq:Dyson-TMat}) allows to determine the $\mathbf{T}(z)$ self-consistently. 

We define the self-energy as $\mathbf\Sigma(z):=(\mathbf H_1 - z \mathbf S_1) \, \mathbf T(z)$, and obtain the Dyson equation for the 
self-energy:
\begin{eqnarray}
  \mathbf\Sigma(z) = (\mathbf H_1 - z \mathbf S_1) \, (z \mathbf S_0 - \mathbf H_0 - \mathbf\Sigma(z) )^{-1} \, (\mathbf H_1^\dagger - z \mathbf S_1^\dagger) \nonumber\\
\end{eqnarray}
We will now see that this self-energy is indeed identical to the one defined for the right lead 
in eq. (\ref{eq:DysonR}), i.e. $\mathbf\Sigma(z)\equiv{\mathbf\Sigma}_r(E)$. By plugging in the 
definition of the transfer matrix, eq. (\ref{eq:T-matrix}), into eq. (\ref{eq:g11}) for determining
the surface GF, ${\mathbf g}_{1,1}$ we find:
\begin{eqnarray}
 (z \mathbf S_0 - \mathbf H_0) \, {\mathbf g}_{1,1}(z) + (z \mathbf S_1 - \mathbf H_1) \, \mathbf T(z) \, {\mathbf g}_{1,1}(z)  &=&  \mathbf 1  \nonumber
\end{eqnarray}
Plugging in the definition of the self-energy we obtain: 
\begin{eqnarray}
 \Rightarrow (z \mathbf S_0 - \mathbf H_0 + \mathbf\Sigma(z) ) \, {\mathbf g}_{1,1}(z)  &=&  \mathbf 1
\end{eqnarray}

Thus we obtain for the surface GF of the right lead:
\begin{eqnarray}
  {\mathbf g}_{1,1}(z) &=& (z \mathbf S_0 - \mathbf H_0 + \mathbf\Sigma(z) )^{-1}.
\end{eqnarray}
And vice-versa the self-energy can be expressed in terms of the surface GF:
\begin{eqnarray}
  \mathbf\Sigma(z) = (\mathbf H_1 - z \mathbf S_1) \, {\mathbf g}_{1,1}(z) \, (\mathbf H_1^\dagger - z \mathbf S_1^\dagger).
\end{eqnarray}
This proofs that the self-energy $\mathbf\Sigma(z)$ defined above in terms of the transfer matrix is identical 
to the self-energy ${\mathbf\Sigma}_r(z)$ defined earlier in Sec.\ref{sec:nanowire} so that
the self-energy ${\mathbf\Sigma}_r(z)$ can be calculated iteratively by the Dyson equation (\ref{eq:DysonR}).

The proof for the left lead runs completely analogously. The surface GF of the left lead is now:
\begin{eqnarray}
  {\mathbf g}_{-1,-1}(z) &=& (z \mathbf S_0 - \mathbf H_0 + {\mathbf\Sigma}_l(z) )^{-1}.
\end{eqnarray}

\end{appendix}

\bibliography{matcon,footnotes}

\begin{thebibliography}{90}
\expandafter\ifx\csname natexlab\endcsname\relax\def\natexlab#1{#1}\fi
\expandafter\ifx\csname bibnamefont\endcsname\relax
  \def\bibnamefont#1{#1}\fi
\expandafter\ifx\csname bibfnamefont\endcsname\relax
  \def\bibfnamefont#1{#1}\fi
\expandafter\ifx\csname citenamefont\endcsname\relax
  \def\citenamefont#1{#1}\fi
\expandafter\ifx\csname url\endcsname\relax
  \def\url#1{\texttt{#1}}\fi
\expandafter\ifx\csname urlprefix\endcsname\relax\def\urlprefix{URL }\fi
\providecommand{\bibinfo}[2]{#2}
\providecommand{\eprint}[2][]{\url{#2}}

\bibitem[{\citenamefont{Agra{\"{i}}t et~al.}(2003)\citenamefont{Agra{\"{i}}t,
  Yeyati, and Ruitenbeek}}]{Agrait:pr:03}
\bibinfo{author}{\bibfnamefont{N.}~\bibnamefont{Agra{\"{i}}t}},
  \bibinfo{author}{\bibfnamefont{A.~L.} \bibnamefont{Yeyati}},
  \bibnamefont{and} \bibinfo{author}{\bibfnamefont{J.~M.~v.}
  \bibnamefont{Ruitenbeek}}, \bibinfo{journal}{Physics Reports}
  \textbf{\bibinfo{volume}{377}}, \bibinfo{pages}{81} (\bibinfo{year}{2003}),
  \bibinfo{note}{and references therein}.

\bibitem[{\citenamefont{Lang}(1995)}]{Lang:prb:95}
\bibinfo{author}{\bibfnamefont{N.~D.} \bibnamefont{Lang}},
  \bibinfo{journal}{Phys.\ Rev.\ B} \textbf{\bibinfo{volume}{52}},
  \bibinfo{pages}{5335} (\bibinfo{year}{1995}).

\bibitem[{\citenamefont{Di~Ventra et~al.}(2000)\citenamefont{Di~Ventra,
  Pantelides, and Lang}}]{PhysRevLett.84.979}
\bibinfo{author}{\bibfnamefont{M.}~\bibnamefont{Di~Ventra}},
  \bibinfo{author}{\bibfnamefont{S.~T.} \bibnamefont{Pantelides}},
  \bibnamefont{and} \bibinfo{author}{\bibfnamefont{N.~D.} \bibnamefont{Lang}},
  \bibinfo{journal}{Phys. Rev. Lett.} \textbf{\bibinfo{volume}{84}},
  \bibinfo{pages}{979} (\bibinfo{year}{2000}).

\bibitem[{\citenamefont{Taylor et~al.}(2001)\citenamefont{Taylor, Guo, and
  Wang}}]{PhysRevB.63.121104}
\bibinfo{author}{\bibfnamefont{J.}~\bibnamefont{Taylor}},
  \bibinfo{author}{\bibfnamefont{H.}~\bibnamefont{Guo}}, \bibnamefont{and}
  \bibinfo{author}{\bibfnamefont{J.}~\bibnamefont{Wang}},
  \bibinfo{journal}{Phys. Rev. B} \textbf{\bibinfo{volume}{63}},
  \bibinfo{pages}{121104} (\bibinfo{year}{2001}).

\bibitem[{\citenamefont{Palacios et~al.}(2001)\citenamefont{Palacios,
  P{\'{e}}rez-Jim{\'{e}}nez, Louis, and Verg{\'{e}}s}}]{Palacios:prb:01}
\bibinfo{author}{\bibfnamefont{J.~J.} \bibnamefont{Palacios}},
  \bibinfo{author}{\bibfnamefont{A.~J.}
  \bibnamefont{P{\'{e}}rez-Jim{\'{e}}nez}},
  \bibinfo{author}{\bibfnamefont{E.}~\bibnamefont{Louis}}, \bibnamefont{and}
  \bibinfo{author}{\bibfnamefont{J.~A.} \bibnamefont{Verg{\'{e}}s}},
  \bibinfo{journal}{Phys.\ Rev.\ B} \textbf{\bibinfo{volume}{64}},
  \bibinfo{pages}{115411} (\bibinfo{year}{2001}).

\bibitem[{\citenamefont{Xue et~al.}(2001)\citenamefont{Xue, Datta, and
  Ratner}}]{Xue:jcp:01}
\bibinfo{author}{\bibfnamefont{Y.}~\bibnamefont{Xue}},
  \bibinfo{author}{\bibfnamefont{S.}~\bibnamefont{Datta}}, \bibnamefont{and}
  \bibinfo{author}{\bibfnamefont{M.~A.} \bibnamefont{Ratner}},
  \bibinfo{journal}{J.\ Chem.\ Phys.} \textbf{\bibinfo{volume}{115}},
  \bibinfo{pages}{4292} (\bibinfo{year}{2001}).

\bibitem[{\citenamefont{Palacios et~al.}(2002)\citenamefont{Palacios,
  P{\'{e}}rez-Jim{\'{e}}nez, Louis, SanFabi{\'{a}}n, and
  Verg{\'{e}}s}}]{Palacios:prb:02}
\bibinfo{author}{\bibfnamefont{J.~J.} \bibnamefont{Palacios}},
  \bibinfo{author}{\bibfnamefont{A.~J.}
  \bibnamefont{P{\'{e}}rez-Jim{\'{e}}nez}},
  \bibinfo{author}{\bibfnamefont{E.}~\bibnamefont{Louis}},
  \bibinfo{author}{\bibfnamefont{E.}~\bibnamefont{SanFabi{\'{a}}n}},
  \bibnamefont{and} \bibinfo{author}{\bibfnamefont{J.~A.}
  \bibnamefont{Verg{\'{e}}s}}, \bibinfo{journal}{Phys.\ Rev.\ B}
  \textbf{\bibinfo{volume}{66}}, \bibinfo{pages}{035322}
  (\bibinfo{year}{2002}).

\bibitem[{\citenamefont{Brandbyge et~al.}(2002)\citenamefont{Brandbyge, Mozos,
  Ordej{\'{o}}n, Taylor, and Stokbro}}]{Brandbyge:prb:02}
\bibinfo{author}{\bibfnamefont{M.}~\bibnamefont{Brandbyge}},
  \bibinfo{author}{\bibfnamefont{J.~L.} \bibnamefont{Mozos}},
  \bibinfo{author}{\bibfnamefont{P.}~\bibnamefont{Ordej{\'{o}}n}},
  \bibinfo{author}{\bibfnamefont{J.}~\bibnamefont{Taylor}}, \bibnamefont{and}
  \bibinfo{author}{\bibfnamefont{K.}~\bibnamefont{Stokbro}},
  \bibinfo{journal}{Phys.\ Rev.\ B} \textbf{\bibinfo{volume}{65}},
  \bibinfo{pages}{165401} (\bibinfo{year}{2002}).

\bibitem[{\citenamefont{Heurich et~al.}(2002)\citenamefont{Heurich, Cuevas,
  Wenzel, and Sch{\"{o}}n}}]{Heurich:prl:02}
\bibinfo{author}{\bibfnamefont{J.}~\bibnamefont{Heurich}},
  \bibinfo{author}{\bibfnamefont{J.~C.} \bibnamefont{Cuevas}},
  \bibinfo{author}{\bibfnamefont{W.}~\bibnamefont{Wenzel}}, \bibnamefont{and}
  \bibinfo{author}{\bibfnamefont{G.}~\bibnamefont{Sch{\"{o}}n}},
  \bibinfo{journal}{Phys.\ Rev.\ Lett.} \textbf{\bibinfo{volume}{88}},
  \bibinfo{pages}{256803} (\bibinfo{year}{2002}).

\bibitem[{\citenamefont{Ventra and Lang}(2002)}]{Di-Ventra:prb:02}
\bibinfo{author}{\bibfnamefont{M.~D.} \bibnamefont{Ventra}} \bibnamefont{and}
  \bibinfo{author}{\bibfnamefont{N.~D.} \bibnamefont{Lang}},
  \bibinfo{journal}{Phys. Rev. B} \textbf{\bibinfo{volume}{65}},
  \bibinfo{pages}{045402} (\bibinfo{year}{2002}).

\bibitem[{\citenamefont{Palacios et~al.}(2003)\citenamefont{Palacios,
  P{\'{e}}rez-Jim{\'{e}}nez, Louis, SanFabi{\'{a}}n, and
  Verg{\'{e}}s}}]{PhysRevLett.90.106801}
\bibinfo{author}{\bibfnamefont{J.~J.} \bibnamefont{Palacios}},
  \bibinfo{author}{\bibfnamefont{A.~J.}
  \bibnamefont{P{\'{e}}rez-Jim{\'{e}}nez}},
  \bibinfo{author}{\bibfnamefont{E.}~\bibnamefont{Louis}},
  \bibinfo{author}{\bibfnamefont{E.}~\bibnamefont{SanFabi{\'{a}}n}},
  \bibnamefont{and} \bibinfo{author}{\bibfnamefont{J.~A.}
  \bibnamefont{Verg{\'{e}}s}}, \bibinfo{journal}{Phys. Rev. Lett.}
  \textbf{\bibinfo{volume}{90}}, \bibinfo{pages}{106801}
  (\bibinfo{year}{2003}).

\bibitem[{\citenamefont{Fujimoto and Hirose}(2003)}]{Fujimoto:prb:03}
\bibinfo{author}{\bibfnamefont{Y.}~\bibnamefont{Fujimoto}} \bibnamefont{and}
  \bibinfo{author}{\bibfnamefont{K.}~\bibnamefont{Hirose}},
  \bibinfo{journal}{Phys. Rev. B} \textbf{\bibinfo{volume}{67}},
  \bibinfo{pages}{195315} (\bibinfo{year}{2003}).

\bibitem[{\citenamefont{Louis et~al.}(2003)\citenamefont{Louis, Verg{\'{e}}s,
  Palacios, P{\'{e}}rez-Jim{\'{e}}nez, and SanFabi{\'{a}}n}}]{Louis:prb:03}
\bibinfo{author}{\bibfnamefont{E.}~\bibnamefont{Louis}},
  \bibinfo{author}{\bibfnamefont{J.~A.} \bibnamefont{Verg{\'{e}}s}},
  \bibinfo{author}{\bibfnamefont{J.~J.} \bibnamefont{Palacios}},
  \bibinfo{author}{\bibfnamefont{A.~J.}
  \bibnamefont{P{\'{e}}rez-Jim{\'{e}}nez}}, \bibnamefont{and}
  \bibinfo{author}{\bibfnamefont{E.}~\bibnamefont{SanFabi{\'{a}}n}},
  \bibinfo{journal}{Phys.\ Rev.\ B} \textbf{\bibinfo{volume}{67}},
  \bibinfo{pages}{155321} (\bibinfo{year}{2003}).

\bibitem[{\citenamefont{Xue and Ratner}(2003{\natexlab{a}})}]{Xue:prb:03:I}
\bibinfo{author}{\bibfnamefont{Y.}~\bibnamefont{Xue}} \bibnamefont{and}
  \bibinfo{author}{\bibfnamefont{M.~A.} \bibnamefont{Ratner}},
  \bibinfo{journal}{Phys. Rev. B} \textbf{\bibinfo{volume}{68}},
  \bibinfo{pages}{115406} (\bibinfo{year}{2003}{\natexlab{a}}).

\bibitem[{\citenamefont{Xue and Ratner}(2003{\natexlab{b}})}]{Xue:prb:03:II}
\bibinfo{author}{\bibfnamefont{Y.}~\bibnamefont{Xue}} \bibnamefont{and}
  \bibinfo{author}{\bibfnamefont{M.~A.} \bibnamefont{Ratner}},
  \bibinfo{journal}{Phys. Rev. B} \textbf{\bibinfo{volume}{68}},
  \bibinfo{pages}{115407} (\bibinfo{year}{2003}{\natexlab{b}}).

\bibitem[{\citenamefont{Basch and Ratner}(2003)}]{Basch2003a}
\bibinfo{author}{\bibfnamefont{H.}~\bibnamefont{Basch}} \bibnamefont{and}
  \bibinfo{author}{\bibfnamefont{M.~A.} \bibnamefont{Ratner}},
  \bibinfo{journal}{J. Chem. Phys.} \textbf{\bibinfo{volume}{119}},
  \bibinfo{pages}{11926} (\bibinfo{year}{2003}).

\bibitem[{\citenamefont{Jel{\'{i}}nek et~al.}(2003)\citenamefont{Jel{\'{i}}nek,
  P{\'{e}}rez, Ortega, and Flores}}]{Jelinek:prb:03}
\bibinfo{author}{\bibfnamefont{P.}~\bibnamefont{Jel{\'{i}}nek}},
  \bibinfo{author}{\bibfnamefont{R.}~\bibnamefont{P{\'{e}}rez}},
  \bibinfo{author}{\bibfnamefont{J.}~\bibnamefont{Ortega}}, \bibnamefont{and}
  \bibinfo{author}{\bibfnamefont{F.}~\bibnamefont{Flores}},
  \bibinfo{journal}{Phys.\ Rev.\ B} \textbf{\bibinfo{volume}{68}},
  \bibinfo{pages}{085403} (\bibinfo{year}{2003}).

\bibitem[{\citenamefont{Ke et~al.}(2004)\citenamefont{Ke, Baranger, and
  Yang}}]{Ke:prb:04}
\bibinfo{author}{\bibfnamefont{S.-H.} \bibnamefont{Ke}},
  \bibinfo{author}{\bibfnamefont{H.~U.} \bibnamefont{Baranger}},
  \bibnamefont{and} \bibinfo{author}{\bibfnamefont{W.}~\bibnamefont{Yang}},
  \bibinfo{journal}{Phys. Rev. B} \textbf{\bibinfo{volume}{70}},
  \bibinfo{pages}{085410} (\bibinfo{year}{2004}).

\bibitem[{\citenamefont{Hirose et~al.}(2004)\citenamefont{Hirose, Kobayashi,
  and Tsukada}}]{Hirose:prb:04}
\bibinfo{author}{\bibfnamefont{K.}~\bibnamefont{Hirose}},
  \bibinfo{author}{\bibfnamefont{N.}~\bibnamefont{Kobayashi}},
  \bibnamefont{and} \bibinfo{author}{\bibfnamefont{M.}~\bibnamefont{Tsukada}},
  \bibinfo{journal}{Phys. Rev. B} \textbf{\bibinfo{volume}{69}},
  \bibinfo{pages}{245412} (\bibinfo{year}{2004}).

\bibitem[{\citenamefont{Xue and Ratner}(2004)}]{Xue:prb:04:70:8}
\bibinfo{author}{\bibfnamefont{Y.}~\bibnamefont{Xue}} \bibnamefont{and}
  \bibinfo{author}{\bibfnamefont{M.~A.} \bibnamefont{Ratner}},
  \bibinfo{journal}{Phys. Rev. B} \textbf{\bibinfo{volume}{70}},
  \bibinfo{pages}{081404} (\bibinfo{year}{2004}).

\bibitem[{\citenamefont{Liang et~al.}(2004)\citenamefont{Liang, Ghosh,
  Paulsson, and Datta}}]{Liang:prb:04}
\bibinfo{author}{\bibfnamefont{G.~C.} \bibnamefont{Liang}},
  \bibinfo{author}{\bibfnamefont{A.~W.} \bibnamefont{Ghosh}},
  \bibinfo{author}{\bibfnamefont{M.}~\bibnamefont{Paulsson}}, \bibnamefont{and}
  \bibinfo{author}{\bibfnamefont{S.}~\bibnamefont{Datta}},
  \bibinfo{journal}{Phys. Rev. B} \textbf{\bibinfo{volume}{69}},
  \bibinfo{pages}{115302} (\bibinfo{year}{2004}).

\bibitem[{\citenamefont{Rocha and Sanvito}(2004)}]{Rocha:prb:04}
\bibinfo{author}{\bibfnamefont{A.~R.} \bibnamefont{Rocha}} \bibnamefont{and}
  \bibinfo{author}{\bibfnamefont{S.}~\bibnamefont{Sanvito}},
  \bibinfo{journal}{Phys.\ Rev.\ B} \textbf{\bibinfo{volume}{70}},
  \bibinfo{pages}{094406} (\bibinfo{year}{2004}).

\bibitem[{\citenamefont{Frederiksen et~al.}(2004)\citenamefont{Frederiksen,
  Brandbyge, Lorente, and Jauho}}]{Frederiksen:prl:04}
\bibinfo{author}{\bibfnamefont{T.}~\bibnamefont{Frederiksen}},
  \bibinfo{author}{\bibfnamefont{M.}~\bibnamefont{Brandbyge}},
  \bibinfo{author}{\bibfnamefont{N.}~\bibnamefont{Lorente}}, \bibnamefont{and}
  \bibinfo{author}{\bibfnamefont{A.~P.} \bibnamefont{Jauho}},
  \bibinfo{journal}{Phys. Rev. Lett.} \textbf{\bibinfo{volume}{93}},
  \bibinfo{pages}{256601} (\bibinfo{year}{2004}).

\bibitem[{\citenamefont{Thygesen and Jacobsen}(2005)}]{Thygesen:prb:05}
\bibinfo{author}{\bibfnamefont{K.~S.} \bibnamefont{Thygesen}} \bibnamefont{and}
  \bibinfo{author}{\bibfnamefont{K.~W.} \bibnamefont{Jacobsen}},
  \bibinfo{journal}{Phys. Rev. B} \textbf{\bibinfo{volume}{72}},
  \bibinfo{pages}{033401} (\bibinfo{year}{2005}).

\bibitem[{\citenamefont{Evers et~al.}(2004)\citenamefont{Evers, Weigend, and
  Koentopp}}]{Evers:prb:04}
\bibinfo{author}{\bibfnamefont{F.}~\bibnamefont{Evers}},
  \bibinfo{author}{\bibfnamefont{F.}~\bibnamefont{Weigend}}, \bibnamefont{and}
  \bibinfo{author}{\bibfnamefont{M.}~\bibnamefont{Koentopp}},
  \bibinfo{journal}{Phys. Rev. B} \textbf{\bibinfo{volume}{69}},
  \bibinfo{pages}{235411} (\bibinfo{year}{2004}).

\bibitem[{\citenamefont{Tada et~al.}(2004)\citenamefont{Tada, Kondo, and
  Yoshizawa}}]{Tada04}
\bibinfo{author}{\bibfnamefont{T.}~\bibnamefont{Tada}},
  \bibinfo{author}{\bibfnamefont{M.}~\bibnamefont{Kondo}}, \bibnamefont{and}
  \bibinfo{author}{\bibfnamefont{K.}~\bibnamefont{Yoshizawa}},
  \bibinfo{journal}{J. Chem. Phys.} \textbf{\bibinfo{volume}{121}},
  \bibinfo{pages}{8050} (\bibinfo{year}{2004}).

\bibitem[{\citenamefont{Nara et~al.}(2004)\citenamefont{Nara, Geng, Kino,
  Kobayashi, and Ohno}}]{Nara2004}
\bibinfo{author}{\bibfnamefont{J.}~\bibnamefont{Nara}},
  \bibinfo{author}{\bibfnamefont{W.~T.} \bibnamefont{Geng}},
  \bibinfo{author}{\bibfnamefont{H.}~\bibnamefont{Kino}},
  \bibinfo{author}{\bibfnamefont{N.}~\bibnamefont{Kobayashi}},
  \bibnamefont{and} \bibinfo{author}{\bibfnamefont{T.}~\bibnamefont{Ohno}},
  \bibinfo{journal}{J. Chem. Phys.} \textbf{\bibinfo{volume}{121}},
  \bibinfo{pages}{6485} (\bibinfo{year}{2004}).

\bibitem[{\citenamefont{Ferretti et~al.}(2005)\citenamefont{Ferretti,
  Calzolari, Felice, Manghi, Caldas, Nardelli, and Molinari}}]{Ferretti:prl:05}
\bibinfo{author}{\bibfnamefont{A.}~\bibnamefont{Ferretti}},
  \bibinfo{author}{\bibfnamefont{A.}~\bibnamefont{Calzolari}},
  \bibinfo{author}{\bibfnamefont{R.~D.} \bibnamefont{Felice}},
  \bibinfo{author}{\bibfnamefont{F.}~\bibnamefont{Manghi}},
  \bibinfo{author}{\bibfnamefont{M.~J.} \bibnamefont{Caldas}},
  \bibinfo{author}{\bibfnamefont{M.~B.} \bibnamefont{Nardelli}},
  \bibnamefont{and} \bibinfo{author}{\bibfnamefont{E.}~\bibnamefont{Molinari}},
  \bibinfo{journal}{Phys. Rev. Lett.} \textbf{\bibinfo{volume}{94}},
  \bibinfo{pages}{116802} (\bibinfo{year}{2005}).

\bibitem[{\citenamefont{Toher et~al.}(2005)\citenamefont{Toher, Filippetti,
  Sanvito, and Burke}}]{Toher:prl:05}
\bibinfo{author}{\bibfnamefont{C.}~\bibnamefont{Toher}},
  \bibinfo{author}{\bibfnamefont{A.}~\bibnamefont{Filippetti}},
  \bibinfo{author}{\bibfnamefont{S.}~\bibnamefont{Sanvito}}, \bibnamefont{and}
  \bibinfo{author}{\bibfnamefont{K.}~\bibnamefont{Burke}},
  \bibinfo{journal}{Phys. Rev. Lett.} \textbf{\bibinfo{volume}{95}},
  \bibinfo{pages}{146402} (\bibinfo{year}{2005}).

\bibitem[{\citenamefont{Fujimoto et~al.}(2005)\citenamefont{Fujimoto, Asari,
  Kondo, Nara, and Ohno}}]{Fujimoto:prb:05}
\bibinfo{author}{\bibfnamefont{Y.}~\bibnamefont{Fujimoto}},
  \bibinfo{author}{\bibfnamefont{Y.}~\bibnamefont{Asari}},
  \bibinfo{author}{\bibfnamefont{H.}~\bibnamefont{Kondo}},
  \bibinfo{author}{\bibfnamefont{J.}~\bibnamefont{Nara}}, \bibnamefont{and}
  \bibinfo{author}{\bibfnamefont{T.}~\bibnamefont{Ohno}},
  \bibinfo{journal}{Phys. Rev. B} \textbf{\bibinfo{volume}{72}},
  \bibinfo{pages}{113407} (\bibinfo{year}{2005}).

\bibitem[{\citenamefont{Asari et~al.}(2005)\citenamefont{Asari, Nara,
  Kobayashi, and Ohno}}]{Asari:prb:05}
\bibinfo{author}{\bibfnamefont{Y.}~\bibnamefont{Asari}},
  \bibinfo{author}{\bibfnamefont{J.}~\bibnamefont{Nara}},
  \bibinfo{author}{\bibfnamefont{N.}~\bibnamefont{Kobayashi}},
  \bibnamefont{and} \bibinfo{author}{\bibfnamefont{T.}~\bibnamefont{Ohno}},
  \bibinfo{journal}{Phys. Rev. B} \textbf{\bibinfo{volume}{72}},
  \bibinfo{pages}{035459} (\bibinfo{year}{2005}).

\bibitem[{\citenamefont{Wu et~al.}(2005)\citenamefont{Wu, Li, Huang, and
  Yang}}]{Wu05}
\bibinfo{author}{\bibfnamefont{X.}~\bibnamefont{Wu}},
  \bibinfo{author}{\bibfnamefont{Q.}~\bibnamefont{Li}},
  \bibinfo{author}{\bibfnamefont{J.}~\bibnamefont{Huang}}, \bibnamefont{and}
  \bibinfo{author}{\bibfnamefont{J.}~\bibnamefont{Yang}}, \bibinfo{journal}{J.
  Chem. Phys.} \textbf{\bibinfo{volume}{123}}, \bibinfo{pages}{184712}
  (\bibinfo{year}{2005}).

\bibitem[{\citenamefont{Choi et~al.}(2005)\citenamefont{Choi, Kim, Park,
  Tarakeshwar, Kim, Kim, and Lee}}]{Choi2005}
\bibinfo{author}{\bibfnamefont{Y.~C.} \bibnamefont{Choi}},
  \bibinfo{author}{\bibfnamefont{W.~Y.} \bibnamefont{Kim}},
  \bibinfo{author}{\bibfnamefont{K.~S.} \bibnamefont{Park}},
  \bibinfo{author}{\bibfnamefont{P.}~\bibnamefont{Tarakeshwar}},
  \bibinfo{author}{\bibfnamefont{K.~S.} \bibnamefont{Kim}},
  \bibinfo{author}{\bibfnamefont{T.~S.} \bibnamefont{Kim}}, \bibnamefont{and}
  \bibinfo{author}{\bibfnamefont{J.~Y.} \bibnamefont{Lee}},
  \bibinfo{journal}{J. Chem. Phys.} \textbf{\bibinfo{volume}{122}},
  \bibinfo{pages}{094706} (\bibinfo{year}{2005}).

\bibitem[{\citenamefont{Ke et~al.}(2005{\natexlab{a}})\citenamefont{Ke,
  Baranger, and Yang}}]{Ke2005b}
\bibinfo{author}{\bibfnamefont{S.~H.} \bibnamefont{Ke}},
  \bibinfo{author}{\bibfnamefont{H.~U.} \bibnamefont{Baranger}},
  \bibnamefont{and} \bibinfo{author}{\bibfnamefont{W.~T.} \bibnamefont{Yang}},
  \bibinfo{journal}{J. Chem. Phys.} \textbf{\bibinfo{volume}{123}},
  \bibinfo{pages}{114701} (\bibinfo{year}{2005}{\natexlab{a}}).

\bibitem[{\citenamefont{Ke et~al.}(2005{\natexlab{b}})\citenamefont{Ke,
  Baranger, and Yang}}]{Ke2005a}
\bibinfo{author}{\bibfnamefont{S.~H.} \bibnamefont{Ke}},
  \bibinfo{author}{\bibfnamefont{H.~U.} \bibnamefont{Baranger}},
  \bibnamefont{and} \bibinfo{author}{\bibfnamefont{W.~T.} \bibnamefont{Yang}},
  \bibinfo{journal}{J. Chem. Phys.} \textbf{\bibinfo{volume}{122}},
  \bibinfo{pages}{074704} (\bibinfo{year}{2005}{\natexlab{b}}).

\bibitem[{\citenamefont{Bagrets et~al.}(2006)\citenamefont{Bagrets,
  Papanikolaou, and Mertig}}]{Bagrets:prb:06}
\bibinfo{author}{\bibfnamefont{A.}~\bibnamefont{Bagrets}},
  \bibinfo{author}{\bibfnamefont{N.}~\bibnamefont{Papanikolaou}},
  \bibnamefont{and} \bibinfo{author}{\bibfnamefont{I.}~\bibnamefont{Mertig}},
  \bibinfo{journal}{Phys. Rev. B} \textbf{\bibinfo{volume}{73}},
  \bibinfo{pages}{045428} (\bibinfo{year}{2006}).

\bibitem[{\citenamefont{Smogunov et~al.}(2006)\citenamefont{Smogunov, DalCorso,
  and Tosatti}}]{Smogunov:prb:06}
\bibinfo{author}{\bibfnamefont{A.}~\bibnamefont{Smogunov}},
  \bibinfo{author}{\bibfnamefont{A.}~\bibnamefont{DalCorso}}, \bibnamefont{and}
  \bibinfo{author}{\bibfnamefont{E.}~\bibnamefont{Tosatti}},
  \bibinfo{journal}{Phys. Rev. B} \textbf{\bibinfo{volume}{73}},
  \bibinfo{pages}{075418} (\bibinfo{year}{2006}).

\bibitem[{\citenamefont{Jiang et~al.}(2006)\citenamefont{Jiang, Kula, and
  Luo}}]{Jiang06}
\bibinfo{author}{\bibfnamefont{J.}~\bibnamefont{Jiang}},
  \bibinfo{author}{\bibfnamefont{M.}~\bibnamefont{Kula}}, \bibnamefont{and}
  \bibinfo{author}{\bibfnamefont{Y.}~\bibnamefont{Luo}}, \bibinfo{journal}{J.
  Chem. Phys.} \textbf{\bibinfo{volume}{124}}, \bibinfo{pages}{034708}
  (\bibinfo{year}{2006}).

\bibitem[{\citenamefont{Havu et~al.}(2006)\citenamefont{Havu, Havu, Puska,
  Hakala, Foster, and Nieminen}}]{Havu2006}
\bibinfo{author}{\bibfnamefont{P.}~\bibnamefont{Havu}},
  \bibinfo{author}{\bibfnamefont{V.}~\bibnamefont{Havu}},
  \bibinfo{author}{\bibfnamefont{M.~J.} \bibnamefont{Puska}},
  \bibinfo{author}{\bibfnamefont{M.~H.} \bibnamefont{Hakala}},
  \bibinfo{author}{\bibfnamefont{A.~S.} \bibnamefont{Foster}},
  \bibnamefont{and} \bibinfo{author}{\bibfnamefont{R.~M.}
  \bibnamefont{Nieminen}}, \bibinfo{journal}{J. Chem. Phys.}
  \textbf{\bibinfo{volume}{124}}, \bibinfo{pages}{054707}
  (\bibinfo{year}{2006}).

\bibitem[{\citenamefont{Hod et~al.}(2006)\citenamefont{Hod, Peralta, and
  Scuseria}}]{Hod2006}
\bibinfo{author}{\bibfnamefont{O.}~\bibnamefont{Hod}},
  \bibinfo{author}{\bibfnamefont{J.~E.} \bibnamefont{Peralta}},
  \bibnamefont{and} \bibinfo{author}{\bibfnamefont{G.~E.}
  \bibnamefont{Scuseria}}, \bibinfo{journal}{J. Chem. Phys.}
  \textbf{\bibinfo{volume}{125}}, \bibinfo{pages}{114704}
  (\bibinfo{year}{2006}).

\bibitem[{\citenamefont{Rocha et~al.}(2006)\citenamefont{Rocha, Garcia-Suarez,
  Bailey, Lambert, Ferrer, and Sanvito}}]{Rocha:prb:06}
\bibinfo{author}{\bibfnamefont{A.~R.} \bibnamefont{Rocha}},
  \bibinfo{author}{\bibfnamefont{V.~M.} \bibnamefont{Garcia-Suarez}},
  \bibinfo{author}{\bibfnamefont{S.}~\bibnamefont{Bailey}},
  \bibinfo{author}{\bibfnamefont{C.}~\bibnamefont{Lambert}},
  \bibinfo{author}{\bibfnamefont{J.}~\bibnamefont{Ferrer}}, \bibnamefont{and}
  \bibinfo{author}{\bibfnamefont{S.}~\bibnamefont{Sanvito}},
  \bibinfo{journal}{Phys. Rev. B} \textbf{\bibinfo{volume}{73}},
  \bibinfo{pages}{085414} (\bibinfo{year}{2006}).

\bibitem[{\citenamefont{Nakamura and Yamashita}(2006)}]{Nakamura2006a}
\bibinfo{author}{\bibfnamefont{H.}~\bibnamefont{Nakamura}} \bibnamefont{and}
  \bibinfo{author}{\bibfnamefont{K.}~\bibnamefont{Yamashita}},
  \bibinfo{journal}{J. Chem. Phys.} \textbf{\bibinfo{volume}{125}},
  \bibinfo{pages}{194106} (\bibinfo{year}{2006}).

\bibitem[{\citenamefont{Prociuk et~al.}(2006)\citenamefont{Prociuk, Van~Kuiken,
  and Dunietz}}]{Prociuk2006}
\bibinfo{author}{\bibfnamefont{A.}~\bibnamefont{Prociuk}},
  \bibinfo{author}{\bibfnamefont{B.}~\bibnamefont{Van~Kuiken}},
  \bibnamefont{and} \bibinfo{author}{\bibfnamefont{B.~D.}
  \bibnamefont{Dunietz}}, \bibinfo{journal}{J. Chem. Phys.}
  \textbf{\bibinfo{volume}{125}}, \bibinfo{pages}{204717}
  (\bibinfo{year}{2006}).

\bibitem[{\citenamefont{Mera et~al.}(2007)\citenamefont{Mera, Bokes, and
  Godby}}]{Mera2007}
\bibinfo{author}{\bibfnamefont{H.}~\bibnamefont{Mera}},
  \bibinfo{author}{\bibfnamefont{P.}~\bibnamefont{Bokes}}, \bibnamefont{and}
  \bibinfo{author}{\bibfnamefont{R.~W.} \bibnamefont{Godby}},
  \bibinfo{journal}{Phys. Rev. B} \textbf{\bibinfo{volume}{76}},
  \bibinfo{pages}{125319} (\bibinfo{year}{2007}).

\bibitem[{\citenamefont{Mizuseki et~al.}(2007)\citenamefont{Mizuseki,
  Belosludov, Uehara, Lee, and Kawazoe}}]{Mizuseki2007}
\bibinfo{author}{\bibfnamefont{H.}~\bibnamefont{Mizuseki}},
  \bibinfo{author}{\bibfnamefont{R.~V.} \bibnamefont{Belosludov}},
  \bibinfo{author}{\bibfnamefont{T.}~\bibnamefont{Uehara}},
  \bibinfo{author}{\bibfnamefont{S.~U.} \bibnamefont{Lee}}, \bibnamefont{and}
  \bibinfo{author}{\bibfnamefont{Y.}~\bibnamefont{Kawazoe}}, in
  \emph{\bibinfo{booktitle}{4th Conference of the
  Asian-Consortium-on-Computational-Materials-Science}}
  (\bibinfo{address}{Seoul, South Korea}, \bibinfo{year}{2007}), pp.
  \bibinfo{pages}{1197--1201}.

\bibitem[{\citenamefont{Nakamura}(2007)}]{Nakamura2007}
\bibinfo{author}{\bibfnamefont{H.}~\bibnamefont{Nakamura}}, in
  \emph{\bibinfo{booktitle}{12th International Conference on Vibrations at
  Surfaces}} (\bibinfo{address}{Erice, Italy}, \bibinfo{year}{2007}).

\bibitem[{\citenamefont{Qian et~al.}(2007)\citenamefont{Qian, Li, Hou, Xue, and
  Sanvito}}]{Qian2007b}
\bibinfo{author}{\bibfnamefont{Z.}~\bibnamefont{Qian}},
  \bibinfo{author}{\bibfnamefont{R.}~\bibnamefont{Li}},
  \bibinfo{author}{\bibfnamefont{S.}~\bibnamefont{Hou}},
  \bibinfo{author}{\bibfnamefont{Z.}~\bibnamefont{Xue}}, \bibnamefont{and}
  \bibinfo{author}{\bibfnamefont{S.}~\bibnamefont{Sanvito}},
  \bibinfo{journal}{J. Chem. Phys.} \textbf{\bibinfo{volume}{127}},
  \bibinfo{pages}{194710} (\bibinfo{year}{2007}).

\bibitem[{\citenamefont{Thygesen and Rubio}(2007)}]{Thygesen2007}
\bibinfo{author}{\bibfnamefont{K.~S.} \bibnamefont{Thygesen}} \bibnamefont{and}
  \bibinfo{author}{\bibfnamefont{A.}~\bibnamefont{Rubio}}, \bibinfo{journal}{J.
  Chem. Phys.} \textbf{\bibinfo{volume}{126}}, \bibinfo{pages}{091101}
  (\bibinfo{year}{2007}).

\bibitem[{\citenamefont{Pauly et~al.}(2008)\citenamefont{Pauly, Viljas, Huniar,
  Hafner, Wohlthat, Burkle, Cuevas, and Schon}}]{Pauly08}
\bibinfo{author}{\bibfnamefont{F.}~\bibnamefont{Pauly}},
  \bibinfo{author}{\bibfnamefont{J.~K.} \bibnamefont{Viljas}},
  \bibinfo{author}{\bibfnamefont{U.}~\bibnamefont{Huniar}},
  \bibinfo{author}{\bibfnamefont{M.}~\bibnamefont{Hafner}},
  \bibinfo{author}{\bibfnamefont{S.}~\bibnamefont{Wohlthat}},
  \bibinfo{author}{\bibfnamefont{M.}~\bibnamefont{Burkle}},
  \bibinfo{author}{\bibfnamefont{J.~C.} \bibnamefont{Cuevas}},
  \bibnamefont{and} \bibinfo{author}{\bibfnamefont{G.}~\bibnamefont{Schon}},
  \bibinfo{journal}{New J. Phys.} \textbf{\bibinfo{volume}{10}},
  \bibinfo{pages}{125019} (\bibinfo{year}{2008}).

\bibitem[{\citenamefont{Liu et~al.}(2008)\citenamefont{Liu, Wang, Zhao, Guo,
  Yin, Boey, and Zhang}}]{ISI:000257822200012}
\bibinfo{author}{\bibfnamefont{H.}~\bibnamefont{Liu}},
  \bibinfo{author}{\bibfnamefont{N.}~\bibnamefont{Wang}},
  \bibinfo{author}{\bibfnamefont{J.}~\bibnamefont{Zhao}},
  \bibinfo{author}{\bibfnamefont{Y.}~\bibnamefont{Guo}},
  \bibinfo{author}{\bibfnamefont{X.}~\bibnamefont{Yin}},
  \bibinfo{author}{\bibfnamefont{F.~Y.~C.} \bibnamefont{Boey}},
  \bibnamefont{and} \bibinfo{author}{\bibfnamefont{H.}~\bibnamefont{Zhang}},
  \bibinfo{journal}{Chem. Phys. Chem.} \textbf{\bibinfo{volume}{9}},
  \bibinfo{pages}{1416} (\bibinfo{year}{2008}).

\bibitem[{\citenamefont{Bernholc et~al.}(2008)\citenamefont{Bernholc, Hodak,
  and Lu}}]{ISI:000257325900010}
\bibinfo{author}{\bibfnamefont{J.}~\bibnamefont{Bernholc}},
  \bibinfo{author}{\bibfnamefont{M.}~\bibnamefont{Hodak}}, \bibnamefont{and}
  \bibinfo{author}{\bibfnamefont{W.}~\bibnamefont{Lu}}, \bibinfo{journal}{J.
  Phys.: Condens. Matter} \textbf{\bibinfo{volume}{20}},
  \bibinfo{pages}{294205} (\bibinfo{year}{2008}).

\bibitem[{\citenamefont{Li et~al.}(2008{\natexlab{a}})\citenamefont{Li, Yin,
  Yao, and Zhao}}]{ISI:000256777300014}
\bibinfo{author}{\bibfnamefont{Y.}~\bibnamefont{Li}},
  \bibinfo{author}{\bibfnamefont{G.}~\bibnamefont{Yin}},
  \bibinfo{author}{\bibfnamefont{J.}~\bibnamefont{Yao}}, \bibnamefont{and}
  \bibinfo{author}{\bibfnamefont{J.}~\bibnamefont{Zhao}},
  \bibinfo{journal}{Comput. Mater. Sci.} \textbf{\bibinfo{volume}{42}},
  \bibinfo{pages}{638} (\bibinfo{year}{2008}{\natexlab{a}}).

\bibitem[{\citenamefont{Oetzel et~al.}(2008)\citenamefont{Oetzel, Preuss,
  Ortmann, Hannewald, and Bechstedt}}]{ISI:000256242300016}
\bibinfo{author}{\bibfnamefont{B.}~\bibnamefont{Oetzel}},
  \bibinfo{author}{\bibfnamefont{M.}~\bibnamefont{Preuss}},
  \bibinfo{author}{\bibfnamefont{F.}~\bibnamefont{Ortmann}},
  \bibinfo{author}{\bibfnamefont{K.}~\bibnamefont{Hannewald}},
  \bibnamefont{and}
  \bibinfo{author}{\bibfnamefont{F.}~\bibnamefont{Bechstedt}},
  \bibinfo{journal}{Phys. Stat. Sol. B} \textbf{\bibinfo{volume}{245}},
  \bibinfo{pages}{854} (\bibinfo{year}{2008}).

\bibitem[{\citenamefont{Jelinek et~al.}(2008)\citenamefont{Jelinek, Perez,
  Ortega, and Flores}}]{ISI:000254542800199}
\bibinfo{author}{\bibfnamefont{P.}~\bibnamefont{Jelinek}},
  \bibinfo{author}{\bibfnamefont{R.}~\bibnamefont{Perez}},
  \bibinfo{author}{\bibfnamefont{J.}~\bibnamefont{Ortega}}, \bibnamefont{and}
  \bibinfo{author}{\bibfnamefont{F.}~\bibnamefont{Flores}},
  \bibinfo{journal}{Phys. Rev. B} \textbf{\bibinfo{volume}{77}},
  \bibinfo{pages}{115447} (\bibinfo{year}{2008}).

\bibitem[{\citenamefont{Bredow et~al.}(2008)\citenamefont{Bredow, Tegenkamp,
  Pfnur, Meyer, Maslyuk, and Mertig}}]{Bredow2008}
\bibinfo{author}{\bibfnamefont{T.}~\bibnamefont{Bredow}},
  \bibinfo{author}{\bibfnamefont{C.}~\bibnamefont{Tegenkamp}},
  \bibinfo{author}{\bibfnamefont{H.}~\bibnamefont{Pfnur}},
  \bibinfo{author}{\bibfnamefont{J.}~\bibnamefont{Meyer}},
  \bibinfo{author}{\bibfnamefont{V.~V.} \bibnamefont{Maslyuk}},
  \bibnamefont{and} \bibinfo{author}{\bibfnamefont{I.}~\bibnamefont{Mertig}},
  \bibinfo{journal}{J. Chem. Phys.} \textbf{\bibinfo{volume}{128}},
  \bibinfo{pages}{064704} (\bibinfo{year}{2008}).

\bibitem[{\citenamefont{Hyldgaard}(2008)}]{Hyldgaard2008}
\bibinfo{author}{\bibfnamefont{P.}~\bibnamefont{Hyldgaard}},
  \bibinfo{journal}{Phys. Rev. B} \textbf{\bibinfo{volume}{78}},
  \bibinfo{pages}{165109} (\bibinfo{year}{2008}).

\bibitem[{\citenamefont{Kondo et~al.}(2008)\citenamefont{Kondo, Nara, Kin, and
  Ohno}}]{Kondo2008a}
\bibinfo{author}{\bibfnamefont{H.}~\bibnamefont{Kondo}},
  \bibinfo{author}{\bibfnamefont{J.}~\bibnamefont{Nara}},
  \bibinfo{author}{\bibfnamefont{H.}~\bibnamefont{Kin}}, \bibnamefont{and}
  \bibinfo{author}{\bibfnamefont{T.}~\bibnamefont{Ohno}},
  \bibinfo{journal}{Jap. J. Appl. Phys.} \textbf{\bibinfo{volume}{47}},
  \bibinfo{pages}{4792} (\bibinfo{year}{2008}).

\bibitem[{\citenamefont{Li et~al.}(2008{\natexlab{b}})\citenamefont{Li, Yin,
  Yao, and Zhao}}]{Li2008}
\bibinfo{author}{\bibfnamefont{Y.~W.} \bibnamefont{Li}},
  \bibinfo{author}{\bibfnamefont{G.~P.} \bibnamefont{Yin}},
  \bibinfo{author}{\bibfnamefont{J.~H.} \bibnamefont{Yao}}, \bibnamefont{and}
  \bibinfo{author}{\bibfnamefont{J.~W.} \bibnamefont{Zhao}},
  \bibinfo{journal}{Comput. Mater. Sci.} \textbf{\bibinfo{volume}{42}},
  \bibinfo{pages}{638} (\bibinfo{year}{2008}{\natexlab{b}}).

\bibitem[{\citenamefont{Mowbray et~al.}(2008)\citenamefont{Mowbray, Jones, and
  Thygesen}}]{Mowbray2008}
\bibinfo{author}{\bibfnamefont{D.~J.} \bibnamefont{Mowbray}},
  \bibinfo{author}{\bibfnamefont{G.}~\bibnamefont{Jones}}, \bibnamefont{and}
  \bibinfo{author}{\bibfnamefont{K.~S.} \bibnamefont{Thygesen}},
  \bibinfo{journal}{J. Chem. Phys.} \textbf{\bibinfo{volume}{128}},
  \bibinfo{pages}{111103} (\bibinfo{year}{2008}).

\bibitem[{\citenamefont{Smeu et~al.}(2008)\citenamefont{Smeu, Wolkow, and
  DiLabio}}]{Smeu2008}
\bibinfo{author}{\bibfnamefont{M.}~\bibnamefont{Smeu}},
  \bibinfo{author}{\bibfnamefont{R.~A.} \bibnamefont{Wolkow}},
  \bibnamefont{and} \bibinfo{author}{\bibfnamefont{G.~A.}
  \bibnamefont{DiLabio}}, \bibinfo{journal}{J. Chem. Phys.}
  \textbf{\bibinfo{volume}{129}}, \bibinfo{pages}{034707}
  (\bibinfo{year}{2008}).

\bibitem[{\citenamefont{Thygesen}(2008)}]{Thygesen2008a}
\bibinfo{author}{\bibfnamefont{K.~S.} \bibnamefont{Thygesen}},
  \bibinfo{journal}{Phys. Rev. Lett.} \textbf{\bibinfo{volume}{100}}
  (\bibinfo{year}{2008}).

\bibitem[{\citenamefont{Yoshizawa et~al.}(2008)\citenamefont{Yoshizawa, Tada,
  and Staykov}}]{Yoshizawa2008}
\bibinfo{author}{\bibfnamefont{K.}~\bibnamefont{Yoshizawa}},
  \bibinfo{author}{\bibfnamefont{T.}~\bibnamefont{Tada}}, \bibnamefont{and}
  \bibinfo{author}{\bibfnamefont{A.}~\bibnamefont{Staykov}},
  \bibinfo{journal}{Journal of the American Chemical Society}
  \textbf{\bibinfo{volume}{130}}, \bibinfo{pages}{9406} (\bibinfo{year}{2008}).

\bibitem[{\citenamefont{Zhao and Dunietz}(2008)}]{Zhao2008}
\bibinfo{author}{\bibfnamefont{Z.}~\bibnamefont{Zhao}} \bibnamefont{and}
  \bibinfo{author}{\bibfnamefont{B.~D.} \bibnamefont{Dunietz}},
  \bibinfo{journal}{J. Chem. Phys.} \textbf{\bibinfo{volume}{129}},
  \bibinfo{pages}{024702} (\bibinfo{year}{2008}).

\bibitem[{\citenamefont{Garcia-Lekue and Wang}(2009)}]{ISI:000267183200024}
\bibinfo{author}{\bibfnamefont{A.}~\bibnamefont{Garcia-Lekue}}
  \bibnamefont{and} \bibinfo{author}{\bibfnamefont{L.~W.} \bibnamefont{Wang}},
  \bibinfo{journal}{Comput. Mater. Sci.} \textbf{\bibinfo{volume}{45}},
  \bibinfo{pages}{1016} (\bibinfo{year}{2009}).

\bibitem[{\citenamefont{Gutierrez et~al.}(2009)\citenamefont{Gutierrez,
  Caetano, Woiczikowski, Kubar, Elstner, and Cuniberti}}]{Gutierrez2009}
\bibinfo{author}{\bibfnamefont{R.}~\bibnamefont{Gutierrez}},
  \bibinfo{author}{\bibfnamefont{R.~A.} \bibnamefont{Caetano}},
  \bibinfo{author}{\bibfnamefont{B.~P.} \bibnamefont{Woiczikowski}},
  \bibinfo{author}{\bibfnamefont{T.}~\bibnamefont{Kubar}},
  \bibinfo{author}{\bibfnamefont{M.}~\bibnamefont{Elstner}}, \bibnamefont{and}
  \bibinfo{author}{\bibfnamefont{G.}~\bibnamefont{Cuniberti}},
  \bibinfo{journal}{Phys. Rev. Lett.} \textbf{\bibinfo{volume}{102}}
  (\bibinfo{year}{2009}).

\bibitem[{\citenamefont{Lopez-Bezanilla
  et~al.}(2009)\citenamefont{Lopez-Bezanilla, Triozon, Latil, Blase, and
  Roche}}]{Lopez-Bezanilla2009}
\bibinfo{author}{\bibfnamefont{A.}~\bibnamefont{Lopez-Bezanilla}},
  \bibinfo{author}{\bibfnamefont{F.}~\bibnamefont{Triozon}},
  \bibinfo{author}{\bibfnamefont{S.}~\bibnamefont{Latil}},
  \bibinfo{author}{\bibfnamefont{X.}~\bibnamefont{Blase}}, \bibnamefont{and}
  \bibinfo{author}{\bibfnamefont{S.}~\bibnamefont{Roche}},
  \bibinfo{journal}{Nano Letters} \textbf{\bibinfo{volume}{9}},
  \bibinfo{pages}{940} (\bibinfo{year}{2009}).

\bibitem[{\citenamefont{Wang and Guo}(2009)}]{Wang2009}
\bibinfo{author}{\bibfnamefont{J.}~\bibnamefont{Wang}} \bibnamefont{and}
  \bibinfo{author}{\bibfnamefont{H.}~\bibnamefont{Guo}},
  \bibinfo{journal}{Phys. Rev. B} \textbf{\bibinfo{volume}{79}},
  \bibinfo{pages}{045119} (\bibinfo{year}{2009}).

\bibitem[{\citenamefont{Jacob et~al.}(2009)\citenamefont{Jacob, Haule, and
  Kotliar}}]{Jacob:prl:09}
\bibinfo{author}{\bibfnamefont{D.}~\bibnamefont{Jacob}},
  \bibinfo{author}{\bibfnamefont{K.}~\bibnamefont{Haule}}, \bibnamefont{and}
  \bibinfo{author}{\bibfnamefont{G.}~\bibnamefont{Kotliar}},
  \bibinfo{journal}{Phys. Rev. Lett.} \textbf{\bibinfo{volume}{103}},
  \bibinfo{pages}{016803} (\bibinfo{year}{2009}).

\bibitem[{\citenamefont{Frisch et~al.}(2003)\citenamefont{Frisch, Trucks,
  Schlegel, Scuseria, Robb, Cheeseman, Montgomery, ~, Vreven, Kudin
  et~al.}}]{Gaussian:03}
\bibinfo{author}{\bibfnamefont{M.~J.} \bibnamefont{Frisch}},
  \bibinfo{author}{\bibfnamefont{G.~W.} \bibnamefont{Trucks}},
  \bibinfo{author}{\bibfnamefont{H.~B.} \bibnamefont{Schlegel}},
  \bibinfo{author}{\bibfnamefont{G.~E.} \bibnamefont{Scuseria}},
  \bibinfo{author}{\bibfnamefont{M.~A.} \bibnamefont{Robb}},
  \bibinfo{author}{\bibfnamefont{J.~R.} \bibnamefont{Cheeseman}},
  \bibinfo{author}{\bibfnamefont{J.~A.} \bibnamefont{Montgomery}},
  \bibinfo{author}{\bibfnamefont{J.}~\bibnamefont{~}},
  \bibinfo{author}{\bibfnamefont{T.}~\bibnamefont{Vreven}},
  \bibinfo{author}{\bibfnamefont{K.~N.} \bibnamefont{Kudin}},
  \bibnamefont{et~al.}, \bibinfo{howpublished}{GAUSSIAN03, Revision B.01,
  Gaussian, Inc., Pittsburgh PA, 2003} (\bibinfo{year}{2003}).

\bibitem[{\citenamefont{Dovesi et~al.}(2006)\citenamefont{Dovesi, Saunders,
  Roetti, Orlando, Zicovich-Wilson, Pascale, Civalleri, Doll, Harrison, Bush
  et~al.}}]{Crystal:06}
\bibinfo{author}{\bibfnamefont{R.}~\bibnamefont{Dovesi}},
  \bibinfo{author}{\bibfnamefont{V.~R.} \bibnamefont{Saunders}},
  \bibinfo{author}{\bibfnamefont{C.}~\bibnamefont{Roetti}},
  \bibinfo{author}{\bibfnamefont{R.}~\bibnamefont{Orlando}},
  \bibinfo{author}{\bibfnamefont{C.~M.} \bibnamefont{Zicovich-Wilson}},
  \bibinfo{author}{\bibfnamefont{F.}~\bibnamefont{Pascale}},
  \bibinfo{author}{\bibfnamefont{B.}~\bibnamefont{Civalleri}},
  \bibinfo{author}{\bibfnamefont{K.}~\bibnamefont{Doll}},
  \bibinfo{author}{\bibfnamefont{N.~M.} \bibnamefont{Harrison}},
  \bibinfo{author}{\bibfnamefont{I.~J.} \bibnamefont{Bush}},
  \bibnamefont{et~al.}, \bibinfo{howpublished}{CRYSTAL06, Release 1.0.2,
  Theoretical Chemistry Group - Universita{'} Di Torino - Torino (Italy)}
  (\bibinfo{year}{2006}).

\bibitem[{\citenamefont{Ordej{\'{o}}n et~al.}(1996)\citenamefont{Ordej{\'{o}}n,
  Artacho, and Soler}}]{Ordejon:prb:96}
\bibinfo{author}{\bibfnamefont{P.}~\bibnamefont{Ordej{\'{o}}n}},
  \bibinfo{author}{\bibfnamefont{E.}~\bibnamefont{Artacho}}, \bibnamefont{and}
  \bibinfo{author}{\bibfnamefont{J.~M.} \bibnamefont{Soler}},
  \bibinfo{journal}{Phys. Rev. B} \textbf{\bibinfo{volume}{53}},
  \bibinfo{pages}{10441} (\bibinfo{year}{1996}).

\bibitem[{\citenamefont{Trouwborst et~al.}(2008)\citenamefont{Trouwborst,
  Huisman, Bakker, van~der Molen, and van Wees}}]{PhysRevLett.100.175502}
\bibinfo{author}{\bibfnamefont{M.~L.} \bibnamefont{Trouwborst}},
  \bibinfo{author}{\bibfnamefont{E.~H.} \bibnamefont{Huisman}},
  \bibinfo{author}{\bibfnamefont{F.~L.} \bibnamefont{Bakker}},
  \bibinfo{author}{\bibfnamefont{S.~J.} \bibnamefont{van~der Molen}},
  \bibnamefont{and} \bibinfo{author}{\bibfnamefont{B.~J.} \bibnamefont{van
  Wees}}, \bibinfo{journal}{Phys. Rev. Lett.} \textbf{\bibinfo{volume}{100}},
  \bibinfo{pages}{175502} (\bibinfo{year}{2008}).

\bibitem[{\citenamefont{Palacios et~al.}()\citenamefont{Palacios, Jacob,
  P{\'{e}}rez-Jim{\'{e}}nez, Fabi{\'{a}}n, Louis, and Verg{\'{e}}s}}]{ALACANT}
\bibinfo{author}{\bibfnamefont{J.~J.} \bibnamefont{Palacios}},
  \bibinfo{author}{\bibfnamefont{D.}~\bibnamefont{Jacob}},
  \bibinfo{author}{\bibfnamefont{A.~J.}
  \bibnamefont{P{\'{e}}rez-Jim{\'{e}}nez}},
  \bibinfo{author}{\bibfnamefont{E.~S.} \bibnamefont{Fabi{\'{a}}n}},
  \bibinfo{author}{\bibfnamefont{E.}~\bibnamefont{Louis}}, \bibnamefont{and}
  \bibinfo{author}{\bibfnamefont{J.~A.} \bibnamefont{Verg{\'{e}}s}},
  \bibinfo{howpublished}{ALACANT ab-initio quantum transport package},
  \urlprefix\url{http://alacant.dfa.ua.es}.

\bibitem[{\citenamefont{Economou}(1970)}]{Economou:book:83}
\bibinfo{author}{\bibfnamefont{E.~N.} \bibnamefont{Economou}},
  \emph{\bibinfo{title}{Green{'}s functions in Quantum Physics}},
  no.~\bibinfo{number}{7} in \bibinfo{series}{Springer Series in Solid State
  Physics} (\bibinfo{publisher}{Springer},
  \bibinfo{address}{Berlin-Heidelberg-New York-Tokyo}, \bibinfo{year}{1970}).

\bibitem[{\citenamefont{Landauer}(1970)}]{Landauer:philmag:70}
\bibinfo{author}{\bibfnamefont{R.}~\bibnamefont{Landauer}},
  \bibinfo{journal}{Philos. Mag.} \textbf{\bibinfo{volume}{21}},
  \bibinfo{pages}{863} (\bibinfo{year}{1970}).

\bibitem[{\citenamefont{Caroli et~al.}(1971)\citenamefont{Caroli, Combescot,
  and Dederichs}}]{Caroli:jphysc:71}
\bibinfo{author}{\bibfnamefont{C.}~\bibnamefont{Caroli}},
  \bibinfo{author}{\bibfnamefont{R.}~\bibnamefont{Combescot}},
  \bibnamefont{and}
  \bibinfo{author}{\bibfnamefont{P.}~\bibnamefont{Dederichs}},
  \bibinfo{journal}{J. Phys. C: Sol. State Phys.} \textbf{\bibinfo{volume}{4}},
  \bibinfo{pages}{916} (\bibinfo{year}{1971}).

\bibitem[{\citenamefont{Burke et~al.}(2006)\citenamefont{Burke, Koentopp, and
  Evers}}]{Burke:prb:06}
\bibinfo{author}{\bibfnamefont{K.}~\bibnamefont{Burke}},
  \bibinfo{author}{\bibfnamefont{M.}~\bibnamefont{Koentopp}}, \bibnamefont{and}
  \bibinfo{author}{\bibfnamefont{F.}~\bibnamefont{Evers}},
  \bibinfo{journal}{Phys. Rev. B} \textbf{\bibinfo{volume}{73}},
  \bibinfo{pages}{121403} (\bibinfo{year}{2006}).

\bibitem[{\citenamefont{Mehl and Papaconstantopoulos}()}]{Papacon-web}
\bibinfo{author}{\bibfnamefont{M.~J.} \bibnamefont{Mehl}} \bibnamefont{and}
  \bibinfo{author}{\bibfnamefont{D.~A.} \bibnamefont{Papaconstantopoulos}},
  \urlprefix\url{http://cst-www.nrl.navy.mil/bind/}.

\bibitem[{\citenamefont{Leonard and Tersoff}(1999)}]{Leonard:prl:99}
\bibinfo{author}{\bibfnamefont{F.}~\bibnamefont{Leonard}} \bibnamefont{and}
  \bibinfo{author}{\bibfnamefont{J.}~\bibnamefont{Tersoff}},
  \bibinfo{journal}{Phys. Rev. Lett.} \textbf{\bibinfo{volume}{83}},
  \bibinfo{pages}{5174} (\bibinfo{year}{1999}).

\bibitem[{foo()}]{footnote1}
\bibinfo{note}{In the case of a non-orthogonal basis set there is actually an
  ambiguity in the exact definition of the electronic charge of a subspace and
  hence also in the definition of the corresponding electron density.}

\bibitem[{\citenamefont{Hurley et~al.}(1986)\citenamefont{Hurley, Pacios,
  Christiansen, Ross, and Ermler}}]{Hurley:jcp:86}
\bibinfo{author}{\bibfnamefont{M.~M.} \bibnamefont{Hurley}},
  \bibinfo{author}{\bibfnamefont{L.~F.} \bibnamefont{Pacios}},
  \bibinfo{author}{\bibfnamefont{P.~A.} \bibnamefont{Christiansen}},
  \bibinfo{author}{\bibfnamefont{R.~B.} \bibnamefont{Ross}}, \bibnamefont{and}
  \bibinfo{author}{\bibfnamefont{W.~C.} \bibnamefont{Ermler}},
  \bibinfo{journal}{J. Chem. Phys} \textbf{\bibinfo{volume}{84}},
  \bibinfo{pages}{6840} (\bibinfo{year}{1986}).

\bibitem[{\citenamefont{Jacob and Palacios}(2006)}]{Jacob:prb:06}
\bibinfo{author}{\bibfnamefont{D.}~\bibnamefont{Jacob}} \bibnamefont{and}
  \bibinfo{author}{\bibfnamefont{J.~J.} \bibnamefont{Palacios}},
  \bibinfo{journal}{Phys. Rev. B} \textbf{\bibinfo{volume}{73}},
  \bibinfo{pages}{075429} (\bibinfo{year}{2006}).

\bibitem[{\citenamefont{Yanson and Ruitenbeek}(1995)}]{Yanson:prl:97}
\bibinfo{author}{\bibfnamefont{A.~I.} \bibnamefont{Yanson}} \bibnamefont{and}
  \bibinfo{author}{\bibfnamefont{J.~M.} \bibnamefont{Ruitenbeek}},
  \bibinfo{journal}{Phys.\ Rev.\ Lett.} \textbf{\bibinfo{volume}{79}},
  \bibinfo{pages}{2157} (\bibinfo{year}{1995}).

\bibitem[{\citenamefont{Hasmy et~al.}(2005)\citenamefont{Hasmy,
  P{\'{e}}rez-Jim{\'{e}}nez, Palacios, Garc{\'{i}}a-Mochales,
  Costa-Kr{\"{a}}mer, D{\'{i}}az, Medina, and Serena}}]{PhysRevB.72.245405}
\bibinfo{author}{\bibfnamefont{A.}~\bibnamefont{Hasmy}},
  \bibinfo{author}{\bibfnamefont{A.~J.}
  \bibnamefont{P{\'{e}}rez-Jim{\'{e}}nez}},
  \bibinfo{author}{\bibfnamefont{J.~J.} \bibnamefont{Palacios}},
  \bibinfo{author}{\bibfnamefont{P.}~\bibnamefont{Garc{\'{i}}a-Mochales}},
  \bibinfo{author}{\bibfnamefont{J.~L.} \bibnamefont{Costa-Kr{\"{a}}mer}},
  \bibinfo{author}{\bibfnamefont{M.}~\bibnamefont{D{\'{i}}az}},
  \bibinfo{author}{\bibfnamefont{E.}~\bibnamefont{Medina}}, \bibnamefont{and}
  \bibinfo{author}{\bibfnamefont{P.~A.} \bibnamefont{Serena}},
  \bibinfo{journal}{Phys. Rev. B} \textbf{\bibinfo{volume}{72}},
  \bibinfo{pages}{245405} (\bibinfo{year}{2005}).

\bibitem[{\citenamefont{Cuevas et~al.}(1998)\citenamefont{Cuevas, Levy~Yeyati,
  Mart\'\i{}n-Rodero, Rubio~Bollinger, Untiedt, and
  Agra\"\i{}t}}]{Cuevas:prl:98:81}
\bibinfo{author}{\bibfnamefont{J.~C.} \bibnamefont{Cuevas}},
  \bibinfo{author}{\bibfnamefont{A.}~\bibnamefont{Levy~Yeyati}},
  \bibinfo{author}{\bibfnamefont{A.}~\bibnamefont{Mart\'\i{}n-Rodero}},
  \bibinfo{author}{\bibfnamefont{G.}~\bibnamefont{Rubio~Bollinger}},
  \bibinfo{author}{\bibfnamefont{C.}~\bibnamefont{Untiedt}}, \bibnamefont{and}
  \bibinfo{author}{\bibfnamefont{N.}~\bibnamefont{Agra\"\i{}t}},
  \bibinfo{journal}{Phys.\ Rev.\ Lett.} \textbf{\bibinfo{volume}{81}},
  \bibinfo{pages}{2990} (\bibinfo{year}{1998}).

\bibitem[{\citenamefont{Jacob et~al.}(2005)\citenamefont{Jacob,
  Fern{\'{a}}ndez-Rossier, and Palacios}}]{Jacob:prb:05}
\bibinfo{author}{\bibfnamefont{D.}~\bibnamefont{Jacob}},
  \bibinfo{author}{\bibfnamefont{J.}~\bibnamefont{Fern{\'{a}}ndez-Rossier}},
  \bibnamefont{and} \bibinfo{author}{\bibfnamefont{J.~J.}
  \bibnamefont{Palacios}}, \bibinfo{journal}{Phys. Rev. B}
  \textbf{\bibinfo{volume}{71}}, \bibinfo{pages}{220403}
  (\bibinfo{year}{2005}).

\bibitem[{\citenamefont{Calvo et~al.}(2008)\citenamefont{Calvo, Caturla, Jacob,
  Untiedt, and Palacios}}]{Calvo:ieee:08}
\bibinfo{author}{\bibfnamefont{M.~R.} \bibnamefont{Calvo}},
  \bibinfo{author}{\bibfnamefont{M.~J.} \bibnamefont{Caturla}},
  \bibinfo{author}{\bibfnamefont{D.}~\bibnamefont{Jacob}},
  \bibinfo{author}{\bibfnamefont{C.}~\bibnamefont{Untiedt}}, \bibnamefont{and}
  \bibinfo{author}{\bibfnamefont{J.~J.} \bibnamefont{Palacios}},
  \bibinfo{journal}{IEEE Transactions in Nanotechnology}
  \textbf{\bibinfo{volume}{7}}, \bibinfo{pages}{165} (\bibinfo{year}{2008}).

\bibitem[{\citenamefont{Jacob et~al.}(2008)\citenamefont{Jacob,
  Fernandez-Rossier, and Palacios}}]{Jacob08}
\bibinfo{author}{\bibfnamefont{D.}~\bibnamefont{Jacob}},
  \bibinfo{author}{\bibfnamefont{J.}~\bibnamefont{Fernandez-Rossier}},
  \bibnamefont{and} \bibinfo{author}{\bibfnamefont{J.~J.}
  \bibnamefont{Palacios}}, \bibinfo{journal}{Phys. Rev. B}
  \textbf{\bibinfo{volume}{77}}, \bibinfo{pages}{165412}
  (\bibinfo{year}{2008}).

\bibitem[{\citenamefont{Munoz-Rojas et~al.}(2006)\citenamefont{Munoz-Rojas,
  Jacob, Fernandez-Rossier, and Palacios}}]{Munoz-Rojas06-1}
\bibinfo{author}{\bibfnamefont{F.}~\bibnamefont{Munoz-Rojas}},
  \bibinfo{author}{\bibfnamefont{D.}~\bibnamefont{Jacob}},
  \bibinfo{author}{\bibfnamefont{J.}~\bibnamefont{Fernandez-Rossier}},
  \bibnamefont{and} \bibinfo{author}{\bibfnamefont{J.~J.}
  \bibnamefont{Palacios}}, \bibinfo{journal}{Phys. Rev. B}
  \textbf{\bibinfo{volume}{74}}, \bibinfo{pages}{195417}
  (\bibinfo{year}{2006}).

\bibitem[{\citenamefont{Szabo and Ostlund}(1989)}]{Szabo:book:89}
\bibinfo{author}{\bibfnamefont{A.}~\bibnamefont{Szabo}} \bibnamefont{and}
  \bibinfo{author}{\bibfnamefont{N.~S.} \bibnamefont{Ostlund}},
  \emph{\bibinfo{title}{Modern Quantum Chemistry}}
  (\bibinfo{publisher}{McGraw-Hill}, \bibinfo{address}{New York},
  \bibinfo{year}{1989}).

\end{thebibliography}

\end{document}